\documentclass[fleqn,usenatbib]{mnras}

\usepackage[T1]{fontenc}

\usepackage{graphicx}
\usepackage{xspace}
\usepackage{subcaption}
\usepackage{etoolbox} 
\usepackage{amssymb}  
\usepackage{xcolor}   
\usepackage{upgreek}
\usepackage{hyperref}
\usepackage{array,multirow,pbox} 
\usepackage{float}
\usepackage{bm} 

\newcommand{\orcid}[2]{\href{http://orcid.org/#2}{#1}}
\def\h4{\ensuremath{h_4}\xspace}
\newcommand{\mstar}{\ensuremath{M_\star}\xspace}
\newcommand{\msun}{\ensuremath{\mathrm{M_\odot}}\xspace}
\newcommand{\re}{\ensuremath{R_\mathrm{e}}\xspace}
\newcommand{\vse}{\ensuremath{(V/\sigma)_\mathrm{e}}\xspace}
\newcommand{\sigap}{\ensuremath{\sigma_\mathrm{ap}}\xspace}
\newcommand{\kms}{\ensuremath{\mathrm{km\,s^{-1}}}\xspace}

\definecolor{ForestGreen}{HTML}{2e8b21}

\makeatletter
\newcommand\sendemail[3]{
\edef\@tempa{mailto:#1?subject=#2 }%
\edef\@tempb{\expandafter\html@spaces\@tempa\@empty}%
\href{\@tempb}{#3}}

\catcode\%=11
\def\html@spaces#1 #2{#1
\catcode\%=14
\makeatother

\newcommand{\ppxf}{{\sc ppxf}\xspace}

\title[Integrated \h4 in star-forming/quiescent galaxies]{Different higher-order kinematics between star-forming and quiescent galaxies based on the SAMI, MAGPI and LEGA-C surveys}

\author[\orcid{F. D'Eugenio}{0000-0003-2388-8172}~et al.]{\parbox{\textwidth}{
\orcid{Francesco D'Eugenio}{0000-0003-2388-8172}$^{\hyperlink{aff1}{1},\hyperlink{aff2}{2},\hyperlink{aff3}{3}}$\thanks{E-mail: francesco.deugenio@gmail.com},
\orcid{Arjen van der Wel}{0000-0002-5027-0135}$^{\hyperlink{aff3}{3}}$,
\orcid{Caro Derkenne}{0000-0003-3474-3542}$^{\hyperlink{aff4}{4},\hyperlink{aff5}{5}}$,
\orcid{Josha van Houdt}{0000-0003-1888-3705}$^{\hyperlink{aff6}{6}}$,
\orcid{Rachel Bezanson}{0000-0001-5063-8254}$^{\hyperlink{aff7}{7}}$,
\orcid{Edward N. Taylor}{0000-0002-5522-9107}$^{\hyperlink{aff8}{8}}$,
\orcid{Jesse van de Sande}{0000-0003-2552-0021}$^{\hyperlink{aff9}{9},\hyperlink{aff5}{5}}$,
\orcid{William M. Baker}{0000-0003-0215-1104}$^{\hyperlink{aff1}{1},\hyperlink{aff2}{2}}$,
\orcid{Eric F. Bell}{0000-0002-5564-9873}$^{\hyperlink{aff10}{10}}$,
\orcid{Joss Bland-Hawthorn}{0000-0001-7516-4016}$^{\hyperlink{aff9}{9},\hyperlink{aff5}{5}}$,
\orcid{Asa F. L. Bluck}{0000-0001-6395-4504}$^{\hyperlink{aff11}{11}}$,
\orcid{Sarah Brough}{0000-0002-9796-1363}$^{\hyperlink{aff12}{12}}$,
\orcid{Julia J. Bryant}{0000-0003-1627-9301}$^{\hyperlink{aff9}{9},\hyperlink{aff13}{13},\hyperlink{aff5}{5}}$,
\orcid{Matthew Colless}{0000-0001-9552-8075}$^{\hyperlink{aff14}{14},\hyperlink{aff5}{5}}$,
\orcid{Luca Cortese}{0000-0002-7422-9823}$^{\hyperlink{aff15}{15},\hyperlink{aff5}{5}}$,
\orcid{Scott M. Croom}{0000-0003-2880-9197}$^{\hyperlink{aff9}{9},\hyperlink{aff5}{5}}$,
\orcid{Pieter van Dokkum}{0000-0002-8282-9888}$^{\hyperlink{aff16}{16}}$,
\orcid{Deanne Fisher}{0000-0003-0645-5260}$^{\hyperlink{aff8}{8},\hyperlink{aff5}{5}}$,
\orcid{Caroline Foster}{0000-0003-0247-1204}$^{\hyperlink{aff12}{12},\hyperlink{aff5}{5}}$,
\orcid{Amelia Fraser-McKelvie}{0000-0001-9557-5648}$^{\hyperlink{aff15}{15},\hyperlink{aff5}{5}}$,
\orcid{Anna Gallazzi}{0000-0002-9656-1800}$^{\hyperlink{aff17}{17}}$,
\orcid{Anna de~Graaff}{0000-0002-2380-9801}$^{\hyperlink{aff18}{18},\hyperlink{aff6}{6}}$,
\orcid{Brent Groves}{0000-0002-9768-0246}$^{\hyperlink{aff14}{14}}$,
\orcid{Claudia del P. Lagos}{0000-0003-3021-8564}$^{\hyperlink{aff15}{15},\hyperlink{aff5}{5}}$,
\href{https://www.kicc.cam.ac.uk/staff/tobias-looser}{Tobias J. Looser}$^{\hyperlink{aff1}{1},\hyperlink{aff2}{2}}$,
\orcid{Roberto Maiolino}{0000-0002-4985-3819}$^{\hyperlink{aff1}{1},\hyperlink{aff2}{2},\hyperlink{aff19}{19}}$,
\orcid{Michael Maseda}{0000-0003-0695-4414}$^{\hyperlink{aff20}{20}}$,
\orcid{J. Trevor Mendel}{0000-0002-6327-9147}$^{\hyperlink{aff14}{14},\hyperlink{aff5}{5}}$,
\orcid{Angelos Nersesian}{0000-0001-6843-409X}$^{\hyperlink{aff3}{3}}$,
\orcid{Camilla Pacifici}{0000-0003-4196-0617}$^{\hyperlink{aff21}{21}}$,
\orcid{Joanna M. Piotrowska}{0000-0003-1661-2338}$^{\hyperlink{aff1}{1},\hyperlink{aff2}{2}}$,
\orcid{Adriano Poci}{0000-0003-4196-0617}$^{\hyperlink{aff22}{22},\hyperlink{aff4}{4}}$,
\href{https://www.usm.uni-muenchen.de/~rhea/index.html}{Rhea-Silvia Remus}$^{\hyperlink{aff23}{23}}$,
\orcid{Gauri Sharma}{0000-0002-6070-2851}$^{\hyperlink{aff24}{24}}$,
\orcid{Sarah M. Sweet}{0000-0002-1576-2505}$^{\hyperlink{aff25}{25}}$,
\orcid{Sabine Thater}{0000-0003-1820-2041}$^{\hyperlink{aff26}{26}}$,
\orcid{Kim-Vy Tran}{0000-0001-9208-2143}$^{\hyperlink{aff12}{12},\hyperlink{aff5}{5},\hyperlink{aff27}{27}}$,
\orcid{Hannah {\"U}bler}{0000-0003-4891-0794}$^{\hyperlink{aff1}{1},\hyperlink{aff2}{2}}$,
\orcid{Lucas M. Valenzuela}{0000-0002-7972-9675}$^{\hyperlink{aff23}{23}}$,
\orcid{Emily Wisnioski}{0000-0003-1657-7878}$^{\hyperlink{aff14}{14},\hyperlink{aff5}{5}}$
and
\orcid{Stefano Zibetti}{0000-0003-1734-8356}$^{\hyperlink{aff17}{17}}$
}
\vspace{0.4cm}
\\
\parbox{\textwidth}{
\hypertarget{aff1}{$^{1}$}Kavli Institute for Cosmology, University of Cambridge, Madingley Road, Cambridge, CB3 0HA, United Kingdom\\
\hypertarget{aff2}{$^{2}$}Cavendish Laboratory - Astrophysics Group, University of Cambridge, 19 JJ Thomson Avenue, Cambridge, CB3 0HE, United Kingdom\\
\hypertarget{aff3}{$^{3}$}Sterrenkundig Observatorium, Universiteit Gent, Krijgslaan 281 S9, B-9000 Gent, Belgium\\
\hypertarget{aff4}{$^{4}$}Research Centre for Astronomy, Astrophysics and Astrophotonics, School of Mathematical and Physical Sciences, Macquarie University, Sydney, NSW 2109, Australia\\
\hypertarget{aff5}{$^{5}$}ARC Centre of Excellence for All Sky Astrophysics in 3 Dimensions (ASTRO 3D), Australia\\
\hypertarget{aff6}{$^{6}$}Max-Planck-Institut f\"ur Astronomie, K\"onigstuhl 17, D-69117, Heidelberg, Germany\\
\hypertarget{aff7}{$^{7}$}Department of Physics and Astronomy and PITT PACC, University of Pittsburgh, Pittsburgh, PA 15260, USA\\
\hypertarget{aff8}{$^{8}$}Centre for Astrophysics and Supercomputing, Swinburne University of Technology, Hawthorn, VIC 3122, Australia\\
\hypertarget{aff9}{$^{9}$}Sydney Institute for Astronomy, School of Physics, The University of Sydney, NSW, 2006, Australia\\
\hypertarget{aff10}{$^{10}$}Department of Astronomy, University of Michigan, Ann Arbor, MI 48109, USA\\
\bigskip
\emph{\normalsize Remaining affiliations are listed at the end of the paper.}
}
}

\date{Accepted XXX. Received YYY; in original form ZZZ}

\pubyear{2022}

\begin{document}
\label{firstpage}
\pagerange{\pageref{firstpage}--\pageref{lastpage}}
\maketitle

\begin{abstract}
We present the first statistical study of spatially integrated non-Gaussian stellar
kinematics spanning 7~Gyr in cosmic time. We use deep, rest-frame optical spectroscopy
of massive galaxies (stellar mass $M_\star > 10^{10.5} \, \mathrm{M_\odot}$)
at redshifts $z=0.05$, 0.3 and 0.8 from the SAMI, MAGPI and LEGA-C surveys,
to measure the excess kurtosis $h_4$ of the stellar velocity distribution, the
latter parametrised as a Gauss-Hermite series. We find that at all redshifts
where we have large enough samples, $h_4$ anti-correlates with the ratio between
rotation and dispersion, highlighting the physical connection between these two
kinematic observables.
In addition, and independently from the anti-correlation with rotation-to-dispersion
ratio, we also find a correlation between $h_4$ and $M_\star$, potentially connected
to the assembly history of galaxies. In contrast, after controlling for mass, we find
no evidence of independent correlation between $h_4$ and aperture velocity
dispersion or galaxy size. These results hold for both star-forming and
quiescent galaxies.
For quiescent galaxies, $h_4$ also correlates with projected shape, even after
controlling for the rotation-to-dispersion ratio.
At any given redshift, star-forming galaxies have lower $h_4$ compared to
quiescent galaxies, highlighting the link between kinematic structure and
star-forming activity.
\end{abstract}




\section{Introduction}

Galaxies form stars in a fairly regular manner, with their star-formation rate
proportional to their stellar mass
\citep{brinchmann+2004, noeske+2007}.
Below this `star-forming sequence', lies a continuous distribution of galaxies
with lower \citep[or undetected][]{feldmann2017, eales+2018} star-formation
rate. The star-forming sequence thus enables
us to divide galaxies between `star-forming' and `quiescent', a classification
that maps onto other physical properties of galaxies. If we consider galaxies at
or above $10^{10} \, \msun$\footnote{Below this stellar mass threshold,
star-forming galaxies appear to have rotation-to-dispersion ratios that decrease
with decreasing stellar mass \citep{falcon-barroso+2019}.}, star-forming galaxies have
flatter intrinsic shapes \citep{sandage+1970, lambas+1992}
less-concentrated light profiles \citep[e.g.][]{driver+2006, simard+2011,
kelvin+2012, bell+2012, mendel+2014}
lower bulge fractions \citep[e.g.][]{cameron+2009, simard+2011, mendel+2014,
bluck+2014},
lower velocity dispersion \citep[e.g.][]{bell+2012, bluck+2016, falcon-barroso+2019}
and higher rotation-to-dispersion ratios \citep[$V/\sigma$; e.g.][]{
vandesande+2018, graham+2018, falcon-barroso+2019}.
The overlap between star-formation status and other galaxy properties gives
us clues on what drives galaxy quenching. For example, the fact that quiescent
galaxies have larger bulge mass and higher stellar velocity dispersion has been
interpreted as evidence for quenching due to feedback from
supermassive black holes \citep{brownson+2022, piotrowska+2022, bluck+2022}.

Kinematically, star-forming galaxies have larger $V/\sigma$, but otherwise
form a continuous distribution with quiescent galaxies, most of which
\citep[60-80~per cent,][]{cappellari+2011b, vandesande+2017b} are also `fast
rotators' \citep{emsellem+2007, cappellari+2007}, albeit with lower average
$V/\sigma$. At the high-mass end of the quiescent population, we find a
distinct kinematic family of `slow rotators' \citep{brough+2007, emsellem+2011,
graham+2018, vandesande+2021a}, characterised by round or
triaxial intrinsic shapes and no net rotation.

These classifications rely primarily on modelling the stellar velocity distribution
as a Gaussian, completely specified by its first three moments\footnote{But note
the non-Gaussian analysis of \citet{krajnovic+2011}.}. However,
stellar velocity distributions are known to deviate from a Gaussian
\citep{bender1990, rix+white1992}. These deviations contain information about the
assembly history of galaxies \citep{naab+2014}; they can be
measured by parametrising the velocity distribution as a Gauss-Hermite
series \citep{vandermarel+franx1993, gerhard1993}; the Gauss-Hermite
coefficients effectively measure the higher-order moments of the distribution.
The coefficient of the fourth-order term of the Hermite polynomial, \h4, is
related to the excess kurtosis of the
velocity distribution: $\h4 > 0$ indicates a leptokurtic distribution (with
broader wings compared to a Gaussian), while $\h4 < 0$ corresponds to a
platykurtic distribution (with less prominent wings). Physically, positive \h4
is associated with radial anisotropy, which causes a lack of stars near the
local circular velocity \citep[e.g.][]{gerhard1993}. Given that \textit{in-situ}
star formation occurs predominantly in discs,
radial anisotropy is linked to gas-poor mergers and \textit{ex-situ} stars, and
should provide insight on the assembly history of a galaxy, at least up until the
last major merger \citep[which may erase the previous kinematic record,][]{lynden-bell1967}.
\citet{vandesande+2017a} have used spatially resolved higher-order kinematics
from the SAMI Galaxy Survey \citep{croom+2012} to investigate the assembly history
of nearby galaxies and to match it to the predictions of numerical simulations
\citep{naab+2014}. However, \h4 also contains information about other kinematic
structures, like bars \citep{seidel+2015, li+2018}, including peanut-shaped bulges
\citep{debattista+2005, mendez-abreu+2008}, so the physical interpretation of
the results is not straightforward.

Because measurements of \h4 require higher signal-to-noise data, until now they
have been restricted to relatively nearby galaxies \citep[$z\lesssim 0.1$, e.g.][]{
emsellem+2007, vandesande+2017a}. However, new large, ultra-deep spectroscopy
surveys enable us, for the first time, to extend these measurements to larger
look-back times.

In this work, we use high-quality optical spectroscopy from the local SAMI
Galaxy Survey, from the MAGPI Survey \citep[][redshift $z=0.3$]{foster+2021} and
from the LEGA-C Survey \citep[][$z=0.8$]{vanderwel+2014}, to investigate the
link between star-forming status and higher-order kinematics. We start by showing
the relation between spatially resolved \h4 and the value integrated inside an
aperture (\S~\ref{s.tmods}). We then introduce
the data (\S~\ref{s.das}) and the sample (\S~\ref{s.samp}). In \S~\ref{s.r}, we
show that \h4 correlates primarily with $V/\sigma$ and stellar mass; in addition,
we also find that at any redshift, star-forming galaxies have lower \h4 than
quiescent galaxies. We conclude this work with a discussion (\S~\ref{s.d}) and
with a summary of our findings (\S~\ref{s.c}).

Throughout this article, we assume a flat $\Lambda$CDM cosmology with $H_0 = 70 \;
\kms \, \mathrm{Mpc}^{-1}$ and $\Omega_m = 0.3$. 
All stellar mass measurements assume a Chabrier initial mass function
\citep{chabrier2003}.

\section{Local vs integrated measurements}\label{s.tmods}

Given the signal-to-noise ratio ($S/N$) of some of our data (see \S~\ref{s.das}), we
propose to obtain only \textit{integrated} \h4, measured from adding the
light inside a given aperture. To understand the relation between this measurement
of \h4 and the local, spatially resolved value used in the literature, we use a toy
kinematic model.
The model consists of a thin-disc with arctan velocity field, uniform velocity
dispersion and an exponential light profile\footnote{Using S{\`e}rsic models
with index $n$ higher than one adds weight to the central, low-velocity regions;
for this reason, the exponential model is the most conservative in the
S{\`e}rsic family with $n\geq1$.}.
The velocity field has root-mean
square velocity $v_\mathrm{rms} = 300$~\kms and uniform value of
the spatially resolved \h4, which we call $h_4(\mathrm{local})$, as opposed to
$h_4(\mathrm{integrated})$. We
create a grid by varying the rotation-to-dispersion ratio \vse (calculated at
one effective radius \re\footnote{This measurement is conceptually similar to
the corresponding measurement for the LEGA-C survey,
\S~\ref{s.das.ss.anc.sss.vse}.}) and
apparent axis ratio $q$, then add the stellar continuum
using the C3K/MIST library \citep{conroy+2019, choi+2016}, convolved with the
appropriate line-of-sight velocity distribution (LOSVD) at each spaxel. From
these mock datacubes, we extract the 1-d spectrum from an elliptical aperture
centred at one \re, as we did for SAMI (see \S~\ref{s.das.ss.ds.sss.sami}). We
then measure the integrated \h4 using \href{https://pypi.org/project/ppxf/}{\ppxf},
the penalised pixel fitting algorithm of \citet{cappellari2017, cappellari2022}.
We created seven grids of models, corresponding to seven values of the input,
$\h4(\mathrm{local})$: -0.03, -0.015, 0, 0.015, 0.03, 0.045 and 0.06. These
values are chosen to span the range of values we measure in real data (\S~\ref{s.r}).

\begin{figure}
  \includegraphics[type=pdf,ext=.pdf,read=.pdf,width=1.\columnwidth]{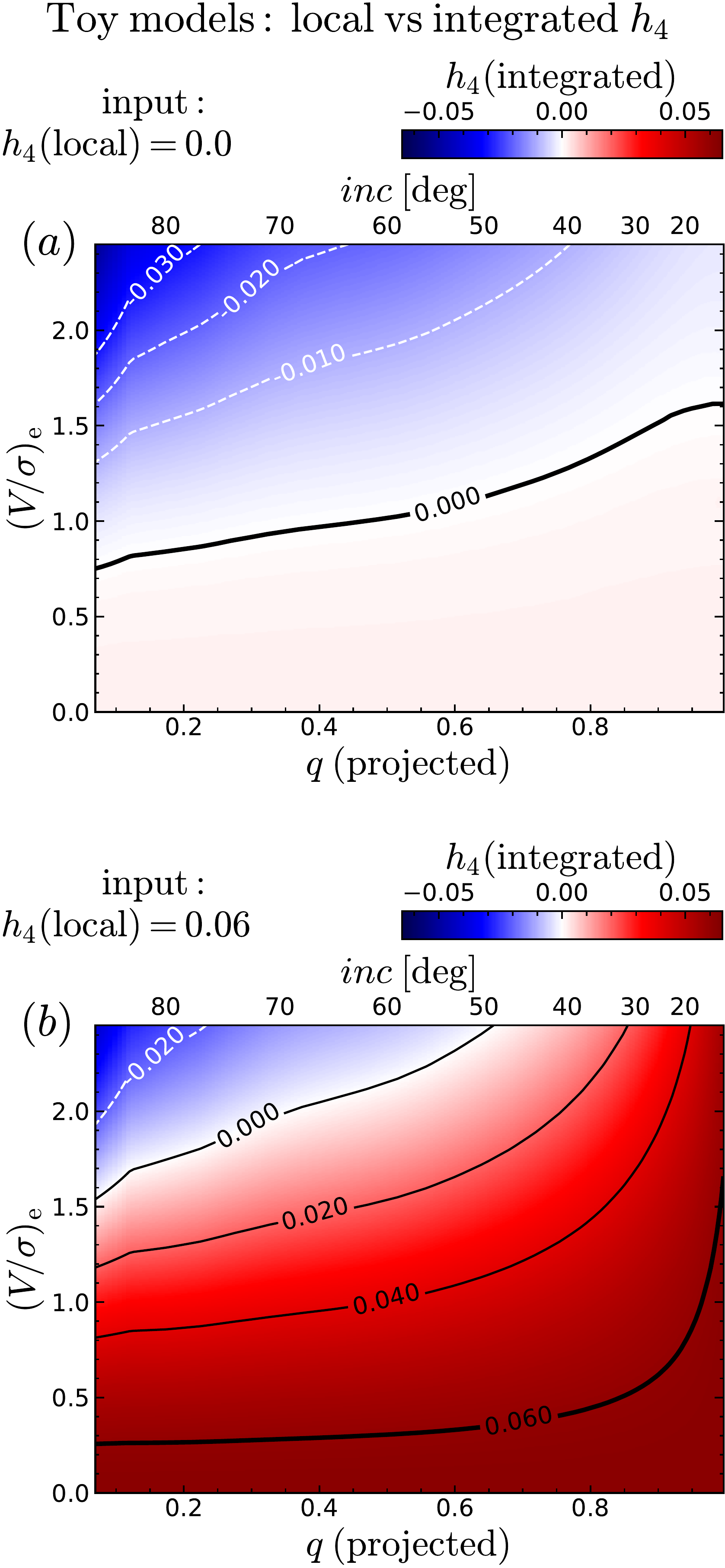}
  {\phantomsubcaption\label{f.tmods.models.a}
   \phantomsubcaption\label{f.tmods.models.b}}
  \caption{Spatially integrated \h4 as a function of \vse and $q$ for our toy
  models. Panel~\subref{f.tmods.models.a} shows the model with local
  (input) $\h4 = 0$; panel~\subref{f.tmods.models.b} shows the model
  with local $\h4 = 0.06$. The dashed/solid contours show loci of
  constant negative/non-negative $\h4 \, (\mathrm{integrated})$; the thick solid
  line is the locus where 
  $\h4 \, (\mathrm{integrated}) = \h4 \, (\mathrm{local})$. For round shapes
  and/or low \vse, $\h4 \, (\mathrm{integrated})$ reflects
  $\h4 \, (\mathrm{local})$; elsewhere, 
  $\h4 \, (\mathrm{integrated})$ also depends on $q$ and \vse; these trends
  are quantified in Fig.~\ref{f.tmods.resvsint} and Table~\ref{t.tmods.mod}; see
  Fig.~\ref{f.r.h4_qvs} for a comparison to the observations (but note that
  --- in our observations --- $\h4 \, (\mathrm{local})$ is unknown).}\label{f.tmods.models}
\end{figure}

The results are shown in Fig.~\ref{f.tmods.models}, where the colour (and
contour lines) show the value of the spatially integrated \h4 --- what we
measure for real data in \S~\ref{s.das} --- as a function of the model \vse and
$q$ (on the top axis, we also show the model inclination $inc$).
The two panels differ by the input value of the local \h4: 0 for
panel~\subref{f.tmods.models.a} and 0.06 for panel~\subref{f.tmods.models.b}. 
It is clear that integrated \h4 does not trace only the local \h4,
but conflates together information from $inc$ and \vse too. At the same time,
the fact that the two figures have largely different colours shows that local
\h4 is reflected in the value of integrated \h4.
In the figures, the
locus where integrated and local \h4 are the same is traced by the thick,
solid line; below this line, integrated \h4 tends to be marginally
larger than local \h4, but well within the observational measurements (which
we limit to be $u(\h4) < 0.05$, see \S~\ref{s.samp.ss.qcsel}). Above the line,
integrated \h4 reflects primarily \vse and inclination.

\begin{figure}
  \includegraphics[type=pdf,ext=.pdf,read=.pdf,width=1.\columnwidth]{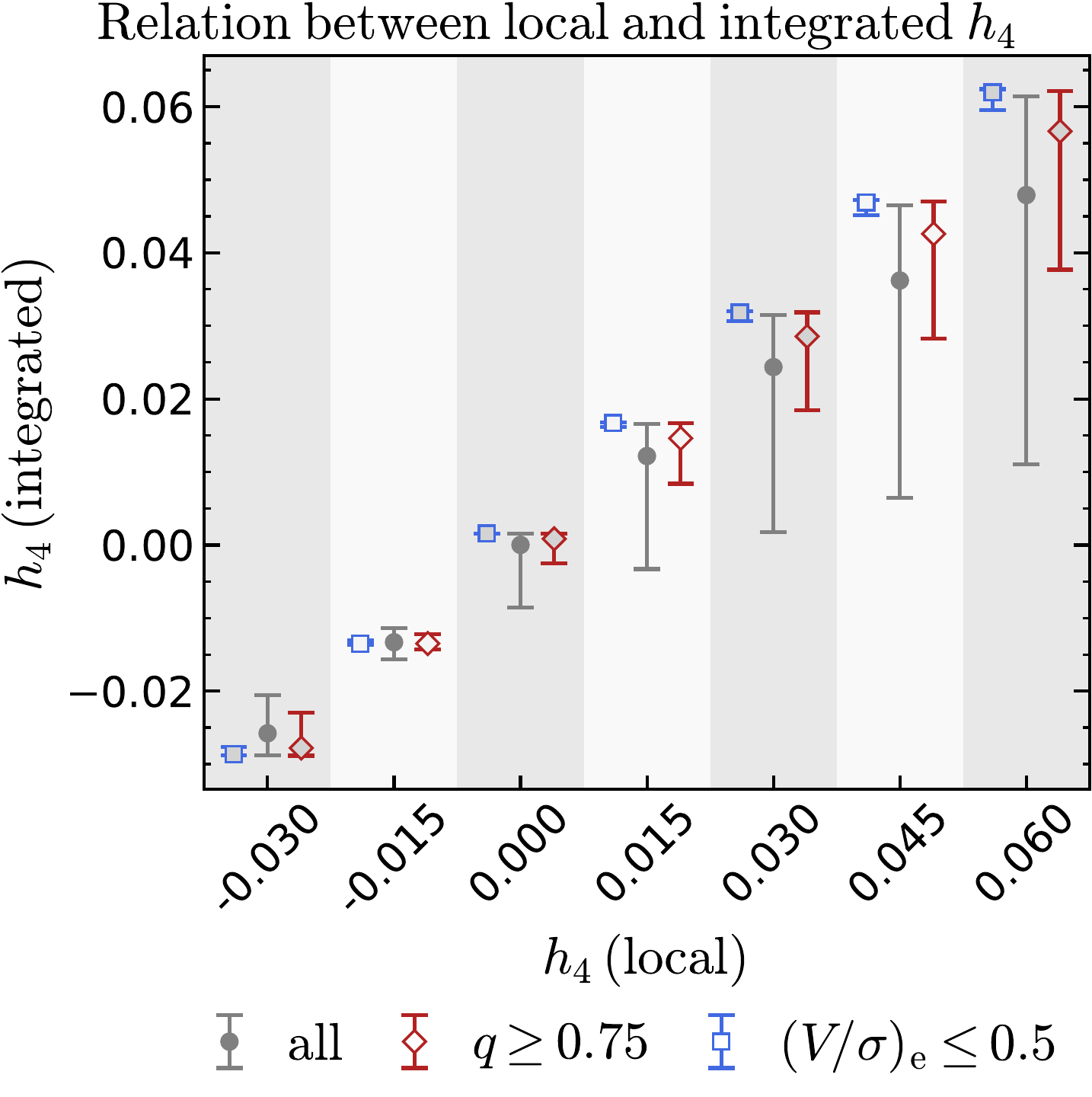}
  \caption{Statistically, $\h4 \, (\mathrm{integrated})$ reflects 
  $\h4 \, (\mathrm{local})$, as traced by the grey circles, which are the
  median values of $\h4 \, (\mathrm{integrated})$ over the \vse--$inc$ grid.
  Selecting models with large $q$ or low \vse reduces both the bias and the
  spread (red diamonds and blue squares, respectively; points inside the same
  shaded regions have the same local \h4: the symbols are offset horizontally
  for clarity).
  }\label{f.tmods.resvsint}
\end{figure}

In Fig.~\ref{f.tmods.resvsint} we consider all seven models, collapsing the grid
of \vse and $inc$: at any value of $\h4 \, (\mathrm{local})$, the grey circle
(and errorbars) represent the median (and 16--84\textsuperscript{th} percentiles)
of the measured $\h4 \; (\mathrm{integrated})$.
If we consider only models with modest rotation support ($\vse
< 0.5$; blue squares), integrated \h4 reflects local \h4 with high
fidelity (Table~\ref{t.tmods.mod}, row~2; the squares in Fig.~\ref{f.tmods.resvsint}
have been offset horizontally for clarity). Similar considerations apply to a
selection based
on apparent axis ratio $q$: rounder models ($q \geq 0.75$; red diamonds)
show a tighter relation than the rest of the models (see also
Table~\ref{t.tmods.mod}, row~4).

We quantify these correlations using the Spearman's rank correlation coefficient
$\rho$ (Table~\ref{t.tmods.mod}; all correlations are statistically significant).
While integrated \h4 correlates with all three of $q$, \vse and local \h4
(row~1), selecting galaxies with low \vse (row~2) or round galaxies (row~4)
reduces the correlations with $q$ and \vse (columns 3--4), while bringing the
correlation with local \h4 to $\rho > 0.9$ (column 5).

    \begin{table}
    \begin{center}
    
    \caption{Toy-model predictions for the correlations of integrated \h4
    with  each of axis ratio $q$,
    rotation-to-dispersion ratio \vse and spatially resolved \h4.
    Selecting round galaxies ($q \geq 0.75$) or galaxies with low
    \vse enhances the correlation between integrated and resolved \h4.
    }\label{t.tmods.mod}
    
    \begin{tabular}{lcccc}
    \hline
        & Subset
        & $q$
        & $(V/\sigma)_\mathrm{e}$
        & $h_4 \, (\mathrm{resolved})$ \\
    (1) & (2) & (3) & (4) & (5) \\
    \hline
    (1) & all                            & 0.26 & -0.51 & 0.68 \\
    (2) & $(V/\sigma)_\mathrm{e}\leq0.5$ & 0.12 & -0.12 & 0.98 \\
    (3) & $(V/\sigma)_\mathrm{e}>0.5$    & 0.33 & -0.49 & 0.63 \\
    (4) & $q\geq0.75$                    & 0.18 & -0.30 & 0.91 \\
    (5) & $q <  0.75$                    & 0.08 & -0.74 & 0.54 \\
    \hline
    \end{tabular}
    \end{center}
    \begin{flushleft}
    Columns: (1) row index (2)~subset of the models used; (3) Spearman's rank
    correlation
    coefficient $\rho$ between integrated \h4 and $q$; (4) same as (3), but for
    \vse; (5) same as (3), but for spatially resolved \h4.
    \end{flushleft}

    \end{table}

These models are only toy models, to help guide the
interpretation of our measurements. In particular, they do not capture the
kinematics of intrinsically round, dispersion-supported galaxies \citep[e.g.,
slow-rotator galaxies,][]{emsellem+2007, cappellari+2007}. It is clear, however,
that for such systems rotation cannot bias \h4, because there is little or no
rotation to start with.
Based on Fig.~\ref{f.tmods.resvsint}, we
expect integrated \h4 to correlate with $q$ and to anti-correlate with \vse.
However, if we select round and/or low-\vse galaxies, integrated \h4 reflects
the local value, which in turn is related to radial anisotropy
\citep{vandermarel+franx1993, gerhard1993}. In the rest of this article, we
generally refer to integrated \h4 simply as `\h4', but we will occasionally
use `integrated \h4' when spatially resolved \h4 is also relevant.

\section{Data}\label{s.das}

In this section, we start by presenting the data (\S~\ref{s.das.ss.ds}), which
we draw from three different surveys: the local SAMI
Galaxy Survey ($z \approx 0$, \S~\ref{s.das.ss.ds.sss.sami}), the MAGPI survey
($z \approx 0.3$, \S~\ref{s.das.ss.ds.sss.magpi}), and the LEGA-C survey
(redshift $z \approx 0.7$, \S~\ref{s.das.ss.ds.sss.legac}). Even though data
from these three surveys are not homogeneous, we only compare our measurements
\textit{within} surveys, not \textit{across} surveys --- the latter is the
subject of a future work. We then explain how
the one-dimensional (1-d) spectra are used to measure \h4
(\S~\ref{s.das.ss.hok}). Finally, in \S~\ref{s.das.ss.anc}, we describe ancillary
measurements obtained from the literature.

\subsection{Data sources}\label{s.das.ss.ds}

\subsubsection{The SAMI Galaxy Survey}\label{s.das.ss.ds.sss.sami}

The SAMI Galaxy Survey (hereafter simply: SAMI) is a large, optical Integral
Field Spectroscopy Survey of local galaxies ($0.04 < z < 0.095$), covering a
broad range of stellar masses ($10^7 < \mstar < 10^{12} \, \msun$),
morphologies and environments \citep[local environment density $0.1 < \Sigma_5 < 100 \,
\mathrm{Mpc^{-2}}$][]{bryant+2015, owers+2017}.
SAMI galaxies were observed with the Sydney-AAO Multi-object Integral field
spectroscopy instrument \citep[hereafter, the SAMI instrument;][]{croom+2012},
formerly placed at the prime focus of the 4-metre Anglo-Australian Telescope.
The SAMI instrument has 13 integral field units (IFUs), deployable inside 
a 1-degree diameter field of view (as well as 26 individual fibres used to sample
the sky background). Each of the 13 IFUs is a lightly-fused fibre bundle
\citep[hexabundle;][]{bland-hawthorn+2011, bryant+2014}, consisting of 61
1.6-arcsec diameter individual fibres, for a total IFU diameter of 15 arcsec.
The fibres are fed to the double-beam AAOmega spectrograph \citep{sharp+2006},
configured with the 570V grating at 3750--5750~\AA\ (blue arm) and with the
R1000 grating at 6300--7400~\AA\ (red arm). With this setup, the resulting
spectral resolutions are $R=1812$ ($\sigma$ = 70.3~\kms) and
$R=4263$ ($\sigma$ = 29.9~\kms) for the blue and red arm,
respectively \citep{vandesande+2017a}. Each galaxy was exposed for approximately
3.5~hours, following a hexagonal dither pattern of seven equal-length exposures
\citep{sharp+2015}. After rejecting observations under inadequate conditions,
the median FWHM seeing of the SAMI datacubes is $2.06 \pm 0.40$~arcsec. The data
reduction is described in \citet{sharp+2015} and \citet{allen+2015}, whereas
subsequent improvements have been described in the public data release papers
\citep{green+2018, scott+2018}. In this work, we use data from the third and
final public data release (Data Release 3, hereafter DR3) consisting of 3068
unique datacubes \citep{croom+2021a}. For our measurements, we use 1-d spectra obtained
by adding the light inside an elliptical aperture. The ellipse is centred on the
centre of the galaxy, its position angle and shape are taken from the best-fit
S{\'e}rsic model, and its semi-major axis is equal to one effective radius \re
(see \S~\ref{s.das.ss.anc.sss.sizeshape} for the size and shape measurements).
The median $S/N$ of these spectra is $24$~\AA$^{-1}$.
Two randomly selected SAMI spectra are shown in Fig.~\ref{f.das.ppxf},
illustrating a quiescent galaxy (SAMI~347471, panel~\subref{f.das.ppxf.a})
and a star-forming galaxy (SAMI~517278, panel~\subref{f.das.ppxf.d}). The galaxy
images (obtained from the datacubes) and the elliptical apertures are illustrated
in panels~\subref{f.das.ppxf.c} and~\subref{f.das.ppxf.f}. Note that the SAMI
wavelength range has been reduced to match the wavelength range of LEGA-C. The
reason is that we find \h4 to depend on the wavelength range, which we will
explore in a future paper (D'Eugenio et~al., in~prep.).

\subsubsection{MAGPI}\label{s.das.ss.ds.sss.magpi}

The Middle Ages Galaxy Properties with Integral Field Spectroscopy survey
\citep[hereafter, \href{https://magpisurvey.org}{MAGPI;}][]{foster+2021} is a
Large Program with the Multi-Unit
Spectroscopic Explorer \citep[MUSE,][]{bacon+2010} on the European Southern
Observatory (ESO) Very Large Telescope (VLT). MAGPI
targets spatially resolved galaxy physics between redshifts $0.15 < z < 0.6$,
the uncharted cosmic `Middle Ages' between `classic' local surveys (e.g. SAMI)
and LEGA-C. The sample consists of 60 central galaxies:
56 drawn from the Galaxy and Mass Assembly survey \citep[GAMA;][]{
driver+2011, liske+2015, baldry+2018}, complemented by four fields chosen from
two legacy programs, targeting clusters Abell~370 (Program ID 096.A-0710; PI: Bauer) and
Abell~2744 (Program IDs: 095.A-0181 and 096.A-0496; PI: Richard). In addition
to the central galaxies, MAGPI will concurrently observe one hundred satellite galaxies
in the target redshift range, plus any background galaxy inside the MUSE field
of view.

MAGPI uses MUSE in the large-field configuration ($1\times1$-arcmin$^2$ field of
view), aided by Ground Layer Adaptive Optics GALACSI \citep{arsenault+2008,
strobele+2012} to achieve a spatial resolution with median FWHM of
0.6--0.8~arcsec (comparable, in physical units, to the spatial resolution of
local surveys such as SAMI). MUSE spectra cover the approximate rest-frame
wavelength range $3600 < \lambda < 7200$~\AA, with a median spectral resolution
FWHM of 1.25~\AA\ (inside one effective radius, the FWHM varies by 3~per cent). The
survey is ongoing, but the program has already obtained fully reduced data for
thirty-five fields, though in this work we use only the first fifteen. An overview of
the observations and data reduction is provided in
the survey paper \citep{foster+2021}, while the full data reduction pipeline
\citetext{based on the MUSE pipeline, \citealp{weilbacher+2020} and on the
Zurich Atmosphere Purge sky-subtraction software, \citealp{soto+2016}}, will
be described in an upcoming work (Mendel~et~al., in~prep.). Each MAGPI cube is
segmented into `minicubes', centred on individual galaxy detections. From these
minicubes, we obtain 1-d spectra by adding up the light inside an aperture,
similar to the approach we used for SAMI.
These spectra have median $S/N = 13$~\AA$^{-1}$, but the subset we use in this
study has larger $S/N$ (see~\ref{s.samp.ss.qcsel}).
Two randomly selected MAGPI galaxies are shown
in Fig.~\ref{f.das.ppxf}: quiescent MAGPI~1203306150
(panels~\subref{f.das.ppxf.g}--\subref{f.das.ppxf.i}) and star-forming
MAGPI~1209136241 (panel~\subref{f.das.ppxf.j}--\subref{f.das.ppxf.l}). Like for
SAMI, the wavelength range has been reduced to match the wavelength range of
LEGA-C.

\subsubsection{LEGA-C}\label{s.das.ss.ds.sss.legac}

The Large Early Galaxy Astrophysics Census is the deepest, large spectroscopy
survey beyond the local Universe \citep{vanderwel+2016}. Targeting 3000
galaxies between $0.6 < z < 1.0$, LEGA-C delivers high-quality absorption
spectra at a look-back time when the Universe was only half its age. The sample
is $K_s$-band selected from the UltraVISTA catalogue \citep{muzzin+2013a},
itself part of the COSMOS field, thus (mostly) covered by the COSMOS HST survey
\citep{scoville+2007b}.
LEGA-C spectra were observed at the ESO VLT using the now decommissioned VIMOS
spectrograph \citep{lefevre+2003} in its multi-object configuration, with
mask-cut slits of 1-arcsec width and length $\geq 8$~arcsec. All slits from the
main survey were oriented in the North-South direction, therefore randomly
aligned with respect to the major axes of the targets. The seeing median
full-width half-maximum (FWHM, measured from a Moffat fit on the slit data) is
0.75~arcsec \citep{vanhoudt+2021}. The
spectral interval varies with the slit position within the relevant mask (but
typically covers the interval $6300 < \lambda < 8800$~\AA), with an
observed-frame spectral resolution $R=2500$ \citep[the effective spectral
resolution is $R=3500$, because the LEGA-C targets underfill the slit;][]{
straatman+2018}. Each target was exposed for 20~h, reaching an integrated
continuum $S/N\approx 20$~\AA$^{-1}$. Given the depth of
the observations, most targets have successful kinematics measurements (93~per cent)
resulting in a mass-completeness limit of $10^{10.5} \, \msun$
\citep{vanderwel+2021}.

\begin{figure*}
  \includegraphics[type=pdf,ext=.pdf,read=.pdf,width=1.\textwidth]{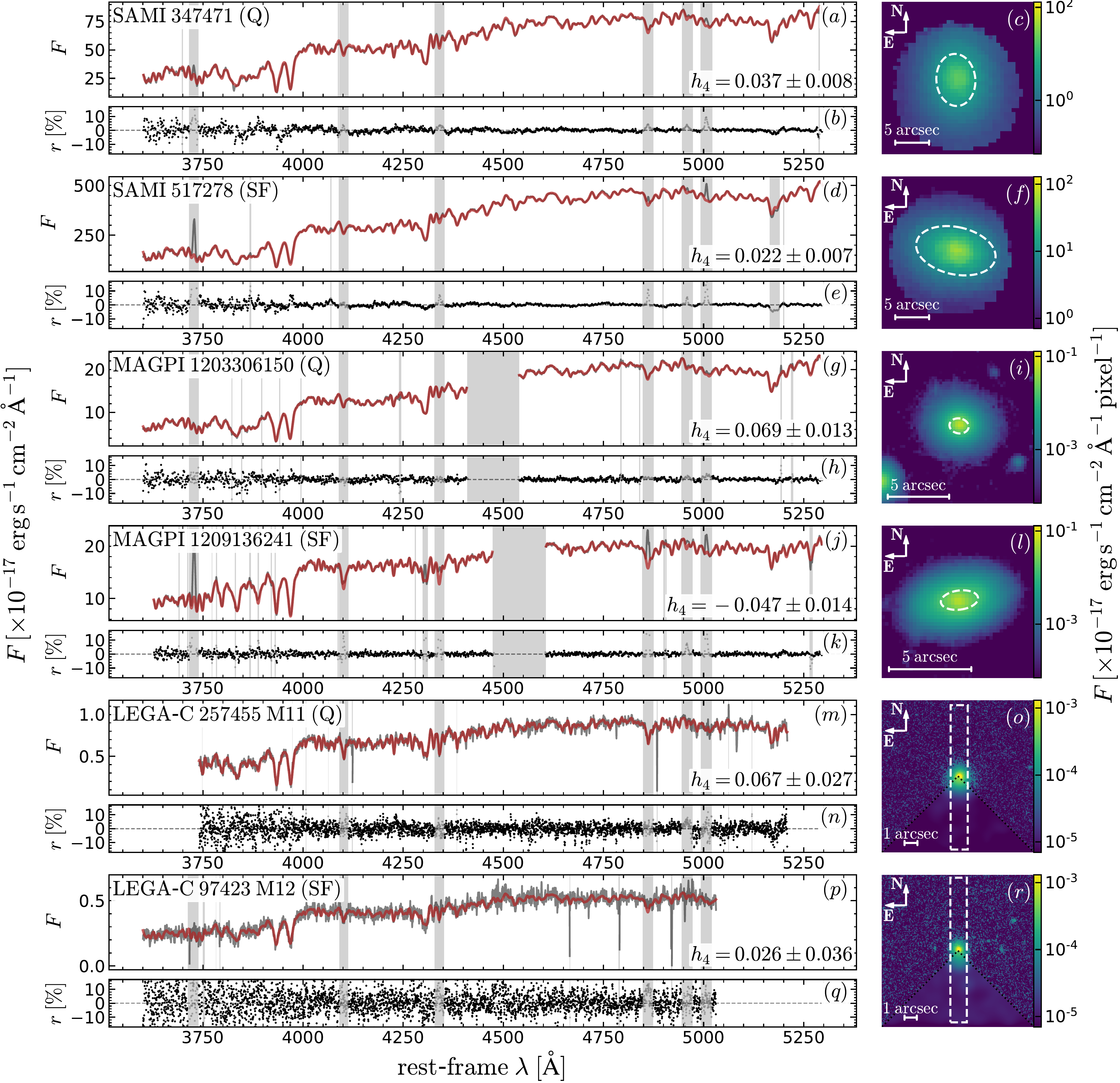}
  {\phantomsubcaption\label{f.das.ppxf.a}
   \phantomsubcaption\label{f.das.ppxf.b}
   \phantomsubcaption\label{f.das.ppxf.c}
   \phantomsubcaption\label{f.das.ppxf.d}
   \phantomsubcaption\label{f.das.ppxf.e}
   \phantomsubcaption\label{f.das.ppxf.f}
   \phantomsubcaption\label{f.das.ppxf.g}
   \phantomsubcaption\label{f.das.ppxf.h}
   \phantomsubcaption\label{f.das.ppxf.i}
   \phantomsubcaption\label{f.das.ppxf.j}
   \phantomsubcaption\label{f.das.ppxf.k}
   \phantomsubcaption\label{f.das.ppxf.l}
   \phantomsubcaption\label{f.das.ppxf.m}
   \phantomsubcaption\label{f.das.ppxf.n}
   \phantomsubcaption\label{f.das.ppxf.o}
   \phantomsubcaption\label{f.das.ppxf.p}
   \phantomsubcaption\label{f.das.ppxf.q}
   \phantomsubcaption\label{f.das.ppxf.r}}
  \caption{
  Comparison between three randomly selected quiescent (Q) galaxies
  and three randomly selected star-forming (SF) galaxies, chosen from SAMI
  (panels~\subref{f.das.ppxf.a}--\subref{f.das.ppxf.f}), MAGPI
  (panels~\subref{f.das.ppxf.g}--\subref{f.das.ppxf.l}) and LEGA-C
  (panels~\subref{f.das.ppxf.m}--\subref{f.das.ppxf.r}).
  For each galaxy, we show the data (dark grey) and best-fit spectra (red),
  alongside the relative residuals (black dots). The galaxy names and their
  \h4 values are reported in the top-left and bottom-right corners of the
  panels with the spectra.
  Vertical lines/regions are masked because of low data quality, or possible
  emission lines (regardless of whether lines were actually detected), or
  because of instrument
  setup (e.g. the GALACSI laser band for MAGPI, panels~\subref{f.das.ppxf.g},
  \subref{f.das.ppxf.h}, \subref{f.das.ppxf.j} and~\subref{f.das.ppxf.k}). The
  inset figures show the galaxy images (derived from the datacubes for SAMI and
  MAGPI, panels~\subref{f.das.ppxf.c}, \subref{f.das.ppxf.f},
  \subref{f.das.ppxf.i} and~\subref{f.das.ppxf.l}; from HST F814W for LEGA-C,
  panels~\subref{f.das.ppxf.o} and~\subref{f.das.ppxf.r}).
  In each of the six galaxy images, we indicate the aperture used to extract
  the spectrum with a dashed white line; these are ellipses with semi-major
  axis equal to the effective radius (for SAMI and MAGPI), or a rectangular
  slit with 1-arcsec width (for LEGA-C). The lowest quadrant of the LEGA-C
  images shows the data convolved to the
  ground-based spatial resolution of LEGA-C.
  }\label{f.das.ppxf}
\end{figure*}

To measure \h4, we use the 1-d LEGA-C spectra from the third
public data release of LEGA-C \citep[DR3,][]{vanderwel+2021}. These were obtained
from optimal extraction \citep{horne1986} of the 2-d spectra. The large physical
width of the LEGA-C slits (7.5~kpc at $z=0.8$) means that the 1-d spectra sample
a representative fraction of the targets' light (the ratio between the slit width
and the circularised galaxy diameter is $1.2\pm0.8$ for our sample, see
\S~\ref{s.samp} for the sample selection). We adopt the method
described in \S~\ref{s.das.ss.hok}, setting the (observed-frame) FWHM to a
wavelength-independent value of 2.12~\AA\
\citep[corresponding to 86~\kms,][]{vanderwel+2021}. Note that we use
emission-line subtracted
spectra \citep{bezanson+2018a}, but the precision and accuracy of the
subtraction do not affect our measured kinematics. This is because we
conservatively mask the spectral regions where gas emission lines may arise in
all galaxies, regardless of whether emission was actually detected (see
\S~\ref{s.das.ss.hok} and Appendix~\ref{app.emlsub}).
Two randomly selected LEGA-C spectra are shown in Fig.~\ref{f.das.ppxf}: a
quiescent galaxy (LEGA-C\ 257455~M11, panel~\subref{f.das.ppxf.m}) and a
star-forming galaxy (LEGA-C\ 97423~M12, panel~\subref{f.das.ppxf.p}). The HST
images and the LEGA-C slits are shown in panels~\subref{f.das.ppxf.o}
and~\subref{f.das.ppxf.r}.

\begin{figure}
  \includegraphics[type=pdf,ext=.pdf,read=.pdf,width=1.\columnwidth]{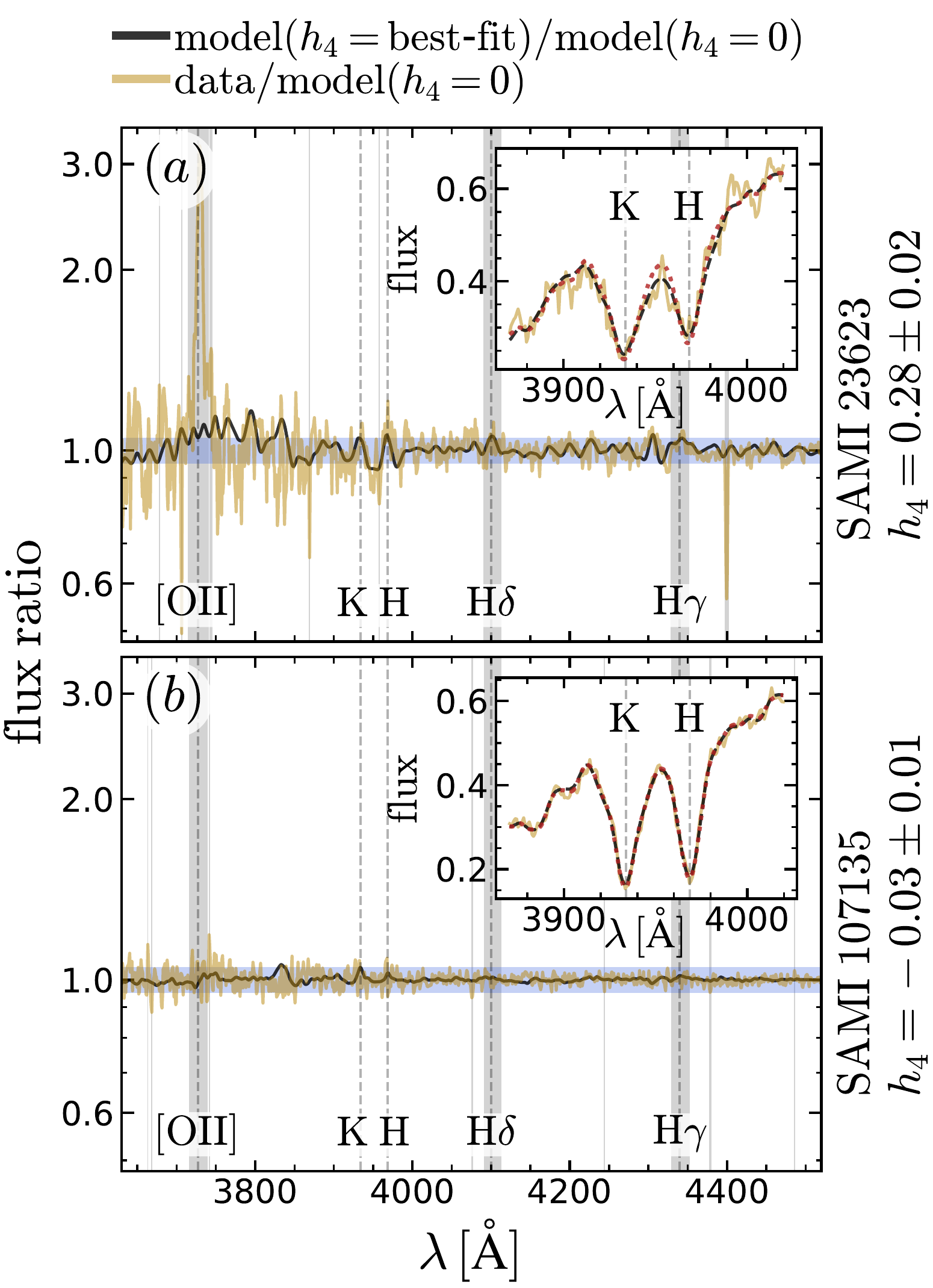}
  {\phantomsubcaption\label{f.das.h4high.a}
   \phantomsubcaption\label{f.das.h4high.b}}
  \caption{Showing the difference between a 2\textsuperscript{nd}- and
  4\textsuperscript{th}-order velocity distribution. In the main panels, the
  sand lines show the ratio between the data and the best-fit
  2\textsuperscript{nd}-order model (labelled `model(\h4=0)'), whereas the
  black lines show the ratio between the best-fit 4\textsuperscript{th}- and
  2\textsuperscript{nd}-order models. 
  Panel~\subref{f.das.h4high.a} shows SAMI galaxy 23623, an extreme system with
  high \h4 (this galaxy appears to be a recent merger, so it is excluded from
  the rest of the study). For this galaxy, outside the noisy region at the blue
  end ($\lambda < 3900$~\AA), the residuals show variations of a few percent
  (the shaded blue region encompasses $\pm$5~per cent from unity); the solid black line
  follows closely the sand line, underscoring the need for a leptokurtic LOSVD.
  Conversely, SAMI galaxy 107135 (panel~\subref{f.das.h4high.b}) has low \h4: the
  difference between the 2\textsuperscript{nd}- and
  4\textsuperscript{th}-order LOSVDs is less pronounced.
  The vertical grey regions are bad pixels, or regions where gas emission lines
  may be located. In each of the inset panels, we focus on the region of the
  spectrum around the H and~K Calcium lines; we show the data (solid sand line),
  the best-fit 4\textsuperscript{th}-order model (dashed black line) and the
  best-fit 2\textsuperscript{nd}-order model (dotted red line). Even for
  SAMI~23623, the two models are barely distinguishable, but comparing the
  the 4\textsuperscript{th}- and 2\textsuperscript{nd}-order models in the
  region between the two lines, it may be observed that the
  4\textsuperscript{th}-order model has broader wings, which follow the data
  more closely.
  }\label{f.das.h4high}
\end{figure}

\subsection{Measuring integrated higher-order kinematics}\label{s.das.ss.hok}

In each of the three datasets, we model the LOSVD
as a 4\textsuperscript{th}-order Gauss-Hermite series \citep{vandermarel+franx1993,
gerhard1993}, because this approach: \textit{i}) provides a compact description
of the non-Gaussianity through the parameters $h_3$ and \h4, the coefficients of
the 3\textsuperscript{rd}- and 4\textsuperscript{th}-order Hermite polynomials,
as well as \textit{ii}) minimises the correlation between the LOSVD parameters
\citep{vandermarel+franx1993}.

Our \h4 measurements are based on one-dimensional spectra spanning rest-frame 
$B$-~and $g$-band, from which we infer the LOSVD using \ppxf. We
model the spectra using a linear combination of simple stellar population (SSP)
spectra from the MILES library \citep{vazdekis+2010, vazdekis+2015}, using BaSTI
isochrones \citep{pietrinferni+2004, pietrinferni+2006} and solar
$[\alpha/\mathrm{Fe}]$.

When necessary and possible, the SSP spectra are matched to the spectral
resolution of the data, using the uniform full-width half-maximum (FWHM)
spectral resolution of 2.51~\AA\ \citep{falcon-barroso+2011}.
However, for some of the SAMI spectra, and for all the MAGPI and LEGA-C spectra,
the instrumental resolution is better than the MILES spectral resolution. In
this case, matching the two resolutions would require broadening the galaxy
spectra, but because this is undesirable, we do not apply any correction. Even
though this introduces a bias in the resulting second moment of the LOSVD,
the MILES SSP library provides consistently the best fits to the galaxy
continuum \citep[surpassed only by the MILES stellar template library, in
agreement with e.g.][]{vandesande+2017a, maseda+2021}.
There are three reasons why a biased measurement of the second moment is not
important in this article.
First, we are not interested in measuring the second moment; when we use second
moment measurements, these values are taken from the literature and are measured
taking into account the appropriate instrument resolution (\S~\ref{s.das.ss.anc}).
Second, our main targets are high-mass galaxies with large physical
dispersion and, finally, our results are unchanged if we repeat our measurements
with the higher resolution SSP spectra from the IndoUS library
\citep{valdes+2004}, or with the synthetic SSPs from the C3K theoretical library
\citep{conroy+2019} using the MIST isochrones \citetext{
\citealp{dotter2016}, \citealp{choi+2016}. See Appendix~\ref{app.libbias}}.
Overall, we deem the fit quality a more desirable property than unbiased
measurements of the second moment (which are available anyway from other
sources). For this reason, our default \h4 measurements are obtained using the
MILES SSP templates.

In addition to the SSP templates, we
also use 12\textsuperscript{th}-order additive Legendre polynomials to fit
residual flux due to flux calibration and background subtraction errors
\citetext{this follows the prescription of \citealp{deugenio+2020} for LEGA-C,
and of \citealp{vandesande+2017a} for SAMI}. The keyword {\sc bias}, which
determines the amount of penalisation against non-Gaussian LOSVDs, is set to
its default value. This choice does not affect our measurements of \h4, because
of the high signal-to-noise ratio ($S/N$) of our spectra (see
\S~\ref{s.das.ss.hok.sss.penal}).

The fit is repeated twice: in the first iteration, we use
uniform weighting for all valid spectral pixels.
After this fit, we rescale the noise spectrum so that the value of the reduced 
$\chi^2$ would be unity.
The second and final fit uses this rescaled noise as well as 3-$\upsigma$
iterative clipping to remove outliers.
\ppxf returns the first (non-trivial) four moments of the LOSVD: mean velocity
$V$, $\sigma$, $h_3$ (a measure of skewness) and \h4 (measuring excess
kurtosis).

The uncertainties on the \h4 measurements are derived from the local curvature
of the $\chi^2$ surface near its minimum. We checked that these formal
uncertainties accurately propagate the observational errors through to the
derived parameter values, by using a Monte-Carlo (MC) approach. For each galaxy,
we created one hundred spectra by randomly shuffling and re-adding the fit
residuals to the best-fit spectrum \citep[see e.g.][]{vandesande+2017a}.
After fitting these random realisations of the data, for each galaxy we obtain a
distribution of one hundred values of \h4; the MC uncertainty is defined as the
standard deviation of this distribution. For SAMI, and for 10~per cent of the LEGA-C
sample, these MC uncertainties are consistent with the default uncertainties, so,
from here on, we always use the formal uncertainties as default.

Example \ppxf fits are shown in Fig.~\ref{f.das.ppxf}; starting from the final
sample (defined in \S~\ref{s.samp.ss.samp}), we randomly selected a quiescent
and a star-forming galaxy from each of the three surveys. Note the different
apparent sizes in the inset images, but the similar wavelength coverage of the
spectra.

In Fig.~\ref{f.das.h4high} we show two example spectra from SAMI: a galaxy
with non-Gaussian, leptokurtic LOSVD ($\h4>0$, top panel) and a galaxy with
(close-to) Gaussian LOSVD (bottom panel). In each panel, we show two spectra: the
sand-coloured line is the ratio between the data and the 4-moments best-fit spectrum,
whereas the black line is the ratio between the 4-moments best-fit spectrum and
the Gaussian best-fit spectrum (vertical grey regions are masked). For galaxy
SAMI~23623, the sand and black lines have several features in common, both
around the Calcium H and K lines as well as around 4200~\AA; in contrast, no such
features are present for galaxy SAMI~107135. This figure demonstrates that
information about the shape of the LOSVD is spectrally `distributed': it is
present both around prominent lines, as well as in less prominent spectral
features.

\subsubsection{Penalisation of non-Gaussian solutions}\label{s.das.ss.hok.sss.penal}

To measure \h4, a critical feature of the \ppxf algorithm is the eponymous
`penalisation' against non-Gaussian LOSVDs. The penalisation is an arbitrary
upscaling of the $\chi^2$, to ensure non-Gaussian solutions (i.e., $h_i \neq 0$)
are accepted only if they come with a `sufficient' decrease in the $\chi^2$
\citep{cappellari+emsellem2004}. In \ppxf, the penalisation is implemented by
the {\sc bias} keyword. To recover \h4 in low-quality data, the value of the
{\sc bias} keyword must be carefully determined using simulations
\citep[see e.g.][their Appendix~A.5]{vandesande+2017a}. For low-$S/N$
spectra, \h4 may depend on the choice of {\sc bias}, but in this
work, we deal with high-$S/N$ data, so the value of {\sc bias}
is not critical. To demonstrate this, we re-measured \h4 setting {\sc bias}=0
and verified that our results do not change. For our sample
(defined in \S~\ref{s.samp.ss.samp}), the difference $\Delta\,\h4$ between
the `non-penalised' \h4 measurement ({\sc bias}=0) and the default `penalised'
\h4 measurement ({\sc bias = None}) is negligible compared to other systematic
errors (which have values of $\approx0.03$, Appendix~\ref{app.agebias}).
For SAMI, we find a median $\Delta\,\h4=0.0005\pm0.0001$ whereas for LEGA-C we
find a median $\Delta\,\h4=0.0002\pm0.0004$ (for MAGPI, the uncertainty on
the median $\Delta\,\h4$ is much larger than the median itself, because of the
small sample size). In all three samples, the standard deviations of
$\Delta\,\h4$ are 3--10 times smaller than the precision threshold for
selecting the sample (see \S~\ref{s.samp.ss.qcsel}).

\subsubsection{Measurement bias}\label{s.das.ss.hok.sss.mbias}

In the following, we aim to compare \h4 between star-forming and quiescent
galaxies; this is subject to possible bias due to the systematic differences in 
the depth of stellar absorption features in these two classes of galaxies: at
fixed luminosity (and so at fixed $S/N$), quiescent galaxies have older stellar 
populations, so have deeper absorption features (except for Balmer lines, which 
we mask as we discuss in \S~\ref{s.samp.ss.qcsel}). Using mock spectra, we find 
that systematics connected to different stellar populations are $\times$10
smaller than the maximum  measurement uncertainties used for the quality cut,
and smaller than the reported difference between star-forming and quiescent
galaxies\footnote{It should be noted, however, that our mock spectra do not
fully capture the complexity of real galaxies, where stellar populations
properties like age vary systematically with kinematics \citetext{e.g.,
\citealp{poci+2019}, \citealp{shetty+2020b}, and \citealp{foster+2023}}.}
(Appendix~\ref{app.agebias}).

Similarly, changing the template library used in \ppxf changes the value of
measured \h4, but we still measure a different \h4 between star-forming and
quiescent galaxies (Appendix~\ref{app.libbias}).

\subsection{Ancillary data}\label{s.das.ss.anc} 

\subsubsection{Stellar masses}

Stellar masses are obtained differently for SAMI and MAGPI compared to LEGA-C.
For the first two surveys, \mstar was derived from S{\'e}rsic-fit $i$-band total
magnitudes, using $g-i$ colour to infer the stellar mass-to-light ratio,
assuming exponentially declining star-formation histories \citep{taylor+2011}. The
actual expression of stellar mass also implements a k-correction
\citep[see e.g.][]{bryant+2015}. For SAMI, $g-i$ colours are derived from
SDSS or VST ground-based photometry \citep[see again][and references therein]{bryant+2015, owers+2017}.
For MAGPI, colours are derived from synthetic MUSE photometry (Taylor~et~al., in~prep.).

In contrast, LEGA-C uses SED fits to observed-frame $BVrizYJ$ photometry drawn
from UltraVISTA \citep{muzzin+2013b}, zero-point corrected as described in the
DR3 article \citep{vanderwel+2021}. The fits are performed using {\sc prospector}
\citep{leja+2019a, johnson+2021}, with the configuration adopted in \citet{
leja+2019b}.

All three methods rely primarily on rest-frame visible photometry, but the
precise bands and the underlying assumptions about dust, star-formation history
and metallicity are different. Despite these differences, however, the mass
measurements are sufficiently close for a qualitative selection in \mstar
(see \S~\ref{s.samp.ss.samp}). To prove this statement, we use a
third set of mass measurements as a `bridge'. These measurements are only
available for a subset of the LEGA-C and SAMI surveys, so they are not
suitable as main mass measurements. \citet{driver+2018} used 
{\sc magphys} \citep{dacunha+2008} to measure stellar masses for the subsets
of the SAMI and LEGA-C samples that fall within the footprint of GAMA.
The SAMI measurements show good agreement with the default measurements we
use here: the median offset between the $g-i$-based (default) and
{\sc magphys} measurements is $0.01$~dex, with a scatter of $0.06$~dex.
For LEGA-C, the median offset between the {\sc prospector} (default) and
{\sc magphys} measurements is $0.03$~dex, with a scatter of $0.07$~dex.

\subsubsection{Sizes and shapes}\label{s.das.ss.anc.sss.sizeshape}

Galaxy sizes and shapes are derived from S{\'e}rsic models. \re is defined as
the half-light semi-major axis and $q$ is the minor-to-major axis ratio of the
best-fit model. For SAMI, we use ground-based $r$-band photometry. For MAGPI, we
use synthetic $r$-band imaging obtained from MUSE. For LEGA-C, we use 
HST F814W images. These heterogeneous data have remarkably similar spatial
resolution in physical units; considering a median point-spread function FWHM of
1.3, 0.6 and 0.12~arcsec respectively for SAMI, MUSE and LEGA-C photometry, the
spatial resolution in physical units is within a factor of three (1.3, 2.7 and
0.9~kpc, respectively).

For SAMI, the models are optimised using either {\sc galfit} \citep[][for the
SAMI subset inside the GAMA regions]{peng+2002}, or {\sc profit}
\citetext{\citealp{robotham+2017}; for the cluster subset}. We refer the reader
to the relevant literature for further information
\citep{croom+2021a, kelvin+2012, owers+2019}.
For both MAGPI and LEGA-C, the S{\'e}rsic models are optimised using {\sc galfit}
\citep[for LEGA-C, see also][]{vanderwel+2011, vanderwel+2021}.

While the measurements (and especially LEGA-C) are not strictly consistent, we use
them only internally to each sample and make no attempt to compare values across
surveys. To test the effect of the different rest-frame wavelength of the
photometry, we replace SAMI $r$-band photometry with $g$-band photometry. This
substitution matches well the rest-frame wavelength of LEGA-C (the effective
wavelength of the SDSS $g$ filter is 4670~\AA; at redshift $z=0.7$, the
rest-frame effective wavelength of the ACS F814W filter is 4710~\AA).
Comparing $g$-band to $r$-band photometry for the subset of our sample that
possess both measurements, we find that the median ratio of $g$-band to $r$-band
axis ratio is 1.01. The median ratio between the effective radii is 1.04. These small
differences are negligible, given the precision of our measurements and our sample
size.
Nevertheless, we tested that replacing the SAMI $r$-band sizes and shapes
with their $g$-band equivalents, does not change our conclusions. In the end,
we prefer to use $r$-band measurements because $g$-band sizes and shapes are only
available for two thirds of the SAMI sample.

\subsubsection{Rotation-to-dispersion ratio and other kinematic quantities}\label{s.das.ss.anc.sss.vse}

For SAMI, MAGPI and LEGA-C, we also use two different measurements of
\vse; for SAMI and MAGPI, this is the observed
ratio averaged inside one~\re, with empirical corrections for seeing and aperture
\citep{vandesande+2017b, harborne+2020, vandesande+2021a, vandesande+2021b}; for
LEGA-C, \vse indicates the value of the best-fit Jeans anisotropic models
\citep{cappellari2008}, evaluated at one~\re \citep[the models and their
optimisation are described in][]{vanhoudt+2021}.
Once again, we remark that these two measurements are not consistent, but we do
not compare them directly.

It is worth noting that dynamical models (and therefore \vse) are only
available for approximately one third of LEGA-C galaxies. This occurs mostly because
galaxies where the slit is misaligned compared to the major axis of the galaxy
were not modelled \citep{vanhoudt+2021}. Fortunately, for the
mass range considered in this article, the galaxies with available models and
\vse represent a random subset of the parent population. We used a
Kolmogorov-Smirnov (KS) test to assess if the mass distribution of our sample is
the same as the mass distribution of the subset with dynamical models; we find a
probability $P_\mathrm{KS} = 0.8$ (for quiescent galaxies) and $P_\mathrm{KS} =
0.6$ (for star-forming galaxies). Similar probabilities are found for the distribution of
\re. In contrast, comparing the distribution of position angles (which determine
the availability of dynamical models) we find $P_\mathrm{KS} = 3\times10^{-13}$
and $P_\mathrm{KS} = 7\times10^{-5}$ for quiescent and star-forming galaxies,
respectively.

We also use integrated velocity dispersions within a fixed aperture, \sigap. For
SAMI and MAGPI, these are calculated inside the ellipse of semi-major axis equal
to one \re; for LEGA-C, these are calculated from the 1-d spectrum.

Finally, for SAMI only, we use the visual kinematic classification of
\citet{vandesande+2021a} to separate dispersion-supported galaxies from
rotation-supported galaxies. We define slow rotators (SR) as having {\sc
kin\_mtype}$<1$, which consists of all `non-obvious rotators' without
kinematic features (e.g., no kinematically decoupled cores), plus intermediate
systems between this class and non-obvious rotators with features. This
definition has good overlap with other definitions of SRs in the literature
\citep{vandesande+2021a}.

\section{Sample selection}\label{s.samp}

In this section, we aim to present the motivation, selection criteria and
properties of our sample.

We propose to study the difference between star-forming and quiescent galaxies,
so the sample is split between these two classes (\S~\ref{s.samp.ss.sfq}). To
ensure that our measurements are reliable, we introduce a quality selection
(\S~\ref{s.samp.ss.qcsel}), and, finally, we introduce a cut in stellar mass
to ensure that our results are representative (\S~\ref{s.samp.ss.samp}).

\subsection{Star-forming and quiescent galaxy separation}\label{s.samp.ss.sfq}

For SAMI, we use the definition of \citet{croom+2021b}: quiescent galaxies have
star-formation rates (SFR) more than 1.6~dex below the star-forming sequence as
defined in \citet{renzini+peng2015}. SFRs are taken from the SAMI DR3 catalogue
\citep{croom+2021a} and are measured from the total, dust-corrected
$\mathrm{H\alpha}$ flux as originally described in \citet{medling+2018}. 

For MAGPI, we use a mixed approach. For galaxies with $z>0.41$, the MUSE spectra
do not include $\mathrm{H\alpha}$, so we used an empirical criterion based on the
equivalent width (EW) of $\mathrm{H\beta}$: galaxies with $\mathrm{EW(H\beta)} <
-1$~\AA\ are
classified as star-forming \citep[see e.g.][]{wu+2018}, the others are classified
as quiescent.
For galaxies with $z<0.41$, the MUSE wavelength range does include
$\mathrm{H\alpha}$. For these targets (the majority of the final sample), we
measure the total $\mathrm{H\alpha}$ and  $\mathrm{H\beta}$ flux inside the
circular aperture with radius equal to three \re (after subtracting the
continuum, using \ppxf). We then apply an attenuation
correction by assuming an intrinsic $\mathrm{H\alpha}/\mathrm{H\beta}$ ratio of
2.86 \citep[case B recombination and $T_\mathrm{e} = 10^4\,\mathrm{K}$][]{osterbrock+ferland2006}
and the \citet{cardelli+1989} dust extinction law. SFRs are measured using the
\citet{kennicutt1998} calibration. When no $\mathrm{H\alpha}$
emission is detected, we classify the galaxy as quiescent.
For galaxies that do have a SFR measurement, we compare our measurements to the
values from GAMA, finding six galaxies in common and a root-mean square difference
of 0.25~dex.
In addition to galaxies with no detected $\mathrm{H}\alpha$ emission, or with
low-EW $\mathrm{H}\beta$ emission, we also consider quiescent all galaxies that
do have a measured SFR, but lie at least 1~dex below the star-forming sequence.
As a reference, we use the empirical, redshift-dependent relation of
\citet{whitaker+2012}.

For LEGA-C, we use only galaxies from the `primary' LEGA-C sample, and adopt
the UVJ diagram \citep{labbe+2005, straatman+2018} to discriminate star-forming
and quiescent galaxies.

The different definitions of star-forming and quiescent galaxies may be a
concern, but, in practice, they are largely equivalent. This has been shown
explicitly for SAMI and LEGA-C \citep{barone+2021}.

\subsection{Quality selection}\label{s.samp.ss.qcsel}

\begin{figure}
  \includegraphics[type=pdf,ext=.pdf,read=.pdf,width=1.\columnwidth]{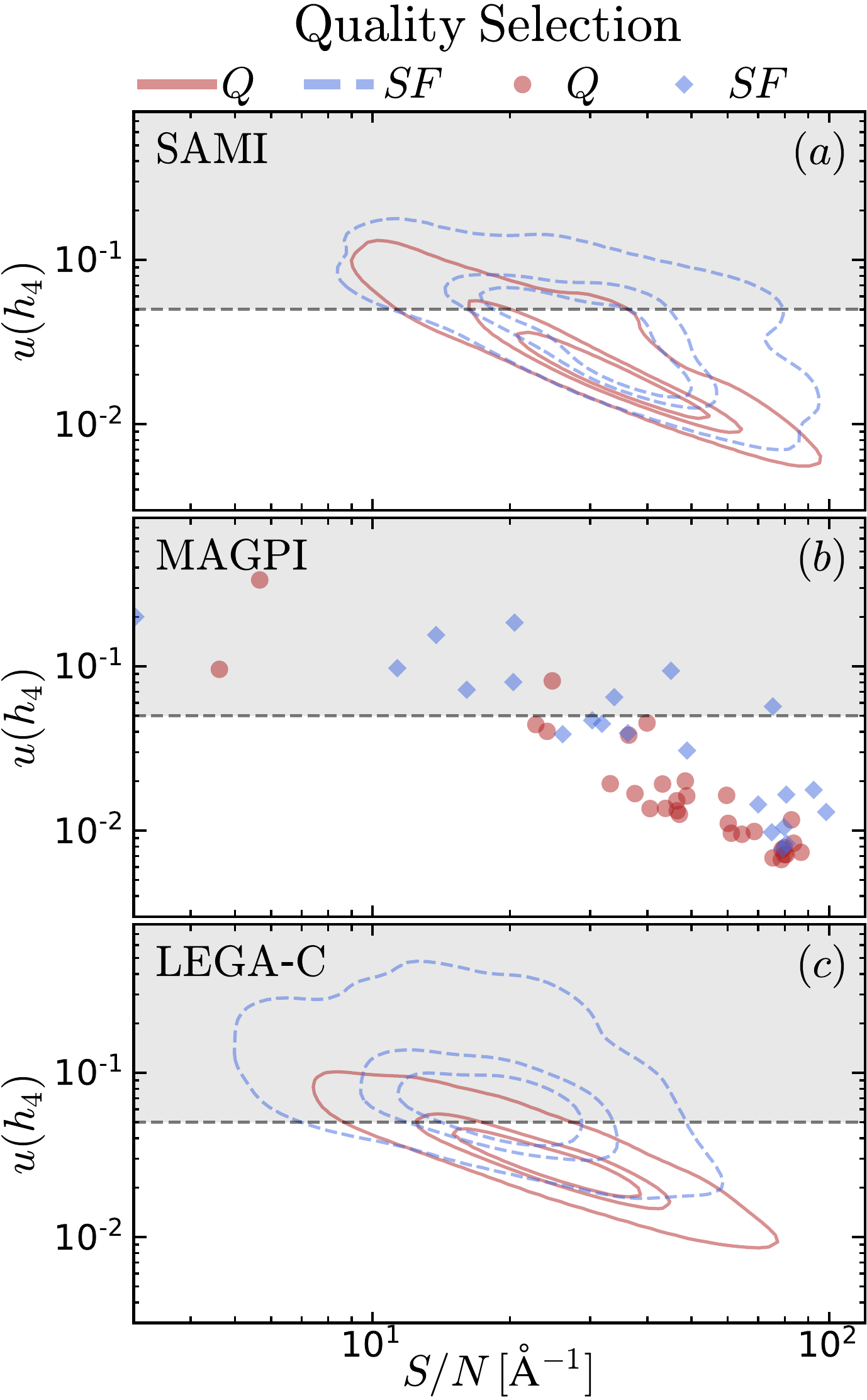}
  {\phantomsubcaption\label{f.samp.qcsel.a}
   \phantomsubcaption\label{f.samp.qcsel.b}
   \phantomsubcaption\label{f.samp.qcsel.c}
  }
  \caption{Our quality selection is based on a cut in the \h4 measurement
  uncertainty, $u(\h4)<0.05$ (horizontal dashed line). There is a clear
  relation between $u(\h4)$ and empirical $S/N$, for SAMI
  (panel~\subref{f.samp.qcsel.a}), MAGPI (panel~\subref{f.samp.qcsel.b}) and
  LEGA-C (panel~\subref{f.samp.qcsel.c}). For SAMI and LEGA-C, we use
  contour lines enclosing the 30\textsuperscript{th}, 50\textsuperscript{th}
  and 90\textsuperscript{th} percentile of the data; dashed blue/solid red
  contours trace star-forming/quiescent galaxies, respectively. For MAGPI, we
  represent individual star-forming/quiescent galaxies with blue diamonds/red
  circles. Note that, at fixed $S/N$, star-forming galaxies
  have larger $u(\h4)$ than quiescent galaxies.}\label{f.samp.qcsel}
\end{figure}

With the default separation between star-forming and quiescent galaxies, we were
able to measure \h4 for a parent sample consisting of 2864 SAMI galaxies (out of
3084 in the DR3 sample), 131 MAGPI galaxies (out of 159), and 2525 (out of 2636)
LEGA-C galaxies.
However, sampling of the galaxy mass
function below $\mstar = 10^{10} \,\msun$ is highly incomplete, so,
in the following, we consider only galaxies above the aforementioned mass
threshold. This sample consists of 1822, 61 and 2475 galaxies for SAMI, MAGPI
and LEGA-C, respectively.

In Fig.~\ref{f.samp.qcsel} we show the relation between the measurement
uncertainty about \h4, labelled $u(\h4)$, and the empirical (\ppxf-derived)
$S/N$ for
the three samples, divided between star-forming and quiescent galaxies
with the criteria described in \S~\ref{s.samp.ss.sfq}. For SAMI and
LEGA-C, we represent the data using dashed blue/solid red contours for
star-forming and quiescent galaxies respectively (panels~\subref{f.samp.qcsel.a}
and~\subref{f.samp.qcsel.c}); these contours enclose the 30\textsuperscript{th},
50\textsuperscript{th} and 90\textsuperscript{th} percentiles of the sample. For
MAGPI, the sample size is
considerably smaller, so we use blue diamonds/red circles that represent
individual galaxies (panel~\subref{f.samp.qcsel.b}). By comparing the locus of
star-forming and quiescent galaxies, it is clear that star-forming galaxies have
larger $u(\h4)$ than quiescent galaxies at fixed $S/N$. This is a reasonable
outcome, because our ability to constrain \h4 depends not only on the continuum
$S/N$, but also on the number and strength of stellar spectral features. These
features are typically weaker in star-forming galaxies than in quiescent
galaxies\footnote{Even though Balmer absorption goes against this trend (i.e.
it is typically strongest for star-forming galaxy spectra), we mask these
spectral regions due to strong contamination from gas emission lines
(Appendix~\ref{app.emlsub}).}
\citep[see e.g.][their fig.~4]{vanderwel+2021}.

Because of the different precision between star-forming and quiescent galaxies
of the same $S/N$,
a selection based solely on $S/N$ would mix together high-precision \h4 values for
one subset of galaxies with low-precision measurements for the other. To avoid
this potential bias, we adopt a quality cut at $u(\h4)<0.05$ (horizontal dashed
line). With this cut, the median $S/N$ values are $31\pm16$~\AA$^{-1}$ (for
SAMI), $45\pm30$~\AA$^{-1}$ (MAGPI) and $20\pm10$~\AA$^{-1}$ (LEGA-C).
Admittedly, this cut is arbitrary, but we note that adopting a threshold between
$0.02$ and $0.1$ does not change our results. If we select $u(\h4)<0.01$, then
the LEGA-C sample is too small to infer any trend with redshift (just nine
galaxies). Similarly, we tested that a cut in $S/N>30$~\AA$^{-1}$ does not
change our results.

\subsection{Stellar mass selection and completeness}\label{s.samp.ss.samp}

\begin{figure}
  \includegraphics[type=pdf,ext=.pdf,read=.pdf,width=1.\columnwidth]{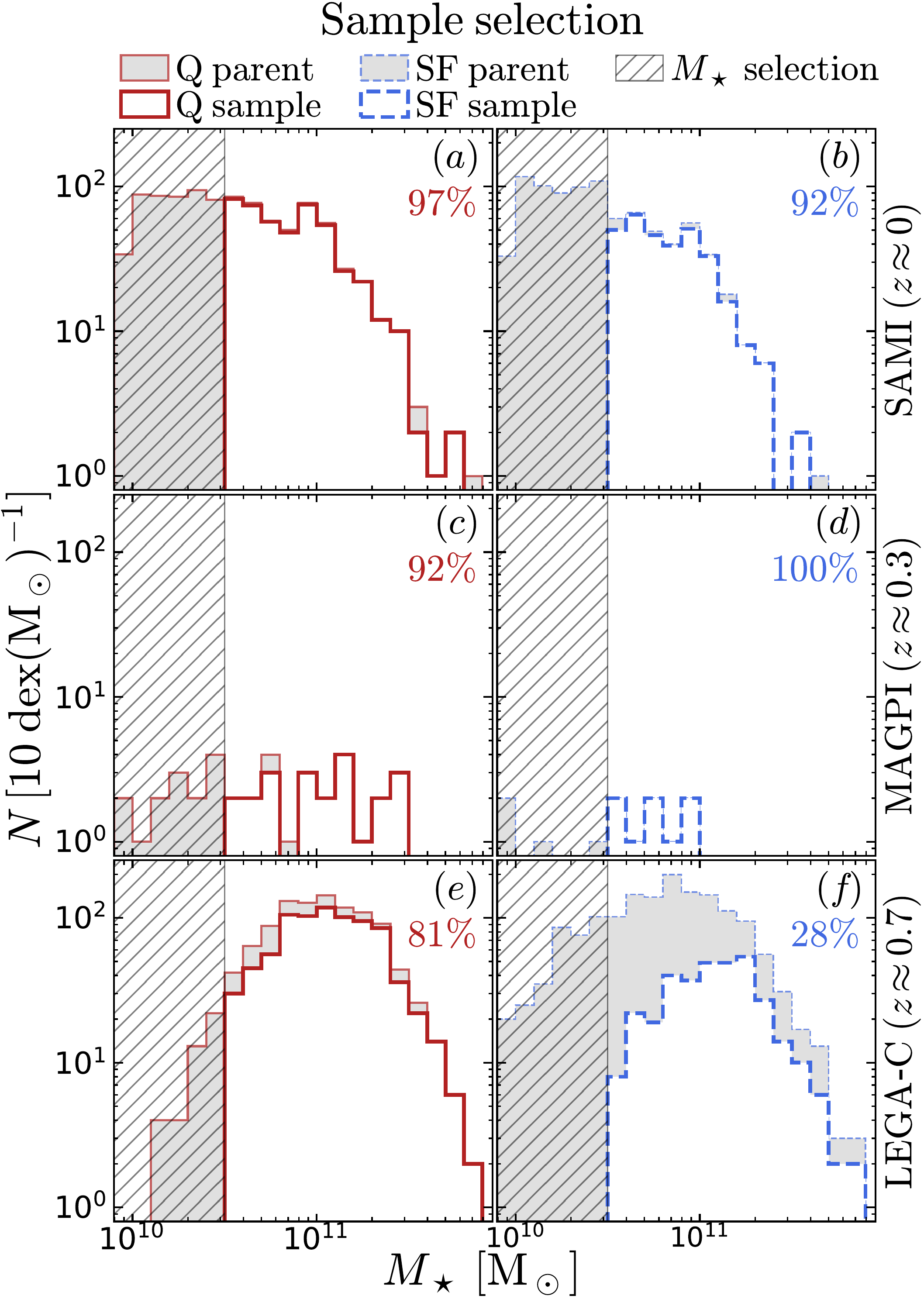}
 {\phantomsubcaption\label{f.samp.samp.a}
  \phantomsubcaption\label{f.samp.samp.b}
  \phantomsubcaption\label{f.samp.samp.c}
  \phantomsubcaption\label{f.samp.samp.d}
  \phantomsubcaption\label{f.samp.samp.e}
  \phantomsubcaption\label{f.samp.samp.f}}
  \caption{The sample selected from SAMI (top row), MAGPI (middle), and
  LEGA-C (bottom). The left/right columns show respectively quiescent and
  star-forming galaxies. In each panel, the filled grey histogram is the mass
  distribution of the parent sample (including galaxies without \h4
  measurements). The empty histograms are our sample, selected to have
  $\mstar \geq 10^{10.5}\, \msun$ and to meet the quality selection
  criteria for \h4 (\S~\ref{s.samp.ss.qcsel}). The percentage in the top right
  corner of each panel is the number ratio between our sample and the parent
  sample, considering only galaxies above the mass limit. For SAMI and LEGA-C,
  the quiescent samples are highly complete; for the star-forming sample, only
  SAMI shows high completeness.
  }\label{f.samp.samp}
\end{figure}

Of the 3083 galaxies in the SAMI DR3 sample, only 1325 meet the quality
selection threshold (defined in \S~\ref{s.samp.ss.qcsel}), giving a completeness
of only 43~per cent. Similar survival rates apply to MAGPI and LEGA-C (27 and 45~per cent,
respectively). To avoid sample incompleteness caused by the quality selection,
we require galaxies in our sample to have $\mstar > 10^{10.5}\,\msun$ (cf. hatched
regions in Fig.~\ref{f.samp.samp}). The sample is thus defined as all galaxies with
$\mstar > 10^{10.5} \,\msun$ and $u(\h4)<0.05$. This particular mass
threshold was chosen as a compromise between sample size and completeness.

In the top row of Fig.~\ref{f.samp.samp} we compare the mass distribution of
the SAMI parent sample to that of our sample, separately between
quiescent (panel~\subref{f.samp.samp.a}) and star-forming galaxies
(panel~\subref{f.samp.samp.b}). Above the mass threshold of $10^{10.5} \,
\msun$, the SAMI DR3 sample contains 821 unique galaxies, of which 
780 meet the quality selection criteria (95~percent). Considering separately
quiescent galaxies, the SAMI DR3 sample and our sample consist of
481 and 465 galaxies (97~percent, cf.~grey filled and red empty histograms
in panel~\subref{f.samp.samp.a}); for star-forming galaxies, the numbers are
340 and 315 (92~percent, cf.~grey filled and blue empty histograms in
panel~\subref{f.samp.samp.b}).
 
In the second row of Fig.~\ref{f.samp.samp}, we provide the mass
distribution for the MAGPI sample. Above the adopted \mstar limit, we have 32
galaxies, of which all but two pass the quality selection (92~per cent). Separating
between quiescent and star-forming galaxies (panels~\subref{f.samp.samp.c} and
\subref{f.samp.samp.d}), we have similar completeness values (22/24 quiescent
galaxies and 8/8 star-forming galaxies meet the quality selection criteria).

Finally, in the bottom row of Fig.~\ref{f.samp.samp}, we compare the mass distribution
of the LEGA-C parent sample to that of our final sample, divided again between
quiescent (panel~\subref{f.samp.samp.e}) and star-forming galaxies
(panel~\subref{f.samp.samp.f}). For quiescent galaxies, the LEGA-C primary
sample consists of 1005 galaxies with $\mstar > 10^{10.5} \, \msun$
and the LEGA-C final sample consists of 818 galaxies (81~percent, cf.~grey
filled and red empty histograms in panel~\subref{f.samp.samp.e}). For
star-forming galaxies, the numbers are 1210 and 339 (28~percent, cf.~grey filled
and blue empty histograms in panel~\subref{f.samp.samp.f}).

Thus, in summary, our sample provides a highly-complete view of the
SAMI galaxies and of the LEGA-C quiescent galaxies, but is considerably skewed to
large \mstar for the LEGA-C star-forming subset. For MAGPI, our selection is highly
representative of the parent sample, but the parent sample is itself skewed to
large values of \mstar, because MAGPI focusses on central galaxies.
Given that we find \h4 to correlate with \mstar, correcting for the selection bias
against low-mass star-forming galaxies in LEGA-C leads to our results becoming even
stronger (cf. Appendix~\ref{app.sel}).

Note that when we compare \h4 to other galaxy observables in
\S~\ref{s.r.ss.mr}--\ref{s.r.ss.mvs},
the actual sample sizes vary according to the availability of the ancillary data
required for each comparison.
In most cases, the change in sample size is small (e.g., only 454/507 quiescent
SAMI galaxies have measurements of \vse). However, we stress again that only one
third of LEGA-C galaxies have measurements of \vse (i.e. only 297/818 quiescent
galaxies and only 132/339 star-forming galaxies), but this selection causes no
bias, as it is a selection by position angle only.

\section{Results}\label{s.r}

In this section, we show that star-forming and quiescent galaxies have different
distributions of \h4, even after matching the samples by stellar mass or $S/N$
(\S~\ref{s.r.ss.sfq}). We then investigate the relation of \h4 with \vse and $q$
(\S~\ref{s.r.ss.qvs}) and find the trends expected from the toy model of
\S~\ref{s.tmods}.
We then move on to study what other galaxy observables are good predictors of
\h4, starting with stellar mass and size (\S~\ref{s.r.ss.mr}), stellar mass and
aperture dispersion (\S~\ref{s.r.ss.ms}), and, finally, stellar mass
and rotation-to-dispersion ratio (\S~\ref{s.r.ss.mvs}), which we find to be the
two most likely drivers of \h4.

Throughout this section we use two statistical tools. To compare the distribution
of \h4 between star-forming and quiescent galaxies (\S~\ref{s.r.ss.sfq})
we use the KS test, for which we quote only the probability value $P_\mathrm{KS}$.

In \S\S~\ref{s.r.ss.qvs} and \ref{s.r.ss.mr}--\ref{s.r.ss.mvs}, we
study how \h4 varies as a function of two other observables. In all cases, these
two observables are correlated (e.g., the mass--size relation, \S~\ref{s.r.ss.mr},
or the stellar-mass Faber--Jackson relation, \S~\ref{s.r.ss.ms}).
As a means to distinguish between primary correlations among related
variables, and secondary correlations that arise as a consequence of primary
correlations, we use partial correlation coefficients
\citep[PCCs; see e.g.,][]{bait+2017, bluck+2019, baker+2022}.
In general, if two variables $x$ and $z$ (e.g., \mstar and \h4) are both
independently correlated with a third variable $y$ (e.g., \re), then this will
induce an apparent correlation between $y$ and $z$ (i.e., \h4 and \re).
PCCs address this issue by quantifying the strength and significance of the
correlation between $y$ and $z$ while controlling for $x$. Similar to the
standard Spearman rank correlation coefficient, a value of zero implies no
correlation, and -/+ 1 implies perfect anti/correlation. In the following, we
denote with $\rho(x, z \vert \, y)$ the partial correlation coefficient between
$x$ and $z$ removing the effect of $y$.
In the context of PCCs, $P$ is the probability that the measured PCC arose by
chance from uncorrelated data.
The numerical values of the PCC and their $P$ values are reported in
Table~\ref{t.r.h4corr}. In addition to these values, in the relevant figures, we
provide the graphical representation of the PCCs as an arrow; the angle and 
direction of this arrow are defined by the $\arctan$ of the ratio between the
PCCs \citep{bluck+2020a}. On the $x$--$y$ plane, an angle equal to 0\textdegree\
means that $z$ correlates with $x$ but not $y$; 90\textdegree~means that
$z$ correlates with $y$ but not $x$, 180\textdegree~means that $z$ anti-correlates
with $x$ but not with $y$, and so on.
Note that, in principle, a meaningful arrow
representation requires that the figures are scaled so that the data have the
same standard deviation along $x$ and $y$. Because this is not always practical
(i.e., to avoid figures with unsavoury aspect ratios), the arrows are always
scaled as if the data have the same standard deviation, even when the figures
are not.

\subsection{Different \texorpdfstring{\h4}{h4} between star-forming and quiescent galaxies}\label{s.r.ss.sfq}

\begin{figure*}
  \includegraphics[type=pdf,ext=.pdf,read=.pdf,width=1.\textwidth]{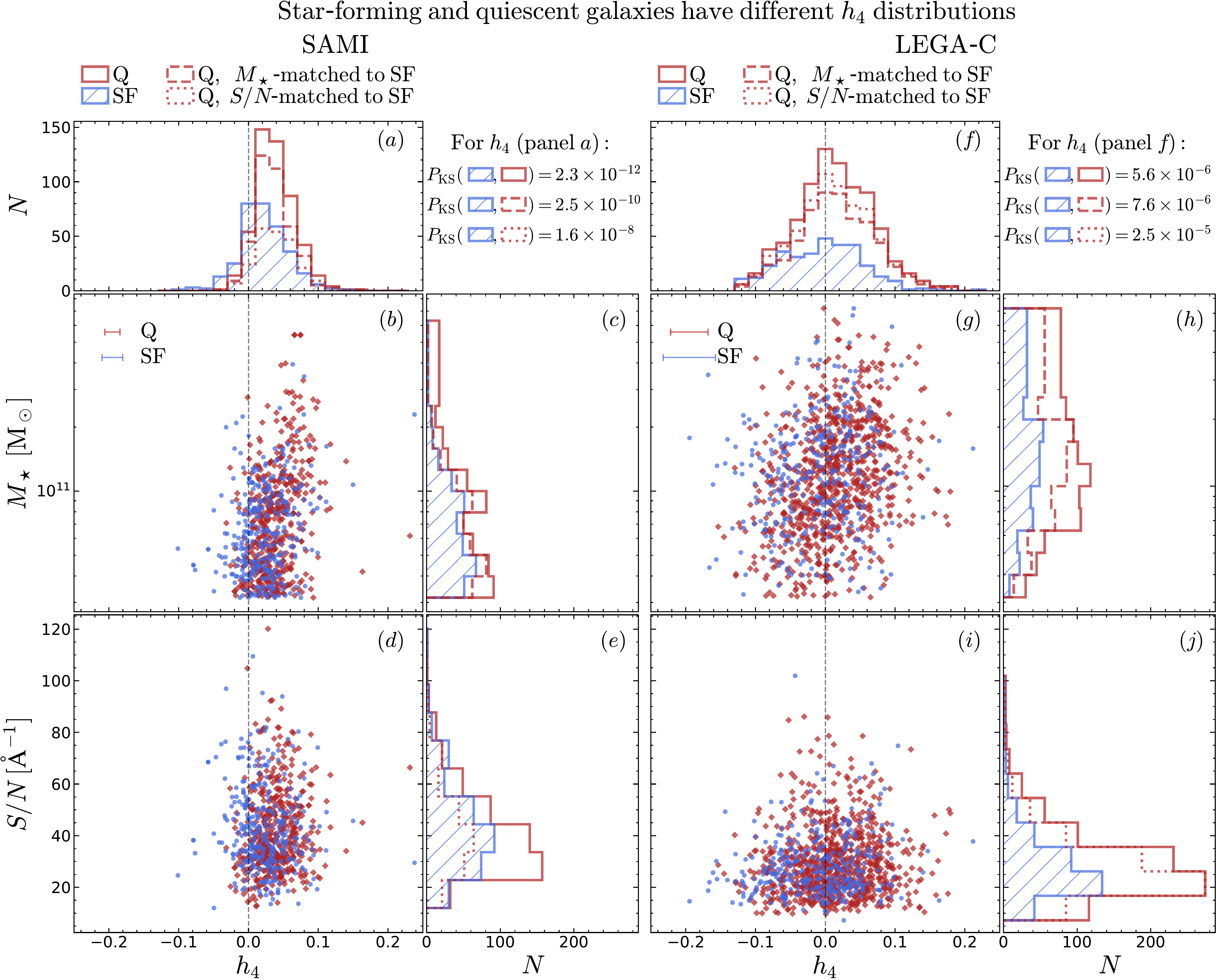}
  {\phantomsubcaption\label{f.r.sfq.a}
   \phantomsubcaption\label{f.r.sfq.b}
   \phantomsubcaption\label{f.r.sfq.c}
   \phantomsubcaption\label{f.r.sfq.d}
   \phantomsubcaption\label{f.r.sfq.e}
   \phantomsubcaption\label{f.r.sfq.f}
   \phantomsubcaption\label{f.r.sfq.g}
   \phantomsubcaption\label{f.r.sfq.h}
   \phantomsubcaption\label{f.r.sfq.i}
   \phantomsubcaption\label{f.r.sfq.j}
  }
  \caption{On average, quiescent (Q) galaxies have larger \h4 compared to
  star-forming (SF) galaxies, even after controlling for stellar mass or $S/N$.
  Panels~\subref{f.r.sfq.a}--\subref{f.r.sfq.e} show SAMI galaxies,
  panels~\subref{f.r.sfq.f}--\subref{f.r.sfq.j} show LEGA-C galaxies.
  In panels~\subref{f.r.sfq.b}, \subref{f.r.sfq.d}, \subref{f.r.sfq.g}
  and~\subref{f.r.sfq.i}, blue circles/red diamonds represent
  star-forming/quiescent galaxies; the errorbars are the median uncertainties
  (for \h4) or a uniform uncertainty of 0.15~dex (for \mstar). All other panels
  show the marginalised
  distributions. The vertical dashed lines trace $\h4 = 0$, corresponding to
  a Gaussian LOSVD.
  Star-forming and quiescent galaxies have
  different \h4 distributions (hatched blue and red histograms in
  panels~\subref{f.r.sfq.a} and~\subref{f.r.sfq.f}), but could this difference
  be due to different \mstar or $S/N$ distributions? (cf.~panels~\subref{f.r.sfq.b},
  \subref{f.r.sfq.d}, \subref{f.r.sfq.g} and~\subref{f.r.sfq.i}).
  The dashed red histograms show a sample of quiescent galaxies randomly drawn
  to match the \mstar distribution of the star-forming sample
  (for SAMI: panels~\subref{f.r.sfq.a} and~\subref{f.r.sfq.c}, for LEGA-C:
  panels~\subref{f.r.sfq.f} and~\subref{f.r.sfq.h}); the dotted red histograms
  show a sample of quiescent galaxies randomly drawn to match the $S/N$
  distribution of the star-forming sample
  (for SAMI: panels~\subref{f.r.sfq.a} and~\subref{f.r.sfq.e}, for LEGA-C:
  panels~\subref{f.r.sfq.f} and~\subref{f.r.sfq.j}). Comparing \h4 of these
  `matched' quiescent samples to the star-forming sample from the same survey, we
  still find they are different (the Kolmogorov-Smirnov test probabilities are
  reported to the right of panels~\subref{f.r.sfq.a} and~\subref{f.r.sfq.f} for SAMI
  and LEGA-C; the labels are the same as the histograms).
  }\label{f.r.sfq}
\end{figure*}

Fig.~\ref{f.r.sfq} shows \h4 for star-forming (SF, blue)
and quiescent galaxies (Q, red), as a function of \mstar and $S/N$.
The first two columns show the SAMI sample
(panels~\subref{f.r.sfq.a}--\subref{f.r.sfq.e}) and the last  two columns show the 
LEGA-C sample (panels~\subref{f.r.sfq.f}--\subref{f.r.sfq.j}); the MAGPI sample is
not shown in this figure. Blue circles/red diamonds represent individual
star-forming/quiescent galaxies.

The top two panels (\subref{f.r.sfq.a} and~\subref{f.r.sfq.f}) show the
distribution of \h4 marginalising over \mstar and $S/N$, for star-forming
(hatched blue histogram) and quiescent galaxies (solid red histogram);
galaxies with $\h4=0$ have a Gaussian LOSVD (vertical dashed line).
For SAMI, we find two star-forming galaxies with $\h4 > 0.1$. These are
SAMI~228105 (at \h4$\approx$0.15, a face-on spiral galaxy with a
strong bar) and SAMI~23623 (at \h4$\approx$0.24, a group central which
underwent a recent merger). The quiescent outlier is SAMI~537467 (at
\h4$\approx$0.23, which has a close neighbour capable of contaminating the
spectrum).
The histograms of star-forming and quiescent galaxies are
different. We report the main statistics in Table~\ref{t.r.sfq}.
Comparing the width of the \h4 distributions to the median uncertainties, we
conclude that the intrinsic scatter is the main driver of the histogram width.
This intrinsic scatter does not disappear if we consider narrow bins in \mstar,
so it seems to reflect a genuine variation in galaxy kinematics.

\begin{table}
\setlength{\tabcolsep}{5pt}
\begin{center}

\caption{Statistical properties of the \h4 distribution for
SAMI and LEGA-C galaxies. For both surveys, quiescent galaxies
have larger \h4 than star-forming galaxies. The difference in
median \h4 is statistically significant, for both SAMI
(6.7~$\upsigma$) and LEGA-C (4.4~$\upsigma$).
}\label{t.r.sfq}

\begin{tabular}{lcccc}
\hline
          Survey         & Subset &       median     & std. dev. & median $u(h_4)$ \\
\hline
\multirow{-0.5}{*}{SAMI}   &   SF   & $\phantom{-}0.019\pm0.002$ &   0.033   &      0.015      \\
\cline{2-5}                                                                        
                           &   Q    & $\phantom{-}0.034\pm0.001$ &   0.027   &      0.011      \\
\hline                                                                             
\multirow{-0.5}{*}{LEGA-C} &   SF   &           $-0.004\pm0.003$ &   0.062   &      0.038      \\
\cline{2-5}                                                                        
                           &   Q    & $\phantom{-}0.012\pm0.002$ &   0.055   &      0.027      \\
\hline
\\
\hline
\multirow{0.0}{*}{SAMI}   & SF\textbackslash SR & $\phantom{-}0.018\pm0.002$ &   0.036   &      0.015      \\
\cline{2-5}                                                                        
                           &  Q\textbackslash SR & $\phantom{-}0.030\pm0.001$ &   0.028   &      0.011      \\
\cline{2-5}                                                                        
                           &   SR   & $\phantom{-}0.063\pm0.002$ &   0.021   &      0.011      \\
\hline
\end{tabular}
\end{center}
\end{table}

Quantitatively, the probability for the null hypothesis that star-forming and
quiescent galaxies have the same \h4 distribution is $P_\mathrm{KS} =
2.3 \times 10^{-12}$ (SAMI) and $5.6\times 10^{-6}$ (LEGA-C). All
$P_\mathrm{KS}$ values are summarised to the right of panel~\subref{f.r.sfq.a}
(for SAMI) and of panel~\subref{f.r.sfq.f} (for LEGA-C). However, the star-forming
and quiescent samples differ not only in \h4, but also
in their \mstar distributions (panels~\subref{f.r.sfq.c} and~\subref{f.r.sfq.h});
besides, \h4 correlates with \mstar (panels~\subref{f.r.sfq.b}
and~\subref{f.r.sfq.g}). Can the difference in \mstar, together with the
\h4--\mstar correlation, explain the observed difference in \h4?
To address this question, we weight the quiescent sample to match the \mstar
distribution of the star-forming sample (dashed red histogram in 
panels~\subref{f.r.sfq.c} and~\subref{f.r.sfq.h}). Yet
even these `mass-matched' quiescent samples have different \h4 than the
corresponding star-forming
samples ($P_\mathrm{KS} = 2.5\times 10^{-10}$ and $7.6\times10^{-6}$ for SAMI and
LEGA-C, respectively).
We conclude that even controlling for \mstar, star-forming and quiescent galaxies have
different \h4 distributions.

In addition to \mstar, another possible concern is represented by the different
mean $S/N$ of star-forming and quiescent galaxies: even though star-forming
galaxies are brighter than quiescent galaxies of the same mass, they have less
prominent absorption features (note that we mask low-order Balmer lines to
avoid gas emission). Similarly to \mstar, we find that
\h4 correlates with $S/N$ (panels~\subref{f.r.sfq.d} and~\subref{f.r.sfq.e}), and
that star-forming and quiescent galaxies have different $S/N$ distributions
(panels~\subref{f.r.sfq.e} and~\subref{f.r.sfq.j}).
This systematic bias is potentially concerning because low-$S/N$ may bias \h4
(\S~\ref{s.das.ss.hok.sss.penal}), but even after matching the quiescent
sample to the $S/N$ distribution of the star-forming sample (dotted red
histogram in panels~\subref{f.r.sfq.e} and~\subref{f.r.sfq.j}), the resulting
\h4 distributions differ from their star-forming counterparts (dotted red
histograms in panels~\subref{f.r.sfq.a} and~\subref{f.r.sfq.f}); we find
$P_\mathrm{KS} = 1.6 \times 10^{-8}$ and $2.5\times10^{-5}$ for SAMI and LEGA-C,
respectively.

For both SAMI and
LEGA-C, and for both quiescent and star-forming galaxies, we find a statistically
significant correlation between \h4 and \mstar; in contrast, correlations between
\h4 and $S/N$ are either not statistically significant, or, when they are, they
are weaker and less significant than the \h4--\mstar correlation.

We conclude that, even accounting for \mstar and $S/N$, star-forming
and quiescent galaxies have different \h4 distributions, both in the local
Universe as well as 7~Gyr ago. Quiescent galaxies have on average higher \h4;
the difference between the median \h4 of quiescent and star-forming galaxies is
$0.015\pm0.003$ (for SAMI) and $0.016\pm0.004$ (for LEGA-C). We do not
compare \h4 between different surveys, because that is the subject of a future
paper.

\subsubsection{Relation with resolved \texorpdfstring{\h4}{h4} and rotation-to-dispersion ratio}\label{s.r.ss.sfq.sss.round}

According to the toy models of \S~\ref{s.tmods}, our integrated \h4 measurements
are physically related to both spatially resolved \h4 as well as \vse.
A physical connection with galaxy shape cannot be ruled out, but our thin-disc
models do not capture this aspect. To find whether the reported differences in
\h4 between star-forming and quiescent galaxies are due to differences in \vse
or in resolved \h4, we repeat the analysis from the previous section for two
subsets: round galaxies ($q \geq 0.75$) and galaxies with
$\vse \leq 0.5$. With these two selections \h4 reflects primarily
resolved \h4 (Table~\ref{t.tmods.mod}, rows~2 and~4; cf. columns~4 and~5).

For galaxies with $q \geq 0.75$, star-forming and quiescent galaxies are still
different in their \h4 (largest $P_\mathrm{KS}$ is 0.01); in contrast, we detect
no difference if we require $\vse \leq 0.5$. Note this does not
necessarily rule out the existence of differences in resolved \h4, but --- if
such differences exist --- they occur together with differences in \vse.

\subsubsection{Relation with the fast- and slow-rotators classification}\label{s.r.ss.sfq.sss.frsr}

\begin{figure}
  \includegraphics[type=pdf,ext=.pdf,read=.pdf,width=1.\columnwidth]{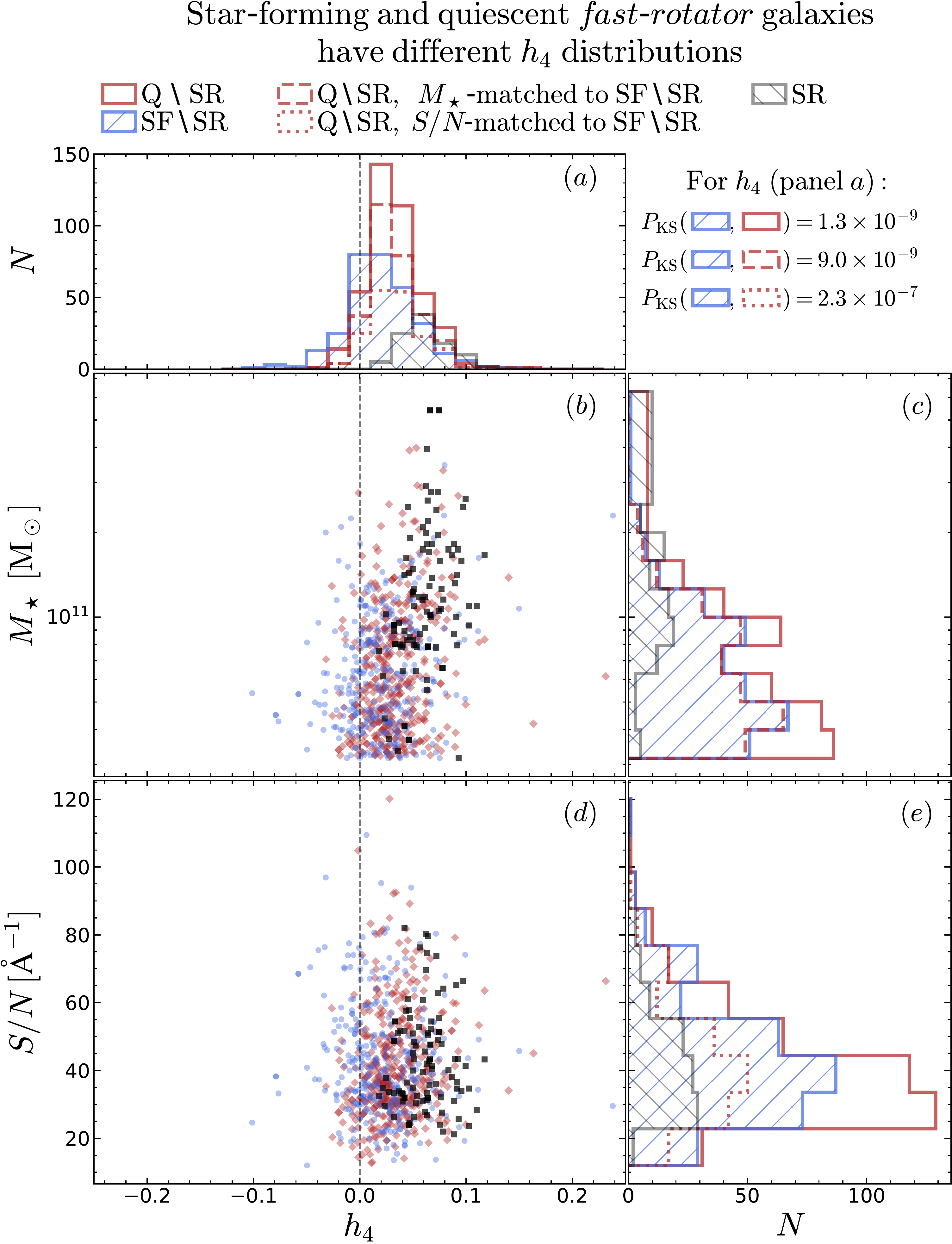}
  {\phantomsubcaption\label{f.r.srfr.a}
   \phantomsubcaption\label{f.r.srfr.b}
   \phantomsubcaption\label{f.r.srfr.c}
   \phantomsubcaption\label{f.r.srfr.d}
   \phantomsubcaption\label{f.r.srfr.e}
  }
  \caption{Same as Fig.~\ref{f.r.sfq}, panels~\subref{f.r.sfq.a}--\subref{f.r.sfq.e},
  but removing slow-rotator (SR) galaxies. The latter are represented by the black squares
  and black hatched histogram. SRs have larger average \h4 than fast-rotator galaxies
  (panel~\subref{f.r.srfr.a}, cf.~empty red and hatched blue histograms vs hatched black
  histogram). However, removing SRs reduces -- but does not remove -- the
  difference in \h4 between star-forming and quiescent galaxies. This is true even after
  matching quiescent galaxies in \mstar or $S/N$ to the star-forming galaxies.
  }\label{f.r.srfr}
\end{figure}

We now investigate the relation between \h4 (and the reported difference between
star-forming and quiescent galaxies) and the kinematic paradigm of slow and fast
rotators \citep{emsellem+2007, cappellari+2007}. We do this by using a definition
of slow-rotator galaxy (SR) based on the SAMI kinematic morphology classification
\citep[see \S~\ref{s.das.ss.anc.sss.vse} and][]{vandesande+2021a}.
Fig.~\ref{f.r.srfr} repeats the SAMI portion of Fig.~\ref{f.r.sfq} (left columns)
but separating SR galaxies (black squares and hatched black histograms).
A KS test confirms that, even after removing SRs, star-forming and fast-rotator (FR)
quiescent galaxies have different distributions of \h4, with quiescent galaxies
having on average larger \h4. This result holds even after matching the
star-forming and quiescent populations in \mstar or $S/N$ (the relevant
$P_\mathrm{KS}$ values are reported in the top right of Fig.~\ref{f.r.srfr}).
Compared to the undivided quiescent population, FR quiescent galaxies are
closer to the star-forming galaxies, as can be seen by comparing the
$P_\mathrm{KS}$ values between Fig.s~\ref{f.r.sfq} and~\ref{f.r.srfr} (the
difference in $P_\mathrm{KS}$ is not due to sample size).
Clearly, because most SRs are quiescent, FR quiescent galaxies
have an \h4 distribution that is more similar to that of star-forming galaxies
(cf.~red vs blue histogram). We can conclude that -- for \h4 -- FR quiescent
galaxies are intermediate between star-forming galaxies and SR quiescent galaxies.

\subsection{Correlations with galaxy projected shape and rotation-to-dispersion ratio}\label{s.r.ss.qvs}

\begin{figure}
  \includegraphics[type=pdf,ext=.pdf,read=.pdf,width=1.\columnwidth]{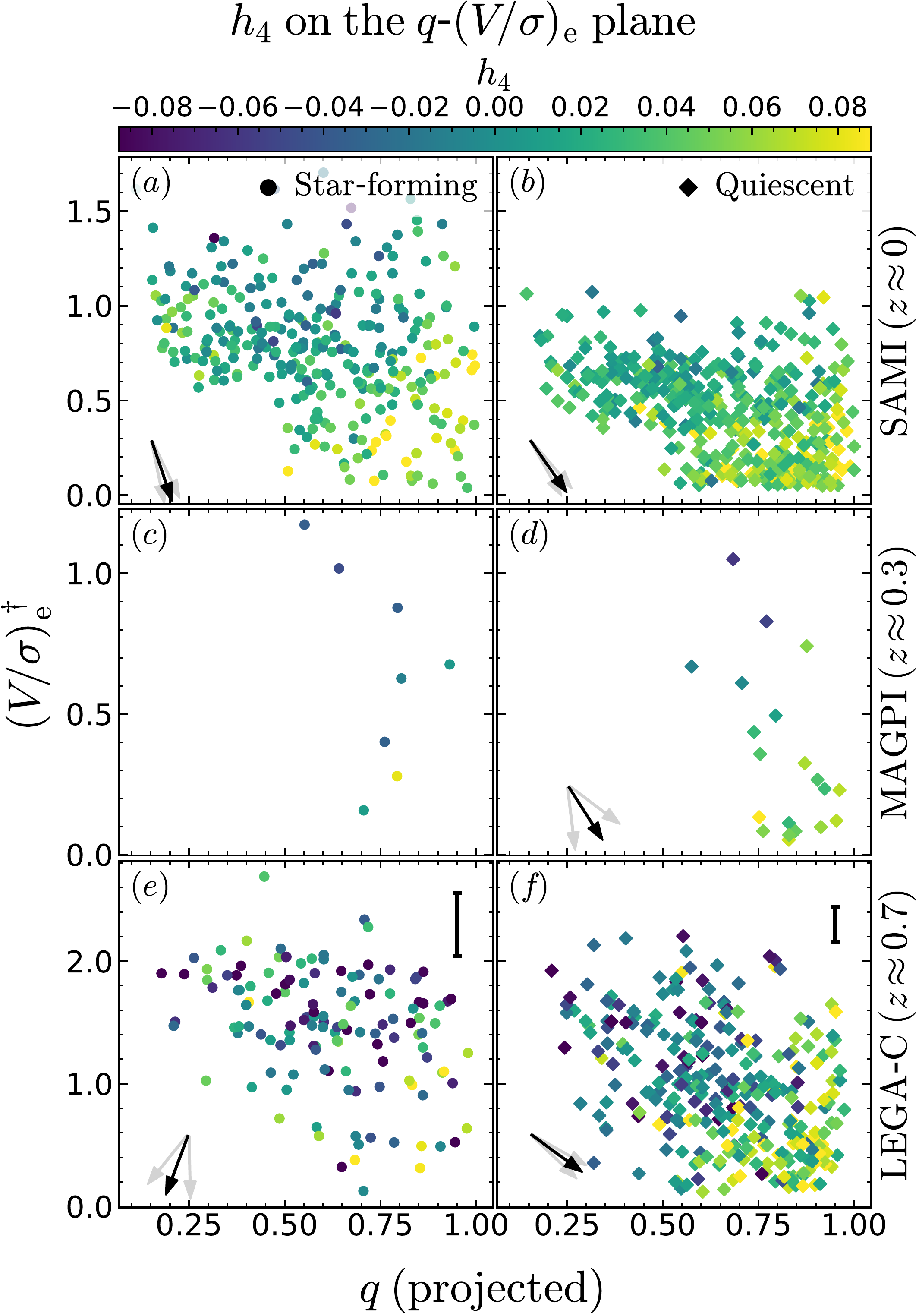}
  {\phantomsubcaption\label{f.r.h4_qvs.a}
   \phantomsubcaption\label{f.r.h4_qvs.b}
   \phantomsubcaption\label{f.r.h4_qvs.c}
   \phantomsubcaption\label{f.r.h4_qvs.d}
   \phantomsubcaption\label{f.r.h4_qvs.e}
   \phantomsubcaption\label{f.r.h4_qvs.f}
  }
  \caption{
  Our samples on the shape--rotation-to-dispersion plane, colour-coded by \h4. The
  left/right columns show star-forming/quiescent galaxies, the top/middle/bottom
  rows show the SAMI/MAGPI/LEGA-C sample. The direction of the black arrows
  indicates the relative strength of the \h4--$q$ and \h4--\vse correlations
  (grey arrows show the 16\textsuperscript{th}-84\textsuperscript{th} range from
  bootstrapping). The numerical value of the PCCs and the angle of the arrows
  are reported in Table~\ref{t.r.h4corr}, rows~1--12. 
  The strong relation between \h4, \vse and projected $q$ highlights
  that these three parameters capture different aspects of the orbital structure
  of a galaxy.\\
  $^\dagger$ \vse has two different meanings for SAMI and MAGPI vs LEGA-C:
  for SAMI and MAGPI, it is the observed value \textit{inside} one~\re, for LEGA-C it is
  the best-fit model value \textit{at} one~\re (\S~\ref{s.das.ss.anc.sss.vse}).
  However, our aim is to show how
  \h4 relates to the degree of rotation support, not to quantify this
  dependence as a function of redshift.
  }\label{f.r.h4_qvs}
\end{figure}

Guided by our toy models (\S~\ref{s.tmods}), we now study how \h4 is related to
projected axis ratio $q$ and to the ratio between rotation and dispersion \vse,
which are two other tracers of orbital structure\footnote{It is understood
that $q$ also depends on inclination, so its value as a probe of orbital
structure is only statistical.}
\citep[e.g.][]{binney1978, davies+1983, cappellari+2007, emsellem+2011}.
In Fig.~\ref{f.r.h4_qvs} we show the $q$--\vse plane, where the symbols
represent individual galaxies, colour coded by \h4. The left/right columns show
star-forming/quiescent galaxies (represented as circles/diamonds), and the three
rows from top to bottom correspond to SAMI, MAGPI and LEGA-C data.
In each panel, the black arrows are a graphical representation of the PCCs (the
grey arrows are the 16\textsuperscript{th} and 84\textsuperscript{th}
percentiles of the distribution of angles after bootstrapping each subset one
thousand times). The value of the PCCs, the associated $P$ values and the
resulting angle $\theta$ are reported in Table~\ref{t.r.h4corr}; rows~1--12,
columns~7--9. We highlight in bold correlations that are statistically
significant, assumed here to have $P < 10^{-3}$ (we recall that \vse has two
different meanings: for SAMI and MAGPI, it is the observed ratio \textit{within}
one~\re, whereas for LEGA-C it is a model-inferred ratio evaluated \textit{at}
one~\re; \S~\ref{s.das.ss.anc.sss.vse}).

For SAMI and
LEGA-C, we have large enough samples to observe the well known anti-correlation
between \vse and $q$ \citep[see e.g.][their fig.~6, where $\epsilon_\mathrm{e}
\sim 1 - q$]{emsellem+2011}. If we look at the correlation coefficient between
\h4 and either of these two parameters, we find in both cases a statistically
significant correlation. 

To disentangle the \vse--$q$ anti-correlation from the relation with \h4,
in the following we always use PCCs. The PCCs between $q$ and \h4 while
controlling for \vse are reported in rows 1, 3, 5, 7, 9 and~11 of
Table~\ref{t.r.h4corr}; the PCCs between \vse and \h4 while controlling
for $q$ are reported in rows 2, 4, 6, 8, 10 and~12.
In general, \h4 anti-correlates with \vse:
this is true for SAMI (rows~2 and~4 in Table~\ref{t.r.h4corr}) and for LEGA-C
quiescent galaxies (row~12).
For MAGPI, none of the PCCs are significant, likely due to the small
sample size, but we highlight that the anti-correlation between \h4 and \vse has
particularly low $P$ (row~8, column~8); this possible anti-correlation can also
be appreciated in the clear gradient of colour hues in panel~\subref{f.r.h4_qvs.c}.
Finally, for LEGA-C and MAGPI star-forming galaxies, we find no correlation with
\vse.

Unlike for \vse, the relation between \h4 and $q$ is related to the star-forming
status of galaxies. For star-forming galaxies, the \h4--$q$ correlation is weak
(SAMI) or not significant (MAGPI and LEGA-C). This is expected from the fact
that the star-forming subsets are dominated by intrinsically flat systems, for
which projected shape $q$ is driven primarily by the random viewing angle.
In contrast, for quiescent galaxies, we find a positive correlation for
both SAMI and LEGA-C (rows~3 and~11). The PCCs are larger than the values found
for star-forming galaxies (column 7, cf. rows 1~and~3, and 9~and~11). Because
projected shape $q$ is also related to intrinsic shape, this correlation
suggests that intrinsically round galaxies tend to have higher \h4.

We remark that, for SAMI, the \h4--\vse correlation has larger PCC and higher
statistical significance than the \h4--$q$ correlation, but the opposite is true
for LEGA-C quiescent galaxies. However, because \vse is calculated differently
between the two samples (\S~\ref{s.das.ss.anc.sss.vse}), we cannot say if the
different importance of the two correlations is meaningful.

In summary, we find statistically significant correlations of \h4 with \vse and
$q$ in three out of six subsets (three out of four subsets excluding MAGPI,
which has <25 galaxies). This confirms that our spatially integrated \h4 contains
kinematic information related to the orbital structure of stars in galaxies, an
independent validation of our measurements.
Because we use PCCs, the strong anti-correlation between \vse and $q$ is always
removed when calculating the correlations with \h4. For this reason, the separate
existence of the \h4--$q$ and \h4--\vse correlations means that, for quiescent
galaxies, \h4 contains independent orbital information that is captured
separately by $q$ and \vse.
We note that none of the above conclusions change if we change the quality selection
criteria from $u(\h4) < 0.05$ to a stricter cut $u(\h4) < 0.03$.

\subsection{Correlations with galaxy mass and size}\label{s.r.ss.mr}

\begin{figure}
  \includegraphics[type=pdf,ext=.pdf,read=.pdf,width=1.\columnwidth]{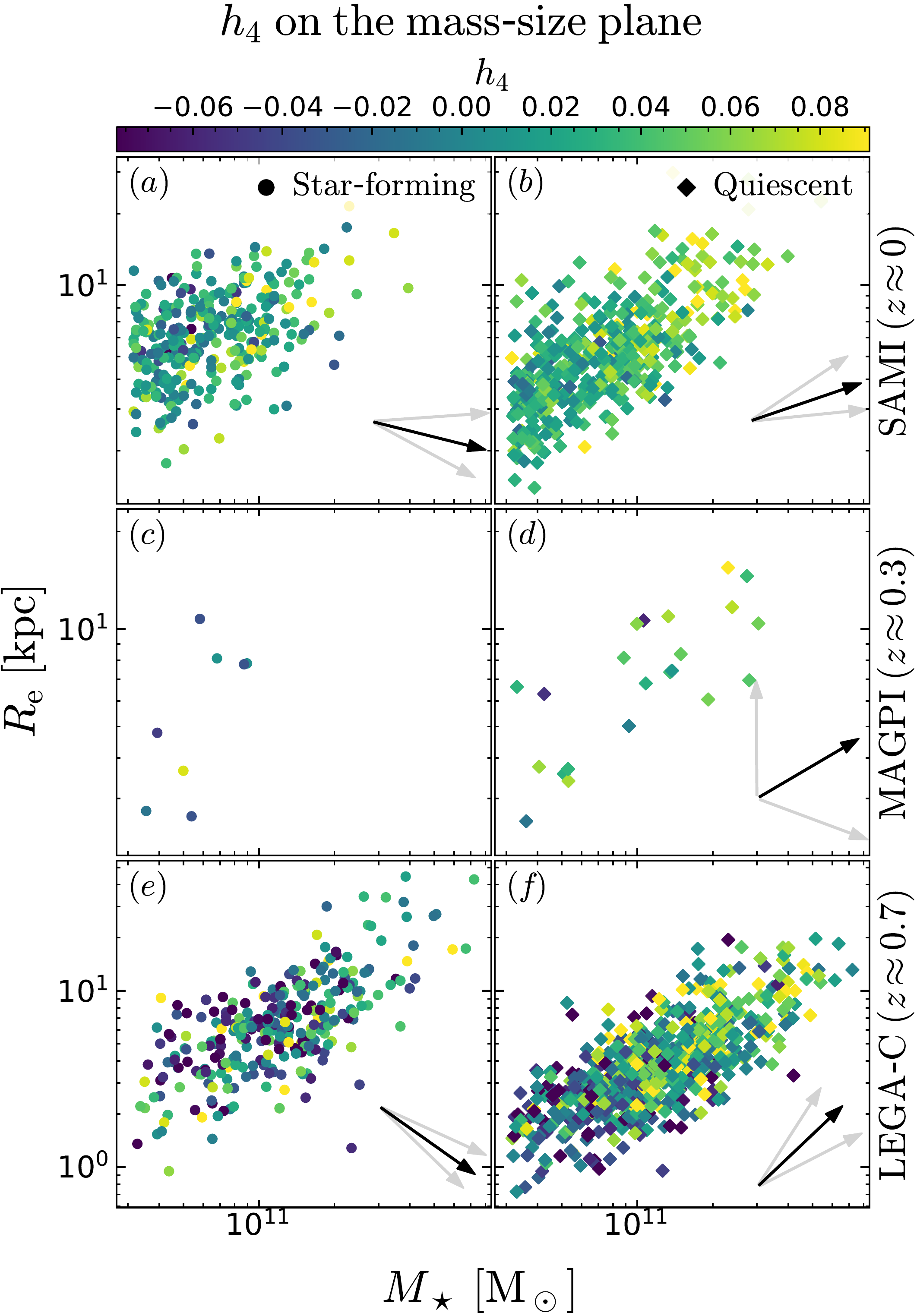}
  {\phantomsubcaption\label{f.r.h4_mr.a}
   \phantomsubcaption\label{f.r.h4_mr.b}
   \phantomsubcaption\label{f.r.h4_mr.c}
   \phantomsubcaption\label{f.r.h4_mr.d}
   \phantomsubcaption\label{f.r.h4_mr.e}
   \phantomsubcaption\label{f.r.h4_mr.f}
  }
  \caption{Our galaxies on the mass--size plane, colour-coded by \h4. The
  symbols are the same as Fig.~\ref{f.r.h4_qvs}. The numerical value of the PCCs
  and the angle of the arrows are reported in Table~\ref{t.r.h4corr}, rows~2--24.
  All the arrows fall in the first or fourth quadrant, consistent
  with a positive \h4--\mstar correlation after controlling for \re. For LEGA-C
  quiescent galaxies, we also find evidence for a correlation between \h4 and \re,
  after controlling for \mstar. The negative \h4--\re correlation for LEGA-C
  star-forming galaxies (panel~\subref{f.r.h4_mr.e}) is not statistically
  significant.
  }\label{f.r.h4_mr}
\end{figure}

The relation of \h4 with stellar mass and size is illustrated in Fig.~\ref{f.r.h4_mr}.
In all panels except~\subref{f.r.h4_mr.c}, we observe a correlation between \mstar
and \re \citep{shen+2003}. We remark that the mass distribution of the three
samples is not the same, with LEGA-C probing on average more massive galaxies
than SAMI, for both the quiescent and star-forming subsets
(cf.~Fig.~\ref{f.samp.samp}). 
Despite this different selection, we can appreciate by eye the increase in
average \re between the look-back times of LEGA-C and SAMI \citep{vanderwel+2014}.
The PCCs and angle are reported in Table~\ref{t.r.h4corr}; rows 13--24, columns 7--9.

For star-forming galaxies (panels~\subref{f.r.h4_mr.a}, \subref{f.r.h4_mr.c}
and~\subref{f.r.h4_mr.e}), the black arrows lie in the fourth quadrant, suggesting
that \h4 correlates with \mstar but anti-correlates with \re. However, quantitatively,
the only statistically significant correlation is the \h4--\mstar correlation for SAMI
star-forming galaxies; the sample size for MAGPI is probably too small, whereas for
LEGA-C, the $P$ value corresponds to three standard deviations, very close but above
the adopted significance threshold ($P<10^{-3}$). In contrast, we find no
statistical evidence for the \h4--\re anti-correlation.

For quiescent galaxies (panels~\subref{f.r.h4_mr.b}, \subref{f.r.h4_mr.d}
and~\subref{f.r.h4_mr.f}), the black arrows lie in the first quadrant,
indicating that \h4 correlates with both \mstar and \re. However, quantitatively,
only the LEGA-C sample shows statistically significant independent correlations of
\h4 with \mstar and \re. For MAGPI, there is no evidence of a correlation with
\mstar either.

\subsection{Correlations with stellar mass and aperture velocity dispersion}\label{s.r.ss.ms}

\begin{figure}
  \includegraphics[type=pdf,ext=.pdf,read=.pdf,width=1.\columnwidth]{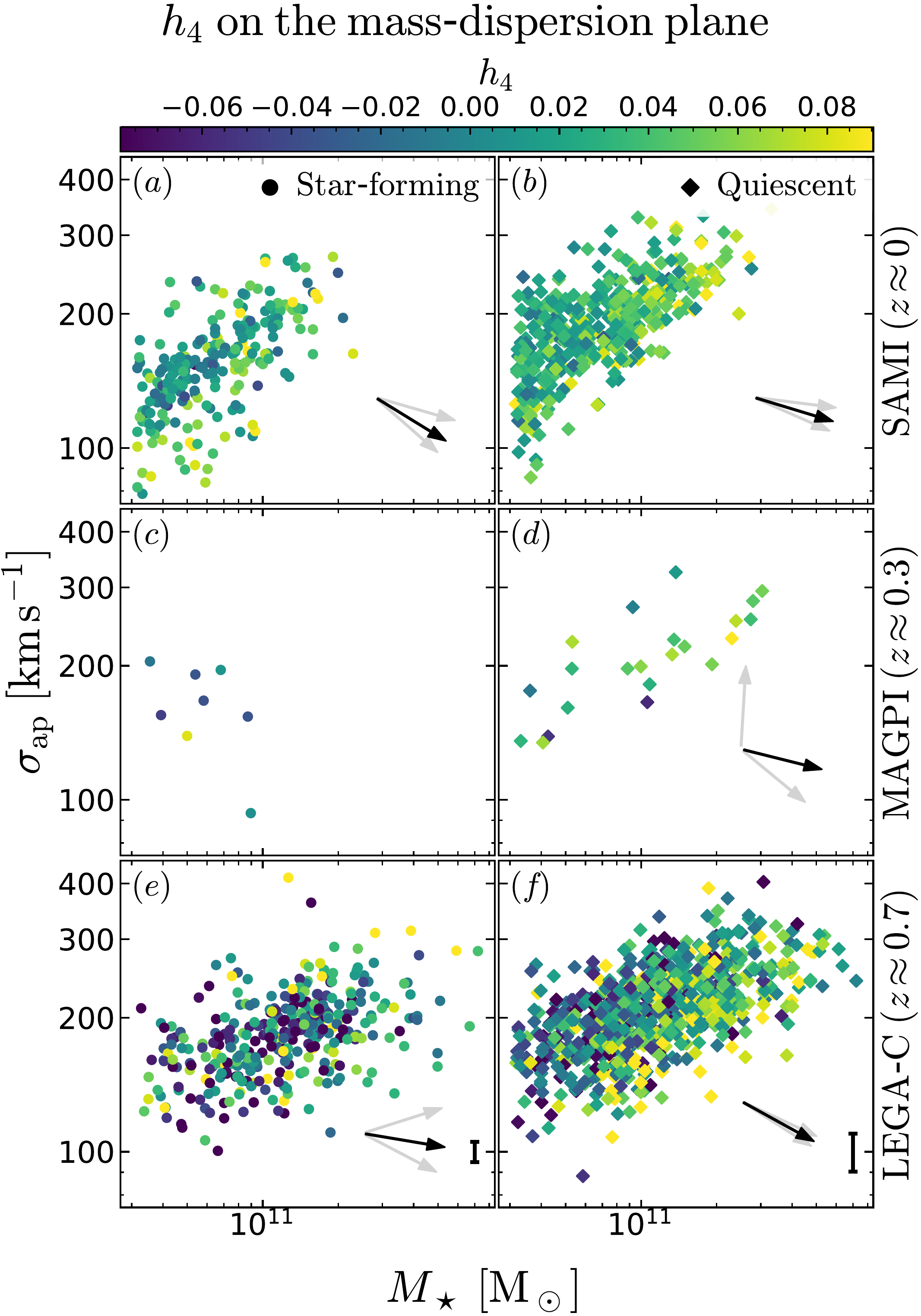}
  {\phantomsubcaption\label{f.r.h4_ms.a}
   \phantomsubcaption\label{f.r.h4_ms.b}
   \phantomsubcaption\label{f.r.h4_ms.c}
   \phantomsubcaption\label{f.r.h4_ms.d}
   \phantomsubcaption\label{f.r.h4_ms.e}
   \phantomsubcaption\label{f.r.h4_ms.f}
  }
  \caption{
  Our galaxies on the mass--dispersion plane, colour-coded by \h4. The
  symbols are the same as Fig.~\ref{f.r.h4_qvs}, the PCCs and the angle of the
  arrows are reported in Table~\ref{t.r.h4corr}, rows~25--36.
  \h4 correlates more strongly with \mstar than with
  aperture velocity dispersion \sigap, both for local quiescent galaxies
  (panel~\subref{f.r.h4_ms.c}) and for quiescent
  galaxies at $z=0.7$ (panel~\subref{f.r.h4_ms.f}).
  After the dissipative formation phase, changes in \sigap are only fractional,
  while \mstar may more than double via gas-poor mergers. Therefore,
  positive correlations of \h4 with \mstar (after controlling for
  \sigap) are suggestive of non-dissipative mass growth.
  }\label{f.r.h4_ms}
\end{figure}

Alongside \mstar, a key galaxy observable is the second moment of the
aperture-integrated LOSVD, \sigap (\S~\ref{s.das.ss.anc.sss.vse}).
This kinematic observable is of particular
interest because, compared to \mstar, relative changes in \sigap are small during
the evolution of a galaxy \citetext{e.g. \citealp{bezanson+2009},
\citealp{naab+2009}, \citealp{oser+2012}, \citealp{vandesande+2013} --- with the
exception of the initial phases of rapid gas accretion and star-formation}.
Our findings are illustrated in Fig.~\ref{f.r.h4_ms} and reported in
Table~\ref{t.r.h4corr}; rows 25--36.

The direction of the arrows indicates clearly that, between \mstar and \sigap,
it is \mstar that correlates most strongly with \h4. However, statistically, we find
significant \h4--\mstar correlations only for SAMI and LEGA-C quiescent galaxies.
For LEGA-C quiescent galaxies, we find a statistically significant anti-correlation
with \sigap; interestingly, this negative correlation disappears if we replace
\sigap with its inclination-corrected equivalent $\sigma_\mathrm{vir}$
\citep{vanderwel+2022}.

    \begin{table*}
    \begin{center}
    
    \caption{
    Partial correlations of \h4 with \vse and $q$ (rows 1--12), \mstar and \re
    (rows~13--24), \mstar and \sigap (rows~25--36), and \mstar and \vse (rows~37--48).
    Statistically significant correlations ($P < 10^{-3}$) are highlighted in bold.
    The strongest and most significant correlations are with \vse, followed by \mstar.
    For MAGPI, the lack of correlation is likely due to the small sample size.
    For SAMI, the reported trends persist if we exclude SR galaxies (but the $P$ values
    are lower).
    }\label{t.r.h4corr}
    
    \begin{tabular}{ccccccccc}
    \hline
        & & survey & subset & $N$ & PCC & $\rho$ & $P$ & $\theta$ \\
       (1) & (2) & (3) & (4) & (5) & (6) & (7) & (8) & (9) \\
    \hline
     & (1) &  &  &  & $\rho(h_4, q \vert (V/\sigma)_\mathrm{e})$ & $0.17 (0.11, 0.26)$ & $4.0 \times 10^{-3}$ & \\
 & (2) &  & \multirow{-2}{*}{SF} & \multirow{-2}{*}{$298$} & $\bm{\rho(h_4, (V/\sigma)_\mathrm{e} \vert q)}$ & $-0.45 (-0.47, -0.51)$ & $\mathbf{1.5 \times 10^{-16}}$ & \multirow{-2}{*}{$-69.9 (-77.2, -62.8)$}\\
\cline{4-9}
 & (3) &  &  &  & $\bm{\rho(h_4, q \vert (V/\sigma)_\mathrm{e})}$ & $0.32 (0.26, 0.39)$ & $\mathbf{3.0 \times 10^{-11}}$ & \\
 & (4) & \multirow{-4}{*}{SAMI} & \multirow{-2}{*}{Q} & \multirow{-2}{*}{$423$} & $\bm{\rho(h_4, (V/\sigma)_\mathrm{e} \vert q)}$ & $-0.42 (-0.42, -0.42)$ & $\mathbf{1.7 \times 10^{-19}}$ & \multirow{-2}{*}{$-53.0 (-58.5, -47.2)$}\\
\cline{3-9}
 & (5) &  &  &  & $\rho(h_4, q \vert (V/\sigma)_\mathrm{e})$ & $0.51 (0.17, 0.59)$ & $2.5 \times 10^{-1}$ & \\
 & (6) &  & \multirow{-2}{*}{SF} & \multirow{-2}{*}{$8$} & $\rho(h_4, (V/\sigma)_\mathrm{e} \vert q)$ & $-0.86 (-0.93, -0.59)$ & $1.2 \times 10^{-2}$ & \multirow{-2}{*}{$-59.6 (-79.4, -45.0)$}\\
\cline{4-9}
 & (7) &  &  &  & $\rho(h_4, q \vert (V/\sigma)_\mathrm{e})$ & $0.38 (0.05, 0.65)$ & $1.1 \times 10^{-1}$ & \\
 & (8) & \multirow{-4}{*}{MAGPI} & \multirow{-2}{*}{Q} & \multirow{-2}{*}{$20$} & $\rho(h_4, (V/\sigma)_\mathrm{e} \vert q)$ & $-0.55 (-0.58, -0.45)$ & $1.6 \times 10^{-2}$ & \multirow{-2}{*}{$-55.4 (-84.8, -34.3)$}\\
\cline{3-9}
 & (9) &  &  &  & $\rho(h_4, q \vert (V/\sigma)_\mathrm{e})$ & $-0.09 (-0.16, 0.01)$ & $3.1 \times 10^{-1}$ & \\
 & (10) &  & \multirow{-2}{*}{SF} & \multirow{-2}{*}{$132$} & $\rho(h_4, (V/\sigma)_\mathrm{e} \vert q)$ & $-0.22 (-0.18, -0.18)$ & $1.2 \times 10^{-2}$ & \multirow{-2}{*}{$-112.0 (-131.0, -87.6)$}\\
\cline{4-9}
 & (11) &  &  &  & $\bm{\rho(h_4, q \vert (V/\sigma)_\mathrm{e})}$ & $0.40 (0.33, 0.47)$ & $\mathbf{1.7 \times 10^{-12}}$ & \\
 \parbox[t]{2mm}{\multirow{-12}{*}{\rotatebox[origin=c]{90}{\hyperref[f.r.h4_qvs]{$q$-\vse plane}}}} & (12) & \multirow{-4}{*}{LEGA-C} & \multirow{-2}{*}{Q} & \multirow{-2}{*}{$297$} & $\bm{\rho(h_4, (V/\sigma)_\mathrm{e} \vert q)}$ & $-0.27 (-0.30, -0.24)$ & $\mathbf{2.7 \times 10^{-6}}$ & \multirow{-2}{*}{$-34.2 (-42.3, -27.0)$}\\
\hline
\multicolumn{9}{c}{}\\[-0.1em]\hline
 & (13) &  &  &  & $\bm{\rho(h_4, M_\star \vert R_\mathrm{e})}$ & $0.19 (0.24, 0.19)$ & $\mathbf{5.4 \times 10^{-4}}$ & \\
 & (14) &  & \multirow{-2}{*}{SF} & \multirow{-2}{*}{$314$} & $\rho(h_4, R_\mathrm{e} \vert M_\star)$ & $-0.04 (-0.12, 0.01)$ & $4.3 \times 10^{-1}$ & \multirow{-2}{*}{$-13.0 (-27.6, 4.5)$}\\
\cline{4-9}
 & (15) &  &  &  & $\bm{\rho(h_4, M_\star \vert R_\mathrm{e})}$ & $0.24 (0.25, 0.19)$ & $\mathbf{2.8 \times 10^{-7}}$ & \\
 & (16) & \multirow{-4}{*}{SAMI} & \multirow{-2}{*}{Q} & \multirow{-2}{*}{$451$} & $\rho(h_4, R_\mathrm{e} \vert M_\star)$ & $0.08 (0.02, 0.12)$ & $1.1 \times 10^{-1}$ & \multirow{-2}{*}{$17.5 (5.2, 31.3)$}\\
\cline{3-9}
 & (17) &  &  &  & $\rho(h_4, M_\star \vert R_\mathrm{e})$ & $0.26 (0.13, -0.37)$ & $5.8 \times 10^{-1}$ & \\
 & (18) &  & \multirow{-2}{*}{SF} & \multirow{-2}{*}{$8$} & $\rho(h_4, R_\mathrm{e} \vert M_\star)$ & $-0.24 (-0.44, 0.75)$ & $6.1 \times 10^{-1}$ & \multirow{-2}{*}{$-42.5 (-73.7, 116.3)$}\\
\cline{4-9}
 & (19) &  &  &  & $\rho(h_4, M_\star \vert R_\mathrm{e})$ & $0.20 (0.19, -0.00)$ & $3.9 \times 10^{-1}$ & \\
 & (20) & \multirow{-4}{*}{MAGPI} & \multirow{-2}{*}{Q} & \multirow{-2}{*}{$21$} & $\rho(h_4, R_\mathrm{e} \vert M_\star)$ & $0.14 (-0.08, 0.33)$ & $5.6 \times 10^{-1}$ & \multirow{-2}{*}{$34.3 (-22.2, 90.5)$}\\
\cline{3-9}
 & (21) &  &  &  & $\rho(h_4, M_\star \vert R_\mathrm{e})$ & $0.18 (0.17, 0.18)$ & $1.3 \times 10^{-3}$ & \\
 & (22) &  & \multirow{-2}{*}{SF} & \multirow{-2}{*}{$326$} & $\rho(h_4, R_\mathrm{e} \vert M_\star)$ & $-0.12 (-0.15, -0.07)$ & $3.6 \times 10^{-2}$ & \multirow{-2}{*}{$-33.3 (-41.9, -22.0)$}\\
\cline{4-9}
 & (23) &  &  &  & $\bm{\rho(h_4, M_\star \vert R_\mathrm{e})}$ & $0.14 (0.19, 0.13)$ & $\mathbf{9.5 \times 10^{-5}}$ & \\
 \parbox[t]{2mm}{\multirow{-12}{*}{\rotatebox[origin=c]{90}{\hyperref[f.r.h4_mr]{\mstar-\re plane}}}} & (24) & \multirow{-4}{*}{LEGA-C} & \multirow{-2}{*}{Q} & \multirow{-2}{*}{$764$} & $\bm{\rho(h_4, R_\mathrm{e} \vert M_\star)}$ & $0.12 (0.09, 0.19)$ & $\mathbf{6.1 \times 10^{-4}}$ & \multirow{-2}{*}{$41.3 (26.1, 55.4)$}\\
\hline
\multicolumn{9}{c}{}\\[-0.1em]\hline
 & (25) &  &  &  & $\rho(h_4, M_\star \vert \sigma_\mathrm{ap})$ & $0.21 (0.15, 0.15)$ & $1.3 \times 10^{-3}$ & \\
 & (26) &  & \multirow{-2}{*}{SF} & \multirow{-2}{*}{$242$} & $\rho(h_4, \sigma_\mathrm{ap} \vert M_\star)$ & $-0.12 (-0.13, -0.05)$ & $6.1 \times 10^{-2}$ & \multirow{-2}{*}{$-30.4 (-40.8, -17.7)$}\\
\cline{4-9}
 & (27) &  &  &  & $\bm{\rho(h_4, M_\star \vert \sigma_\mathrm{ap})}$ & $0.29 (0.35, 0.25)$ & $\mathbf{3.2 \times 10^{-9}}$ & \\
 & (28) & \multirow{-4}{*}{SAMI} & \multirow{-2}{*}{Q} & \multirow{-2}{*}{$409$} & $\rho(h_4, \sigma_\mathrm{ap} \vert M_\star)$ & $-0.07 (-0.14, -0.03)$ & $1.4 \times 10^{-1}$ & \multirow{-2}{*}{$-14.4 (-21.7, -5.7)$}\\
\cline{3-9}
 & (29) &  &  &  & $\rho(h_4, M_\star \vert \sigma_\mathrm{ap})$ & $0.11 (-0.81, 0.89)$ & $8.1 \times 10^{-1}$ & \\
 & (30) &  & \multirow{-2}{*}{SF} & \multirow{-2}{*}{$8$} & $\rho(h_4, \sigma_\mathrm{ap} \vert M_\star)$ & $-0.03 (-0.83, 0.97)$ & $9.5 \times 10^{-1}$ & \multirow{-2}{*}{$-15.7 (-134.2, 47.4)$}\\
\cline{4-9}
 & (31) &  &  &  & $\rho(h_4, M_\star \vert \sigma_\mathrm{ap})$ & $0.30 (0.51, 0.03)$ & $1.8 \times 10^{-1}$ & \\
 & (32) & \multirow{-4}{*}{MAGPI} & \multirow{-2}{*}{Q} & \multirow{-2}{*}{$22$} & $\rho(h_4, \sigma_\mathrm{ap} \vert M_\star)$ & $-0.07 (-0.39, 0.27)$ & $7.6 \times 10^{-1}$ & \multirow{-2}{*}{$-13.0 (-37.7, 83.2)$}\\
\cline{3-9}
 & (33) &  &  &  & $\rho(h_4, M_\star \vert \sigma_\mathrm{ap})$ & $0.14 (0.21, 0.14)$ & $8.7 \times 10^{-3}$ & \\
 & (34) &  & \multirow{-2}{*}{SF} & \multirow{-2}{*}{$339$} & $\rho(h_4, \sigma_\mathrm{ap} \vert M_\star)$ & $-0.02 (-0.10, 0.04)$ & $6.9 \times 10^{-1}$ & \multirow{-2}{*}{$-8.8 (-25.7, 17.4)$}\\
\cline{4-9}
 & (35) &  &  &  & $\bm{\rho(h_4, M_\star \vert \sigma_\mathrm{ap})}$ & $0.37 (0.36, 0.35)$ & $\mathbf{3.0 \times 10^{-27}}$ & \\
 \parbox[t]{2mm}{\multirow{-12}{*}{\rotatebox[origin=c]{90}{\hyperref[f.r.h4_qvs]{\mstar-\sigap plane}}}} & (36) & \multirow{-4}{*}{LEGA-C} & \multirow{-2}{*}{Q} & \multirow{-2}{*}{$818$} & $\bm{\rho(h_4, \sigma_\mathrm{ap} \vert M_\star)}$ & $-0.19 (-0.21, -0.15)$ & $\mathbf{7.6 \times 10^{-8}}$ & \multirow{-2}{*}{$-27.0 (-30.7, -23.2)$}\\
\hline
\multicolumn{9}{c}{}\\[-0.1em]\hline
 & (37) &  &  &  & $\rho(h_4, M_\star \vert (V/\sigma)_\mathrm{e})$ & $0.04 (-0.02, 0.12)$ & $4.6 \times 10^{-1}$ & \\
 & (38) &  & \multirow{-2}{*}{SF} & \multirow{-2}{*}{$298$} & $\bm{\rho(h_4, (V/\sigma)_\mathrm{e} \vert M_\star)}$ & $-0.49 (-0.55, -0.53)$ & $\mathbf{2.6 \times 10^{-19}}$ & \multirow{-2}{*}{$-85.0 (-92.1, -77.4)$}\\
\cline{4-9}
 & (39) &  &  &  & $\bm{\rho(h_4, M_\star \vert (V/\sigma)_\mathrm{e})}$ & $0.19 (0.16, 0.25)$ & $\mathbf{5.9 \times 10^{-5}}$ & \\
 & (40) & \multirow{-4}{*}{SAMI} & \multirow{-2}{*}{Q} & \multirow{-2}{*}{$423$} & $\bm{\rho(h_4, (V/\sigma)_\mathrm{e} \vert M_\star)}$ & $-0.47 (-0.53, -0.45)$ & $\mathbf{2.2 \times 10^{-24}}$ & \multirow{-2}{*}{$-67.5 (-73.4, -61.1)$}\\
\cline{3-9}
 & (41) &  &  &  & $\rho(h_4, M_\star \vert (V/\sigma)_\mathrm{e})$ & $0.36 (-0.33, 0.82)$ & $4.3 \times 10^{-1}$ & \\
 & (42) &  & \multirow{-2}{*}{SF} & \multirow{-2}{*}{$8$} & $\rho(h_4, (V/\sigma)_\mathrm{e} \vert M_\star)$ & $-0.87 (-0.80, -0.88)$ & $1.0 \times 10^{-2}$ & \multirow{-2}{*}{$-67.8 (-112.6, -47.0)$}\\
\cline{4-9}
 & (43) &  &  &  & $\rho(h_4, M_\star \vert (V/\sigma)_\mathrm{e})$ & $0.21 (-0.04, 0.37)$ & $3.9 \times 10^{-1}$ & \\
 & (44) & \multirow{-4}{*}{MAGPI} & \multirow{-2}{*}{Q} & \multirow{-2}{*}{$20$} & $\rho(h_4, (V/\sigma)_\mathrm{e} \vert M_\star)$ & $-0.59 (-0.55, -0.43)$ & $8.3 \times 10^{-3}$ & \multirow{-2}{*}{$-70.2 (-94.2, -49.4)$}\\
\cline{3-9}
 & (45) &  &  &  & $\rho(h_4, M_\star \vert (V/\sigma)_\mathrm{e})$ & $0.15 (0.08, 0.24)$ & $7.7 \times 10^{-2}$ & \\
 & (46) &  & \multirow{-2}{*}{SF} & \multirow{-2}{*}{$132$} & $\rho(h_4, (V/\sigma)_\mathrm{e} \vert M_\star)$ & $-0.20 (-0.26, -0.15)$ & $2.5 \times 10^{-2}$ & \multirow{-2}{*}{$-51.7 (-72.1, -31.1)$}\\
\cline{4-9}
 & (47) &  &  &  & $\bm{\rho(h_4, M_\star \vert (V/\sigma)_\mathrm{e})}$ & $0.25 (0.23, 0.28)$ & $\mathbf{1.4 \times 10^{-5}}$ & \\
 \parbox[t]{2mm}{\multirow{-12}{*}{\rotatebox[origin=c]{90}{\hyperref[f.r.h4_qvs]{\mstar-\vse plane}}}} & (48) & \multirow{-4}{*}{LEGA-C} & \multirow{-2}{*}{Q} & \multirow{-2}{*}{$297$} & $\bm{\rho(h_4, (V/\sigma)_\mathrm{e} \vert M_\star)}$ & $-0.35 (-0.45, -0.31)$ & $\mathbf{3.4 \times 10^{-10}}$ & \multirow{-2}{*}{$-54.8 (-63.0, -47.1)$}\\

    \hline
    \end{tabular}
    \end{center}
    
    \begin{flushleft}
    Columns: (1)~reference to the relevant figure; (2)~unique row identifier;
    (3)~data source; (4)~Q/SF for the quiescent/star-forming subsets; (5)~number
    of galaxies in the current subset, after removing galaxies that do not possess
    the necessary measurements to calculate the PCC; (6)~PCC in this row;
    (7)~value of the PCC, with the numbers in parentheses encompassing the
    16\textsuperscript{th}-84\textsuperscript{th} percentiles from one thousand
    bootstrapping realisations; (8)~$P$ value associated to the PCC (probability that a
    PCC as large as reported arised from uncorrelated data); (9)~angle representation of
    the complementary PCCs, with the number in parentheses encompassing the
    16\textsuperscript{th}-84\textsuperscript{th} percentiles from bootstrapping
    \citetext{see \S~\ref{s.r} and \citealp{bluck+2020a} for a definition}.
    \end{flushleft}
    
    \end{table*}

\subsection{Correlations with galaxy mass and rotation-to-dispersion ratio}\label{s.r.ss.mvs}

\begin{figure}
  \includegraphics[type=pdf,ext=.pdf,read=.pdf,width=1.\columnwidth]{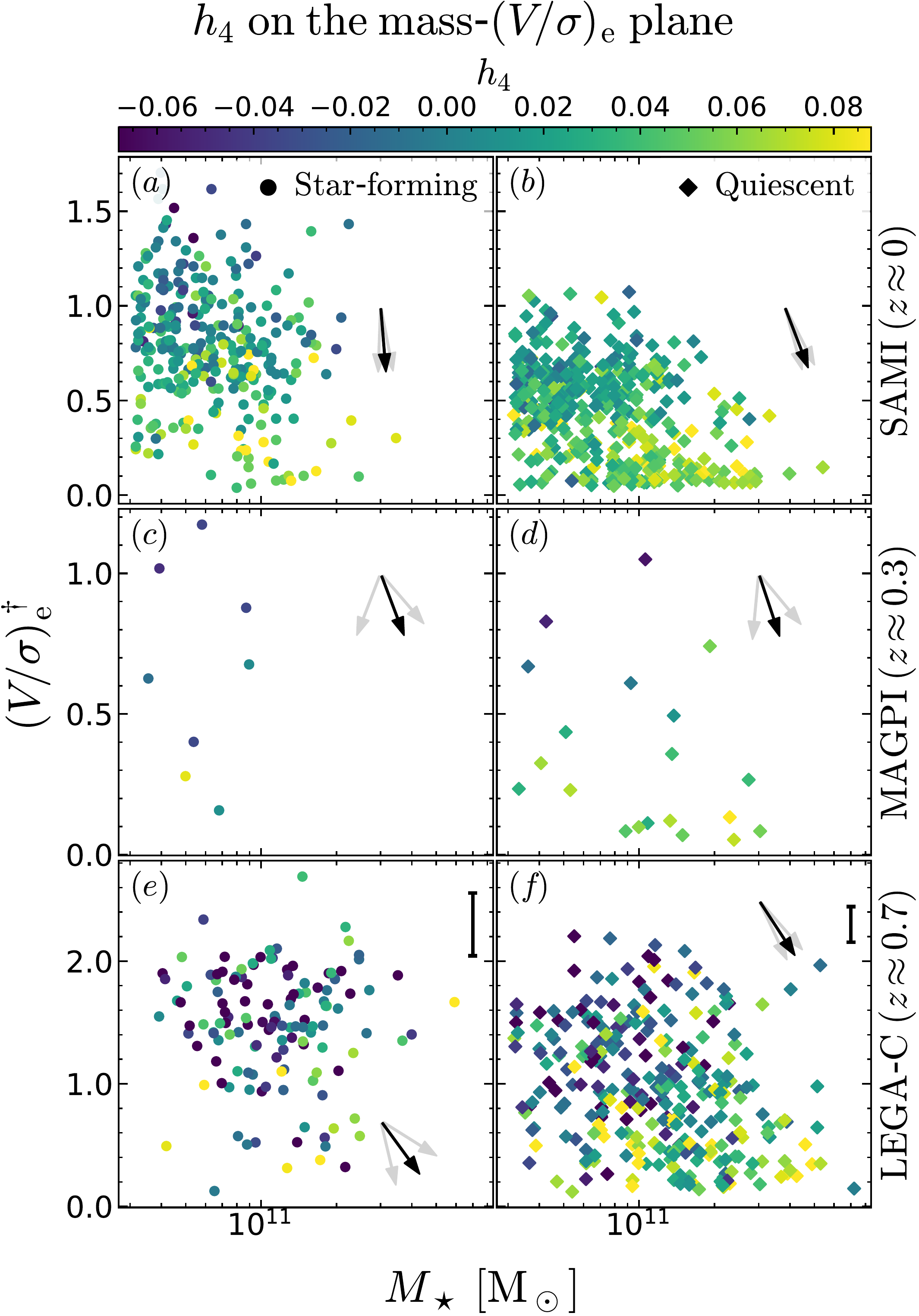}
  {\phantomsubcaption\label{f.r.h4_mvs.a}
   \phantomsubcaption\label{f.r.h4_mvs.b}
   \phantomsubcaption\label{f.r.h4_mvs.c}
   \phantomsubcaption\label{f.r.h4_mvs.d}
   \phantomsubcaption\label{f.r.h4_mvs.e}
   \phantomsubcaption\label{f.r.h4_mvs.f}
  }
  \caption{Our galaxies on the \mstar--\vse plane, colour-coded by \h4. The
  left/right columns show star-forming/quiescent galaxies, the top/middle/bottom
  rows show the SAMI/MAGPI/LEGA-C samples. The symbols are the same as
  Fig.~\ref{f.r.h4_mr}, the PCCs and the angle of the arrows are reported in
  Table~\ref{t.r.h4corr}, rows~37--48.
  In general, the \h4-\vse correlations are stronger than the
  \h4--\mstar correlations; the latter only exist independently for LEGA-C
  quiescent galaxies, and, perhaps, for SAMI quiescent galaxies.\\
  $^\dagger$ \vse has two different meanings for SAMI and MAGPI vs LEGA-C. See
  also note to Fig.~\ref{f.r.h4_qvs} and \S~\ref{s.das.ss.anc.sss.vse}.
  }\label{f.r.h4_mvs}
\end{figure}

Having determined that \mstar and \vse are the best predictors of \h4, we now
compare these two observables directly. The \mstar--\vse plane is shown in
Fig.~\ref{f.r.h4_mvs}, with the usual meaning of the symbols; the relevant PCCs
are reported in Table~\ref{t.r.h4corr}, rows~37--48.

For SAMI star-forming galaxies, after controlling for \vse, there is no evidence
of the \h4--\mstar correlation (cf. rows~37 and~38). In contrast, for quiescent
galaxies, we find evidence of an independent \h4--\mstar correlation for both
SAMI (row~39) and LEGA-C (row~47). In all three cases, the \h4--\vse correlations
have larger magnitude and higher statistical significance than the \h4--\mstar
correlations (rows~38, 40 and~48). In summary, \h4 reflects much more \vse than 
\mstar, but the correlation with \mstar is independent of the \h4--\vse
correlation, highlighting again that integrated \h4 is more than just \vse.

\section{Discussion}\label{s.d}

\subsection{Relation between \texorpdfstring{\h4}{h4} and rotation}\label{s.d.ss.h4rot}

Overall, the almost ubiquitous \h4--\vse anti-correlation (Fig.~\ref{f.r.h4_qvs})
means that our integrated, light-weighted \h4 probes mostly (light-weighted)
rotation-to-dispersion ratio. This strong anti-correlation is found not only in
the data, but is also predicted by our (admittedly simple) kinematic models
(cf.~Fig.s~\ref{f.tmods.models} and~\ref{f.r.h4_qvs}). Indeed, the \h4--\vse
anti-correlation is strong enough that we have tentative evidence even for only
20 galaxies in the MAGPI quiescent sample.
In this light, the different \h4 between star-forming and quiescent galaxies
(Fig.~\ref{f.r.sfq}) reflects to a large degree the known difference in \vse
between these two populations \citep[e.g.][]{vandesande+2018, graham+2018,
falcon-barroso+2019}.

When we consider projected shape, we find that for star-forming galaxies, the
\h4--$q$ correlation is either not detected (for MAGPI and LEGA-C) or, when
detected, the correlation coefficient has smaller magnitude and lower
statistical significance than the \h4--\vse anti-correlation
(Table~\ref{t.r.h4corr}, rows 1--2). These
results are in agreement with our toy models, as expected from the fact that
in the mass range explored here, star-forming galaxies have predominantly
disc-like kinematics.

In contrast, for quiescent galaxies, \h4 has strong independent correlations
with both \vse and projected $q$ --- as expected from the fact that these two
observables capture independent aspects of galaxy kinematics
\citep{illingworth1977, binney1978, davies+1983}. The measured \h4--$q$
correlation (Table~\ref{t.r.h4corr}, rows~3--4 and~11--12) has larger
correlation coefficient than predicted by our thin-disc models
(Table~\ref{t.tmods.mod}).
This is not surprising (quiescent galaxies are not thin-discs) but also suggests
that \h4 measures more than just \vse --- at least for quiescent galaxies.
In fact, for bulge-like structures, $q$ reflects (statistically) the intrinsic
axis ratio, which in turn is related to kinematic properties such as orbital
anisotropy; so, for quiescent galaxies, the \h4--$q$ correlation compounds
inclination effects with physical trends. This combination may explain why we
find the \h4--$q$ correlation in quiescent galaxies to have larger magnitude and
higher statistical significance compared to the same relation for star-forming
galaxies, even after taking into account the different sample sizes.

For SAMI quiescent galaxies, \h4 correlates more strongly with
\vse than with $q$, but for LEGA-C the opposite is true; this fact could be due
to the different spatial resolution between the two surveys (for SAMI, \vse is
measured on kpc scales, whereas LEGA-C spatial resolution is approximately
7~kpc), or to a genuine change in the population demographics over the last
7~Gyr. However, we recall that \vse is defined and measured differently between
the two datasets (\S~\ref{s.das.ss.anc.sss.vse}), so a direct comparison is
unwarranted and we defer it to future work.

\subsection{Fast- and slow-rotator quiescent galaxies}

The reported trends of \h4 with \vse and $q$ persist for the sample of FR
quiescent galaxies. Moreover, we have shown explicitly that star-forming
galaxies and fast-rotator quiescent galaxies have different \h4 distributions,
but that the difference becomes larger if we include SR quiescent galaxies
(\S~\ref{s.r.ss.sfq.sss.frsr}). These facts show that fast-rotator quiescent
galaxies are intermediate between star-forming and SR systems. The existence of
a separate population of SR \citep[i.e. a bimodality in the kinematics of
quiescent galaxies,][]{emsellem+2007, cappellari2016, graham+2018,
vandesande+2021a} is relevant to understanding how stellar orbits transform from
rotation supported (as imprinted from the kinematics of star-forming gas) to
dispersion supported (as observed in SRs). From the point of view of \h4, we
see evidence of a gradual transformation between star-forming galaxies and
fast-rotator quiescent galaxies, which shows that there is some degree of
gradual transformation. We see no evidence for a bimodality in the \h4
distribution of quiescent galaxies, but our number statistics are probably too
low to detect it \citep[see][]{vandesande+2021a}.

\subsection{Relation between \texorpdfstring{\h4}{h4} and dry mergers}\label{s.d.ss.galstr}

When we study \h4 in relation to \mstar and \vse, we find that, once again, \h4
is driven primarily by \vse (Fig.~\ref{f.r.h4_mvs} and Table~\ref{t.r.h4corr},
rows~37--48). However, for quiescent galaxies, \textit{the} \h4--\mstar
\textit{correlation exists independently of} \vse. This fact confirms that
integrated \h4 is not just \vse, and that some of the variability in \h4 at
fixed \vse might be physically associated with processes that increase \mstar.
For LEGA-C quiescent galaxies, we also find evidence for an independent
correlation between \h4 and \re, such that, at fixed \mstar, the largest
galaxies also have the highest \h4. Given the strength of this correlation,
if it was present in SAMI data we should be able to detect it, even after
accounting for the different sample size. Aperture effects do not explain
the difference: measuring the SAMI \h4 inside a slit does not change our
results (cf. the mock-slit spectra from D'Eugenio et~al., in~prep.). A
possible explanation is that SAMI quiescent galaxies are on average
less massive than LEGA-C quiescent galaxies. Indeed, if we resample the latter
to match the \mstar distribution of the SAMI quiescent sample, the \h4--\re
correlation disappears. However, the resulting
\mstar-matched sample is relatively small, so this test is not conclusive.

Of particular interest is the lack of correlation with \sigap
(Fig.~\ref{f.r.h4_ms} and Table~\ref{t.r.h4corr}, rows~25--36). We find only
correlations with \mstar (for SAMI and LEGA-C quiescent galaxies) or marginal
evidence of correlations with \mstar (for SAMI and LEGA-C star-forming galaxies
and for MAGPI quiescent galaxies). For LEGA-C, we even find a negative
correlation between \h4 and \sigap ($P = 7.6\times10^{-8}$), but this may be
due to inclination effects\footnote{As we noted, this negative correlation
disappears if we replace the observed \sigap with $\sigma_\mathrm{vir}$, which
includes a multiplicative inclination correction based on the observed axis
ratio $q$ \citep{vanderwel+2022}.}.

Taken together, these results suggest that \h4 may be related to gas-poor (dry)
mergers: these decrease \vse, increase both \mstar and \re and leave \sigap
unchanged (for major mergers) or lower it very little \citetext{for minor
mergers, \citealp{bezanson+2009, naab+2009}, \citealp{oser+2012},
\citealp{vandesande+2013}}.
In principle, \h4 may reflect dynamical heating due to secular processes, which
could also induce an anti-correlation between resolved \h4 and \vse. However, if
we assume dynamical heating to go together with stellar-population age, we would
then expect an independent correlation between \h4 and \sigap at fixed \mstar
\citep[because \sigap predicts age more accurately than \mstar,][]{
mcdermid+2015, scott+2017, barone+2018, barone+2020}, which is ruled out by our
data (Fig.~\ref{f.r.h4_ms}).

\subsection{Relation between integrated and spatially resolved \texorpdfstring{\h4}{h4}}\label{s.d.ss.h4res}

In addition to the \h4--\vse anti-correlation, our toy models also predict an
equally strong correlation between integrated and spatially resolved \h4
(Fig.~\ref{f.tmods.resvsint}). To probe this correlation,
we used two cuts (in $q$ and \vse) designed to suppress the \h4--\vse
anti-correlation and to enhance the correlation between integrated and resolved
\h4 (Table~\ref{t.tmods.mod}). Comparing integrated \h4 between star-forming and
quiescent galaxies for $q\geq0.75$, we still find a statistically significant
difference, but no difference is found for $\vse\leq0.5$
(\S~\ref{s.r.ss.sfq}). Given that --- with these selections --- there is a
strong correlation between integrated and resolved \h4, the lack of difference
means that either star-forming and quiescent galaxies have the same value of
resolved \h4 (i.e. any difference in integrated \h4 is driven entirely by \vse),
or, alternatively, that resolved \h4 varies jointly with \vse.

Theoretically, resolved \h4 measures radial anisotropy
\citep{vandermarel+franx1993, gerhard1993}, so a link between resolved \h4 and
\vse seems natural: thin discs are by definition low-dispersion systems
generated by dissipative cold-gas accretion, while radial anisotropy is
associated with dissipationless accretion.

\subsection{Insight from spatially resolved spectroscopy}\label{s.d.ss.ifu}

Spatially resolved \h4 measurements from integral-field spectroscopy show
complex structure \citep{emsellem+2004}, which complicates the interpretation of
our spatially integrated measurements.
In fact, our \h4 could be due not only to radial anisotropy vs rotation, but also
to bars \citep[e.g. peanut-shaped bulges,][]{debattista+2005, mendez-abreu+2008}
and kinematically decoupled cores \citep[see e.g.][their fig.~4]{emsellem+2004}.
Indeed, after employing the visual bar classification performed by the SAMI team
on HSC \citep{aihara+2019}, Pan-STARRS \citep{chambers+2016, flewelling+2020}
and Decals \citep{dey+2019} imaging, we find a weak anti-correlation between
integrated \h4 and the probability that a bar is present. However, after
controlling for
\mstar or \vse, this correlation disappears, so it is unlikely that bars are
responsible for the trends we observe. A larger sample, perhaps with better
spatial resolution, may help clarify whether an independent correlation with
integrated \h4 is present.

Following \citet{krajnovic+2011}, \citet{vandesande+2017a} divided the SAMI
galaxies between regular and non-regular rotators, and found that both classes
display the full range of \h4 values (cf.~their fig.~9).
However, regular rotators also show a strong radial \h4 gradient, with their
central regions having $\h4 \lesssim 0$
\citep[][fig.~9, panels~d and~f]{vandesande+2017a}.
Because these central low-\h4 regions are brighter than the high-\h4 outskirts,
it is reasonable to assume that our integrated \h4 measurements are lower for
regular rotators than for non-regular rotators, where the central regions display
$\h4 \approx 0.05$ \citep{vandesande+2017a}.

The fact that on average our star-forming galaxies have lower \h4 than
quiescent galaxies is qualitatively consistent with the
expectation that, in the stellar-mass range explored here, star-forming galaxies
are overwhelmingly regular rotators \citep[or `dynamically cold discs'][]{
fraser-mckelvie+cortese2022}.
On the other hand, even though massive non-regular rotators are predominantly
quiescent, the converse is not true: quiescent galaxies are also dominated by
regular rotators, which account for 60-80~per cent of the local early-type galaxy
population \citep[see e.g.][]{cappellari+2011b, vandesande+2017b}.
The degree to which the fraction of quiescent galaxies that are non-regular
rotators `skews' the \h4 distribution to higher values is hard to determine:
ideally we would like to control for stellar mass by comparing three
\mstar-matched samples: star-forming regular rotators (the control sample),
quiescent regular rotators and quiescent non-regular rotators. However,
non-regular rotators have distinctively higher \mstar \citep{veale+2017,
brough+2017, vandesande+2021b},
so matching in \mstar is not feasible. Nevertheless, a KS test between
star-forming and quiescent regular rotators gives a $P$ value of 0.03, which
is indecisive; larger sample sizes are needed to clarify whether this
difference is significant.

We note that, in principle, the shape of the gravitational potential also
affects the LOSVD. The fact that star-forming and quiescent galaxies have
different light distributions means that they are likely to have different
potentials too. However, in practice, orbital distribution (regardless of the
origin) is the dominant factor in determining \h4, with the detailed shape of
the gravitational potential being a second-order effect \citep{gerhard1993}.

\subsection{Difference between star-forming and quiescent galaxies}\label{s.d.ss.diff}

In \S~\ref{s.r.ss.sfq} we have studied the relation between \h4, \mstar and
$S/N$ using SAMI and LEGA-C data. For both surveys, and for both quiescent and
star-forming galaxies, we find a statistically significant correlation between
\h4 and \mstar.
In comparison, the correlation between \h4 and $S/N$ is either less statistically
significant or not detected. The contrast between the \h4--\mstar correlations
and the weaker or absent correlations with $S/N$ gives us confidence that our \h4
measurements trace a physical property of the galaxies, not the $S/N$ of the
spectra.

At fixed \mstar, the \h4 distribution has large intrinsic scatter, suggesting a
large degree of physical variability, probably reflecting the broad range in $q$,
\vse and integrated \h4 of the sample.

The main result of \S~\ref{s.r.ss.sfq} is that quiescent and star-forming
galaxies have different \h4 distributions, even after matching their \mstar or
$S/N$ distributions; this is true both in the local Universe
(SAMI, Fig.~\ref{f.r.sfq.a}) and 7~Gyr ago\footnote{It would be interesting to
compare \h4 across the three surveys, but this requires careful matching of the
different observing setups, and is the subject of a future work (D'Eugenio
et~al., in~prep.).} (LEGA-C, Fig.~\ref{f.r.sfq.f};
note that, for the ongoing MAGPI Survey, the current sample size is too small for
this analysis).
We also checked that for SAMI the difference persists after matching
the star-forming and quiescent subsets in central velocity dispersion, which is
related to quenching \citep[e.g.][]{bluck+2016, bluck+2020a, bluck+2020b}. If a
connection between \h4 and quenching exists, it must be due to resolved \h4
(i.e. radial anisotropy) and not \vse, because the latter has already been ruled
out as a driver of galaxy quenching \citep{brownson+2022}.
However, the differences in integrated \h4 between star-forming and quiescent
galaxies are small compared to the differences in central dispersion
\citep{bluck+2016, bluck+2020a}. Future work on this subject may clarify the
likelihood of a causal connection with quenching.

Moving to the link between \h4 and other galaxy observables
(Fig.s~\ref{f.r.h4_qvs}--\ref{f.r.h4_mvs}), whenever we find statistically
significant trends for both star-forming and quiescent galaxies, their direction
agrees. This fact suggests that the same physical mechanisms may be determining
\h4 in both star-forming and quiescent galaxies, with the differences in median
value and statistical significance due to different relative contribution.

\subsection{Predicting rotation from integrated \texorpdfstring{\h4}{h4}}\label{s.d.ss.pred}

\begin{figure*}
  \includegraphics[type=pdf,ext=.pdf,read=.pdf,width=1.\textwidth]{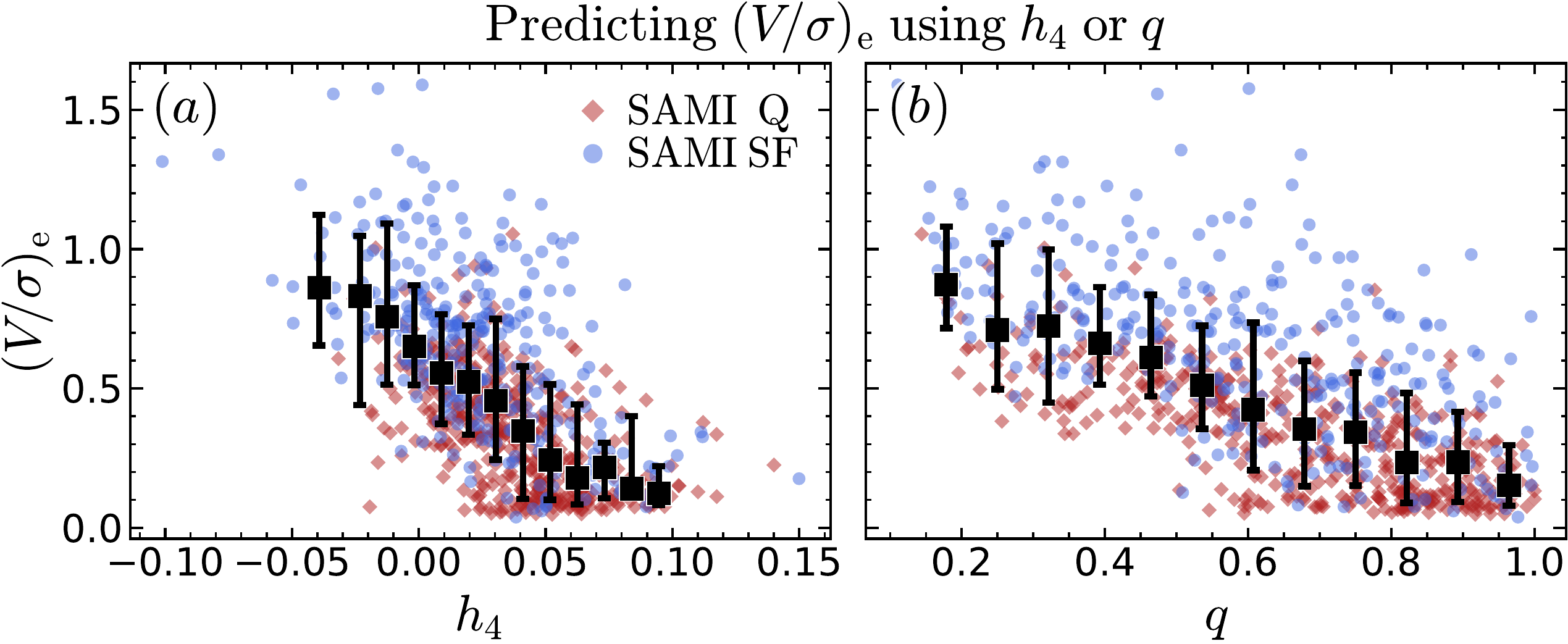}
  {\phantomsubcaption\label{f.d.h4_vs.a}
   \phantomsubcaption\label{f.d.h4_vs.b}}
  \caption{As a predictor of \vse, \h4 (panel~\subref{f.d.h4_vs.a}) performs
  similarly to projected axis ratio $q$ (panel~\subref{f.d.h4_vs.b}).
  The blue circles/red diamonds are SAMI star-forming/quiescent galaxies, the
  black squares are the running median, with the errorbars encompassing the
  16\textsuperscript{th}--84\textsuperscript{th} percentiles of the data.
  Using \h4 alone, one can predict \vse with a typical uncertainty of 0.23, which
  reduces to 0.16 if considering only quiescent galaxies; these results are
  similar to what we would get using galaxy shape $q$ instead of \h4.
  }\label{f.d.h4_vs}
\end{figure*}

The strong \h4--\vse anti-correlation raises the question whether we can use
integrated \h4 (which can be measured from relatively inexpensive single-fibre
observations) to infer \vse (which requires slit or fibre spectroscopy).
The relation between \h4 and \vse for SAMI is shown in Fig.~\ref{f.d.h4_vs.a},
where blue circles/red diamonds represent star-forming/quiescent galaxies. We
used the least-trimmed squares algorithm to fit a linear relation \citep{
rousseeuw+driessen2006, cappellari+2013a}.
The observed scatter about the linear best-fit relation is 0.23 (considering
star-forming and quiescent galaxies together), or 0.16 for quiescent galaxies
alone; this is dominated by intrinsic scatter (respectively 0.15 and 0.10).
While the large intrinsic scatter is in agreement with our hypothesis that \h4
contains more information than just \vse, it also means that we cannot use
single-fibre observations to accurately measure \vse. To make matters worse, we
obtain similar scatter if we replace \h4 with $q$, which can be measured from
imaging alone (Fig.~\ref{f.d.h4_vs.b}). In this case, the observed scatter is
0.20 (0.14 for quiescent galaxies) and the intrinsic scatter is 0.17 (0.11
for quiescent galaxies).
It remains to be seen whether combining \h4 with $q$ or with other observables
can improve our estimate of \vse.

\subsection{Caveats and future outlook}

There are two difficulties with our interpretation of \h4 being related to dry
mergers. First, a key prediction of the dry merger hypothesis is that it affects
the outskirts ($R>R_\mathrm{e}$) more than the central regions. Even though we
observe our galaxies in projection (so part of the outskirts are probed along
the line of sight), the magnitude of the expected signal is unclear. A comparison
with numerical simulations would be beneficial. Resolved
studies have shown that --- for regular rotators --- the central regions have indeed
lower \h4, as expected from this scenario
\citep[][their fig.~9d]{vandesande+2017a}. However, the picture for non-regular
rotators is not as clear, with the centre possibly having higher \h4
\citep[][their fig.~4]{emsellem+2004}. A quantitative study of the radial
distribution of \h4 over a large sample of galaxies could address this question.
A second prediction of the dry merger hypothesis is that it would lead to mass
and even stronger size growth, yet we find almost no evidence of independent correlations
between \h4 and \re (Fig.~\ref{f.r.h4_mr}, Table~\ref{t.r.h4corr}, rows~13--24).
The only exception are LEGA-C quiescent galaxies, where the \h4--\mstar and
\h4--\re correlations have comparable PCCs and significance (rows~23--24). There
are two possible explanations why the \h4--\re correlation is not observed in
the other samples, and especially in the SAMI quiescent sample. The first is that
the mass ranges are different, with LEGA-C probing more massive galaxies than SAMI.
However, other effects may be hiding the \h4--\re correlation. For example, if
star-forming galaxies transition to quiescence without significant changes to
their size and kinematics, this would add newly quiescent galaxies that are large
(because star-forming galaxies are on average
larger than quiescent galaxies of the same mass) but have low \h4. However, after
accounting for \mstar, we find no evidence of an independent correlation between
\h4 and stellar-population age \citep[we used the light-weighted age
measurements from][]{scott+2017}. Future studies based on larger samples may help
answer some of the open questions.

Compared to \vse, \h4 presents the advantage of requiring no spatial information,
so it can be measured for large single-fibre surveys such as the Sloan Digital Sky
Survey \citep{york+2000}. As already noted, it may be interesting to explore if and
to what extent \h4 is related to galaxy quiescence \citetext{expanding on the
work of \citealp{brownson+2022}, \citealp{piotrowska+2022}, and
\citealp{bluck+2022}}. Note that the high $S/N$ required to measure \h4 may
bias the resulting sample, more so than studies based on velocity dispersion
alone.
With this caveat in mind, the current generation of large single-fibre surveys
of the local Universe will give us access to even larger samples than SDSS
\citep[e.g. the Dark Energy Spectroscopic Instrument Survey, DESI;][]{desi+2016a}.
Interestingly, the upcoming generation of large single-fibre surveys will
include deep spectroscopy that is ideal to measure \h4 \citetext{e.g. the 4MOST
Hemisphere Survey, Taylor et~al., in~prep.; the WEAVE-StePS
\citealp{costantin+2019} and 4MOST-StePS surveys; and the DESI Bright Galaxy
Survey, \citealp{ruiz-macias+2021}}, while future high-redshift
programs will enable us to study \h4 for galaxies at cosmic noon
\citep[MOONRISE survey,][]{maiolino+2020}.

\section{Summary and Conclusions}\label{s.c}

In this work, we present the first study of integrated \h4 for a statistical
sample of both star-forming and quiescent galaxies. Using the \ppxf software on
optical spectroscopy data from the SAMI, MAGPI and LEGA-C surveys, we model the
line-of-sight velocity distribution as a Gauss-Hermite series. \h4 is the
coefficient of the 4\textsuperscript{th}-order Hermite polynomial in the
Gauss-Hermite expansion; mathematically, it represents the excess kurtosis.
Physically, integrated \h4 is related to both the rotation-to-dispersion ratio
\vse and to spatially resolved \h4, as we show using a set of simple toy models
(\S~\ref{s.tmods}; Fig.s~\ref{f.tmods.models} and~\ref{f.tmods.resvsint}).
We select galaxies with stellar mass $\mstar > 10^{10.5} \, \msun$,
where, after quality selection (Fig.~\ref{f.samp.qcsel}), we are typically
80--98~per cent complete (Fig.~\ref{f.samp.samp}).
For this sample, we find that
\begin{enumerate}
  \item The largest-magnitude and
  highest-significance correlation is between \h4 and \vse (\S~\ref{s.r.ss.qvs}
  and~\ref{s.r.ss.mvs}, Fig.s~\ref{f.r.h4_qvs} and~\ref{f.r.h4_mvs},
  Table~\ref{t.r.h4corr}, rows 1--12 and 37--48).
  \item After controlling for \vse, the second-largest magnitude correlation is
  between \h4 and \mstar; this suggests that \h4 contains additional information
  compared to \vse, in agreement with the models. This information might be
  connected to mass assembly.
  \item At fixed \vse or \mstar, the \h4 distribution has large intrinsic
  scatter, reflecting physical variability between galaxies (\S~\ref{s.r.ss.sfq}
  and~\ref{s.d.ss.galstr}).
  \item Quiescent galaxies have larger \h4 than coeval star-forming galaxies
  (\S~\ref{s.r.ss.sfq}, Fig.~\ref{f.r.sfq}), both in the local Universe
  (SAMI, $z=0.05$) and 7~Gyr ago (LEGA-C, $z=0.7$).
  \item When controlling for \vse, the difference disappears, suggesting
  that \vse and resolved \h4 vary together.
  \item We interpret \h4 as a combination of \vse and as a measure of the
  fraction of dry mergers. The latter explains the \h4--\mstar correlation at
  fixed \vse, and the lack of independent correlation with \sigap
  (\S~\ref{s.r.ss.ms}--\ref{s.r.ss.mvs}; Fig.s~\ref{f.r.h4_ms}
  and~\ref{f.r.h4_mvs}).
  \item Given our interpretation, we would expect a correlation between \h4
  and~\re, but this is only detected for LEGA-C quiescent galaxies
  (\S~\ref{s.r.ss.mr}). A possible explanation is that the late addition of
  large, low-\h4 star-forming galaxies may hide the \h4--\re correlation.
  \item Our results are qualitatively unchanged if we remove slow-rotator
  galaxies, but the significance of the trends is reduced. Fast-rotator
  quiescent galaxies have intermediate \h4 between star-forming galaxies
  and slow rotators (\S~\ref{s.r.ss.sfq.sss.frsr}, Fig.~\ref{f.r.srfr.c}).
\end{enumerate}

\section*{Acknowledgements}

We thank the anonymous referee for insightful comments that greatly improved this article.
FDE and AvdW acknowledge funding through the H2020 ERC Consolidator Grant 683184.
FDE and RM ackwoledge funding through the ERC Advanced grant 695671 ``QUENCH'' and
support by the Science and Technology Facilities Council (STFC).
JvdS acknowledges support of an Australian Research Council Discovery Early Career Research Award (project number DE200100461) funded by the Australian Government. 
SB acknowledges funding support from the Australian Research Council through a Future Fellowship (FT140101166) ORCID - 0000-0002-9796-1363
JJB acknowledges support of an Australian Research Council Future Fellowship (FT180100231).
CF is the recipient of an Australian Research Council Future Fellowship (project number FT210100168) funded by the Australian Government.
H{\"U} gratefully acknowledges support by the Isaac Newton Trust and by the Kavli Foundation through a Newton-Kavli Junior Fellowship.
LMV acknowledges support by the COMPLEX project from the European Research Council (ERC) under the European Union’s Horizon 2020 research and innovation program grant agreement ERC-2019-AdG 882679.

The SAMI Galaxy Survey is based on observations made at the Anglo-Australian
Telescope. The Sydney-AAO Multi-object Integral-field spectrograph (SAMI) was
developed jointly by the University of Sydney and the Australian Astronomical
Observatory, and funded by ARC grants FF0776384 (Bland-Hawthorn) and
LE130100198. The SAMI input catalog is based on data taken from the Sloan
Digital Sky Survey, the GAMA Survey and the VST ATLAS Survey. The SAMI Galaxy
Survey is funded by the Australian Research Council Centre of Excellence for
All-sky Astrophysics (CAASTRO), through project number CE110001020, and other
participating institutions. The SAMI Galaxy Survey website is
http://sami-survey.org/ .

Funding for SDSS-III has been provided by the Alfred P. Sloan Foundation, the
Participating Institutions, the National Science Foundation, and the U.S.
Department of Energy Office of Science. The SDSS-III web site is
http://www.sdss3.org/ .

GAMA is a joint European-Australasian project based around a spectroscopic
campaign using the Anglo-Australian Telescope. The GAMA input catalogue is
based on data taken from the Sloan Digital Sky Survey and the UKIRT Infrared
Deep Sky Survey. Complementary imaging of the GAMA regions is being obtained
by a number of independent survey programmes including GALEX MIS, VST KiDS,
VISTA VIKING, WISE, Herschel-ATLAS, GMRT and ASKAP providing UV to radio
coverage. GAMA is funded by the STFC (UK), the ARC (Australia), the AAO, and
the participating institutions. The GAMA website is http://www.gama-survey.org/ .

Based on observations made with ESO Telescopes at the La Silla Paranal Observatory
under programme IDs 179.A-2004. Based on observations collected at the European
Organisation for Astronomical Research in the Southern Hemisphere under ESO
programme 1104.B-0536. We wish to thank the ESO staff, and in particular the
staff at Paranal Observatory, for carrying out the MAGPI observations. Part of this 
research was conducted by the Australian Research Council Centre of Excellence
for All Sky Astrophysics in 3 Dimensions (ASTRO 3D), through project number
CE170100013.

The Pan-STARRS1 Surveys (PS1) and the PS1 public science archive have been made
possible through contributions by the Institute for Astronomy, the University of
Hawaii, the Pan-STARRS Project Office, the Max-Planck Society and its
participating institutes, the Max Planck Institute for Astronomy, Heidelberg and
the Max Planck Institute for Extraterrestrial Physics, Garching, The Johns
Hopkins University, Durham University, the University of Edinburgh, the Queen's
University Belfast, the Harvard-Smithsonian Center for Astrophysics, the Las
Cumbres Observatory Global Telescope Network Incorporated, the National Central
University of Taiwan, the Space Telescope Science Institute, the National
Aeronautics and Space Administration under Grant No. NNX08AR22G issued through
the Planetary Science Division of the NASA Science Mission Directorate, the
National Science Foundation Grant No. AST-1238877, the University of Maryland,
Eotvos Lorand University (ELTE), the Los Alamos National Laboratory, and the
Gordon and Betty Moore Foundation.

This work made extensive use of the freely available
\href{http://www.debian.org}{Debian GNU/Linux} operative system. We used the
\href{http://www.python.org}{Python} programming language
\citep{vanrossum1995}, maintained and distributed by the Python Software
Foundation. We further acknowledge direct use of
{\sc \href{https://pypi.org/project/astropy/}{astropy}} \citep{astropyco+2013},
{\sc \href{https://pypi.org/project/ltsfit/}{ltsfit}} \citep{cappellari+2013a},
{\sc \href{https://pypi.org/project/matplotlib/}{matplotlib}} \citep{hunter2007},
{\sc \href{https://pypi.org/project/numpy/}{numpy}} \citep{harris+2020},
{\sc \href{https://pypi.org/project/pathos/}{pathos}} \citep{mckerns+2011},
{\sc \href{https://pypi.org/project/pingouin/}{pingouin}} \citep{vallat2018},
\href{https://pypi.org/project/ppxf/}{\ppxf} \citep{cappellari2017},
{\sc \href{https://pypi.org/project/scipy/}{scipy}} \citep{jones+2001}
and {\sc \href{http://www.star.bris.ac.uk/~mbt/topcat/}{topcat}} \citep{taylor2005}.

\section*{Data availability}

The reduced data used in this work is available in the public domain. For SAMI,
through the \href{https://docs.datacentral.org.au/sami}{SAMI Data Release 3}
\citep{croom+2021a}. Ancillary
data comes from the \href{http://gama-survey.org}{GAMA Data Release 3}
\citep{baldry+2018} and raw data is from
\href{https://classic.sdss.org/dr7/}{SDSS DR7} \citep{abazajian+2009},
\href{https://www.sdss3.org/dr9/}{SDSS DR9} \citep{ahn+2012} and
\href{http://casu.ast.cam.ac.uk/vstsp/imgquery/search}{VST}
\citep{shanks+2013, shanks+2015}.
For MAGPI, the raw data (and a basic data reduction) are available through the
the \href{http://archive.eso.org/cms.html}{ESO Science Archive Facility}.
For LEGA-C, the raw data and a catalogue of basic photometric and kinematic
measurements are available through the
\href{http://archive.eso.org/cms.html}{ESO Science Archive Facility}.

Integrated \h4 measurements are available
\sendemail{francesco.deugenio@gmail.com}{h4 evolution: data request}{
contacting the corresponding author.}

\section*{Affiliations}
\noindent
{\it
\hypertarget{aff11}{$^{11}$}Department of Physics, Florida International University, 11200 SW 8th Street, Miami, FL, USA\\
\hypertarget{aff12}{$^{12}$}School of Physics, University of New South Wales, Sydney, NSW 2052, Australia\\
\hypertarget{aff13}{$^{13}$}Australian Astronomical Optics, Astralis-USydney, School of Physics, University of Sydney, NSW 2006, Australia\\
\hypertarget{aff14}{$^{14}$}Research School of Astronomy and Astrophysics, Australian National University, Canberra, ACT 2611, Australia\\
\hypertarget{aff15}{$^{15}$}International Centre for Radio Astronomy Research (ICRAR), University of Western Australia, Crawley, WA 6009, Australia\\
\hypertarget{aff16}{$^{16}$}Astronomy Department, Yale University, New Haven, CT 06511, USA\\
\hypertarget{aff17}{$^{17}$}INAF-Osservatorio Astrofisico di Arcetri, Largo Enrico Fermi 5, I-50125 Firenze, Italy\\
\hypertarget{aff18}{$^{18}$}Leiden Observatory, Leiden University, P.O. Box 9513, 2300 RA, Leiden, The Netherlands\\
\hypertarget{aff19}{$^{19}$}Department of Physics and Astronomy, University College London, Gower Street, London WC1E 6BT, UK\\
\hypertarget{aff20}{$^{20}$}Department of Astronomy, University of Wisconsin, 475 N. Charter Street, Madison, WI 53706, USA\\
\hypertarget{aff21}{$^{21}$}Space Telescope Science Institute, 3700 San Martin Drive, Baltimore, MD 21218, USA\\
\hypertarget{aff22}{$^{22}$}Centre for Extragalactic Astronomy, University of Durham, Stockton Road, Durham DH1 3LE, United Kingdom\\
\hypertarget{aff23}{$^{23}$}Universit\"ats-Sternwarte, Fakult\"at f\"ur Physik, Ludwig-Maximilians-Universit\"at M\"unchen, Scheinerstr. 1, 81679 M\"unchen, Germany\\
\hypertarget{aff24}{$^{24}$}Department of Physics and Astronomy, University of the Western Cape, Robert Sobukwe Road, Bellville 7535, South Africa\\
\hypertarget{aff25}{$^{25}$}School of Mathematics and Physics, University of Queensland, Brisbane, QLD 4072, Australia\\
\hypertarget{aff26}{$^{26}$}Department of Astrophysics, University of Vienna, T\"urkenschanzstra{\ss}e 17, 1180 Vienna, Austria\\
\hypertarget{aff27}{$^{27}$}George P. and Cynthia Woods Mitchell Institute for Fundamental Physics and Astronomy, Texas A\&M University, College Station, TX 77843-4242, USA\\
}



\bibliographystyle{mnras}
\bibliography{h4dev.bbl}



\appendix

\section{Emission-line subtraction}\label{app.emlsub}

In \S~\ref{s.r.ss.sfq} we compare the value of \h4 between star-forming and
quiescent galaxies. For such a comparison to be meaningful, we have to be careful
about possible measurement bias. In star-forming galaxies, the spectrum is
characterised by strong emission lines due to warm ionised gas. For our data,
these include $[\mathrm{OII}]\lambda\lambda \, 3727,3729$ and the Balmer
series. All of these lines occur in the same spectral regions where stellar
absorption features are present. Subtracting the emission lines may introduce
a bias because, typically, the equivalent width of the emission is larger than
the equivalent width of the absorption lines (in absolute value). This means that
the subtraction residuals may impact the recovered value of \h4.
On the other hand, masking the emission lines means introducing a different bias:
in this case, the affected absorption lines are also masked so they are ignored
in calculating \h4, whereas they would be included for spectra with no emission
lines.

To ensure uniform treatment of all targets, in our default measurements we mask
the spectral regions of possible strong emission lines in all galaxies,
regardless of whether any emission is detected. This way, while ignoring some
information, we ensure the same treatment for both star-forming and quiescent
galaxies.
While we deem this approach to be the least biased, it comes with its own
problem. By definition, the fraction of stars with prominent Balmer absorption
must be different between star-forming and quiescent galaxies, therefore masking
these lines (as we do in our default measurement) has a different effect between
the two galaxy populations.

To address the effect of this bias, we use the SAMI data to test the effect of
two alternative treatments of emission lines, which, unlike the default method,
do not mask regions of possible emission.
For the first test, we model the emission lines simultaneously by using Gaussian
templates \citetext{following \citealp{bezanson+2018a} and the implementation of
\citealp{driver+2022} and Dogruel~et~al., in~prep.}; in this case, we find a
large systematic offset and large scatter compared to the default measurement
(median offset $\Delta \, h_4 = -0.007$, standard deviation 0.029). This result
suggests that subtracting the emission lines has a relatively large impact on
the recovered velocity distribution of the stars, likely because the random
uncertainties on the emission lines are large compared to the flux in the
absorption lines. For the second test, we mask pixels affected by promient
emission using iterative $\sigma$ clipping. This time we find a smaller systematic bias and
modest scatter (median offset $\Delta \, h_4 = 0.002$, standard deviation 0.009).
In both cases, the systematic offset is smaller than the difference between
star-forming and quiescent galaxies we report in \S~\ref{s.r.ss.sfq}, so, for our
purposes, the precise treatment of emission lines does not affect our conclusions.

\section{Effect of alternative spectral template libraries}\label{app.libbias}

\begin{figure}
  \includegraphics[type=pdf,ext=.pdf,read=.pdf,width=1.\columnwidth]{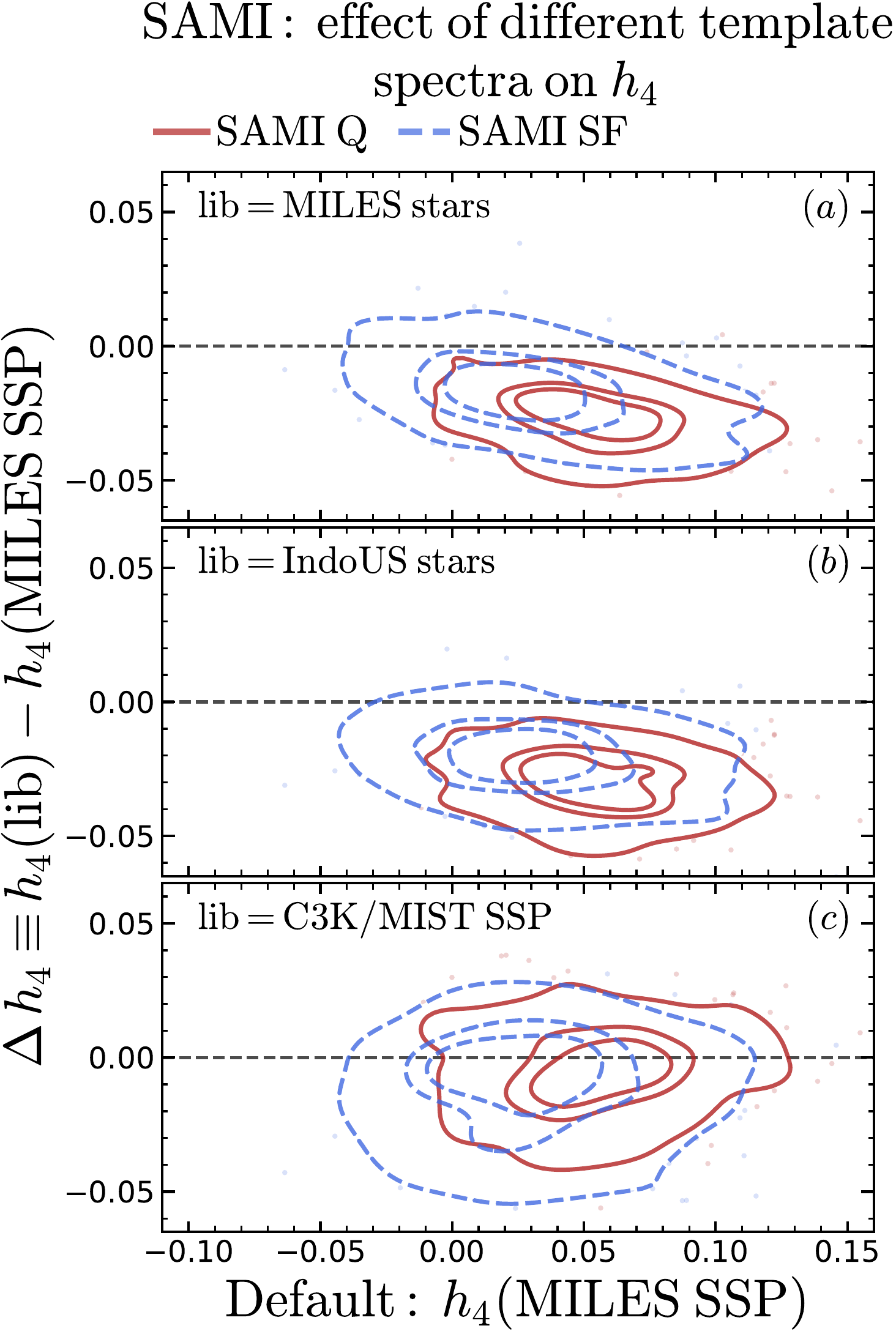}
  {\phantomsubcaption\label{f.app.h4bias.a}
   \phantomsubcaption\label{f.app.h4bias.b}
   \phantomsubcaption\label{f.app.h4bias.c}
  }
  \caption{Effect of different input template spectra on the measured
  value of \h4. The x axis shows the default \h4 measurement, obtained by
  modelling the galaxy spectra as a linear combination of SSP spectra from
  the MILES library. The y-axis shows the difference between alternative
  measurements of \h4 and the default value. Solid red/dashed blue contours
  trace the 30\textsuperscript{th}, 50\textsuperscript{th}
  and~90\textsuperscript{th} percentiles of quiescent/star-forming SAMI galaxies.
  Panels~\subref{f.app.h4bias.a}, \subref{f.app.h4bias.b},
  and~\subref{f.app.h4bias.c} show the effect of replacing the MILES SSP spectra
  with the MILES stellar spectra, the IndoUS stellar spectra, and the C3K/MIST
  SSP spectra. Stellar spectra tend to give lower values of \h4, more so for
  quiescent galaxies; this translates into a smaller difference between
  star-forming and quiescent galaxies. We verified that the difference is
  still statistically significant, regardless of the library used.
  }\label{f.app.h4bias}
\end{figure}

Our default measurements are obtained from \ppxf using the MILES SSP library as
spectral templates. What is the impact of this choice on our results and, more
in general, on the value of \h4?
Here we compare results from the MILES SSP library to three alternatives:
the MILES stellar library \citep{falcon-barroso+2011}, the IndoUS stellar library
\citep{valdes+2004}, and the C3K/MIST SSP library \citep{conroy+2019, choi+2016}.
We repeat our SAMI \h4 measurements using each of these three sets of spectra as
input to \ppxf, leaving everything else unchanged. We then compare the difference
$\Delta\,h_4$ between these new measurements and the default \h4. In
Fig.~\ref{f.app.h4bias}, the solid red/dashed
blue contours trace quiescent/star-forming galaxies (the contours enclose the
30\textsuperscript{th}, 50\textsuperscript{th} and~90\textsuperscript{th}
percentiles of the data). For the MILES stellar library
(panel~\subref{f.app.h4bias.a}), we find a median value $\langle \Delta\,h_4
\rangle = -0.0266\pm0.0005$ for quiescent galaxies, and $\langle \Delta\,h_4
\rangle = -0.0177\pm0.0007$ for star-forming galaxies; the scatter about the
medians are respectively 0.010 and 0.013, smaller than the error cut we
adopted in \S~\ref{s.samp.ss.qcsel}.
For the IndoUS stellar library
(panel~\subref{f.app.h4bias.b}), we find the largest offsets: the median values
are $\langle \Delta\,h_4 \rangle = -0.0298\pm0.0005$ and $-0.0205\pm0.0007$, 
and the scatters are 0.011 and 0.012, respectively for quiescent and
star-forming galaxies. Finally, for the C3K/MIST SSP library
(panel~\subref{f.app.h4bias.c}), we have $\langle \Delta\,h_4 \rangle = -0.0054
\pm0.0007$ and $-0.0081\pm0.0010$, and the scatters are 0.015 and 0.018.
In summary, replacing the SSP libraries with a stellar library, we find a lower
overall \h4, more so for quiescent galaxies; this means that, using these
\h4 measurements instead of our default values, we would infer a \textit{smaller}
--- but still statistically significant --- difference between quiescent and
star-forming galaxies. The scatter about the
median offsets are of the order or smaller than the measurement uncertainties. 
In summary, even though the magnitude of our \h4 measurements is dominated by
systematics, our conclusions about the difference \h4 between star-forming
and quiescent galaxies are unchanged if we use any of the other libraries
considered here.

\section{Age bias}\label{app.agebias}

As discussed in \S~\ref{s.das.ss.hok.sss.mbias}, for a meaningful comparison of \h4 between
star-forming and quiescent galaxies, we need to understand the effect of
systematic differences in the spectra of these two classes of objects.
To this end, we use two mock spectra,
representing an idealised pair of a quiescent and a star-forming galaxy. For
the quiescent galaxy, we assume a constant star-formation rate $SFR>0$ from the
Big Bang until a look-back time of 10.25~Gyr ($z\approx2$), followed by $SFR=0$. For
the star-forming galaxy, we swap the intervals: $SFR=0$ from the Big Bang until
a look-back time of 10.25~Gyr, then constant $SFR>0$. For each of these two
spectra, we create a version with $h_4=0$, and a version with $h_4 = 0.06$,
so we have four models in total. For each model, we then create one thousand
random-noise realisations with $S/N = 20$~\AA$^{-1}$ and Gaussian noise. We then
calculate $\langle
\Delta \,h_4 \rangle$, the median offset between the measured and input values
of \h4. For $h_4=0$, we find $\langle \Delta \,h_4 \rangle = -0.0025\pm0.0004$\footnote{
Note that, for the quiescent mocks, $\langle \Delta \,h_4 \rangle$ is not
consistent with 0, i.e. we recover a biased value of \h4. The reason for this
bias is not clear; we note that we oversampled the spectra by a factor of $\times 10$ to
safeguard against resampling errors, and use the same templates to create the
mock and as input to \ppxf to safeguard against template mismatch.}
and $\langle \Delta \,h_4 \rangle = -0.0008\pm0.0024$ for the quiescent and
star-forming spectra. For $h_4=0.06$, we find
$\langle \Delta \,h_4 \rangle = -0.0028\pm0.0005$ and
$\langle \Delta \,h_4 \rangle = -0.0021\pm0.0015$ for the
quiescent and star-forming spectra. Even though some of these offsets are
statistically significant, their magnitude is $\times$10 smaller than the maximum
measurement uncertainties used for the quality cut ($u(h_4)<0.05$). As for the
scatter, the standard deviation for the quiescent
mocks is 0.010 (for $h_4=0$) and 0.017 (for $h_4=0.06$). These values are
smaller than our threshold measurement uncertainties. For the star-forming
mocks, the values are 0.075 (for $h_4=0$) and 0.046 (for $h_4=0.06$), which
are comparable to our cut in the measurement uncertainties.

We remark here that real galaxies show clear evidence of correlation between
the stellar population properties (age and chemical abundance) and kinematics.
This is true not only in the Milky Way \citep[e.g.][]{freeman+bland-hawthorn2002,
nordstrom+2004, rix+bovy2013, bland-hawthorn+gerhard2016},
but is well established in external galaxies too \citetext{e.g.,
\citealp{poci+2019}, \citealp{shetty+2020b}, and \citealp{foster+2023}}.
For this reason, the tests we performed here are only a first-order approximation
of the \h4 bias between star-forming and quiescent galaxies.

\section{Selection bias against low-mass LEGA-C star-forming galaxies}\label{app.sel}

Our quality selection (\S~\ref{s.samp.ss.qcsel}) results high completeness, except for the
LEGA-C star-forming subset (\S~\ref{s.samp.ss.samp}, Fig.~\ref{f.samp.samp.f}).
Incompleteness is most severe below $M_\star = 10^{11} \, \mathrm{M_\odot}$.
Based on the \h4--\mstar correlation, we expect that --- if it were possible to
include not-selected star-forming galaxies in our analysis, the difference
between star-forming and quiescent galaxies would be even stronger.

However, we cannot exclude that star-forming galaxies that do not meet our
quality cut may be biased in other properties in addition to \mstar; what if
these biases were to act opposite to the \mstar bias?

\begin{figure}
  \includegraphics[type=pdf,ext=.pdf,read=.pdf,width=1.\columnwidth]{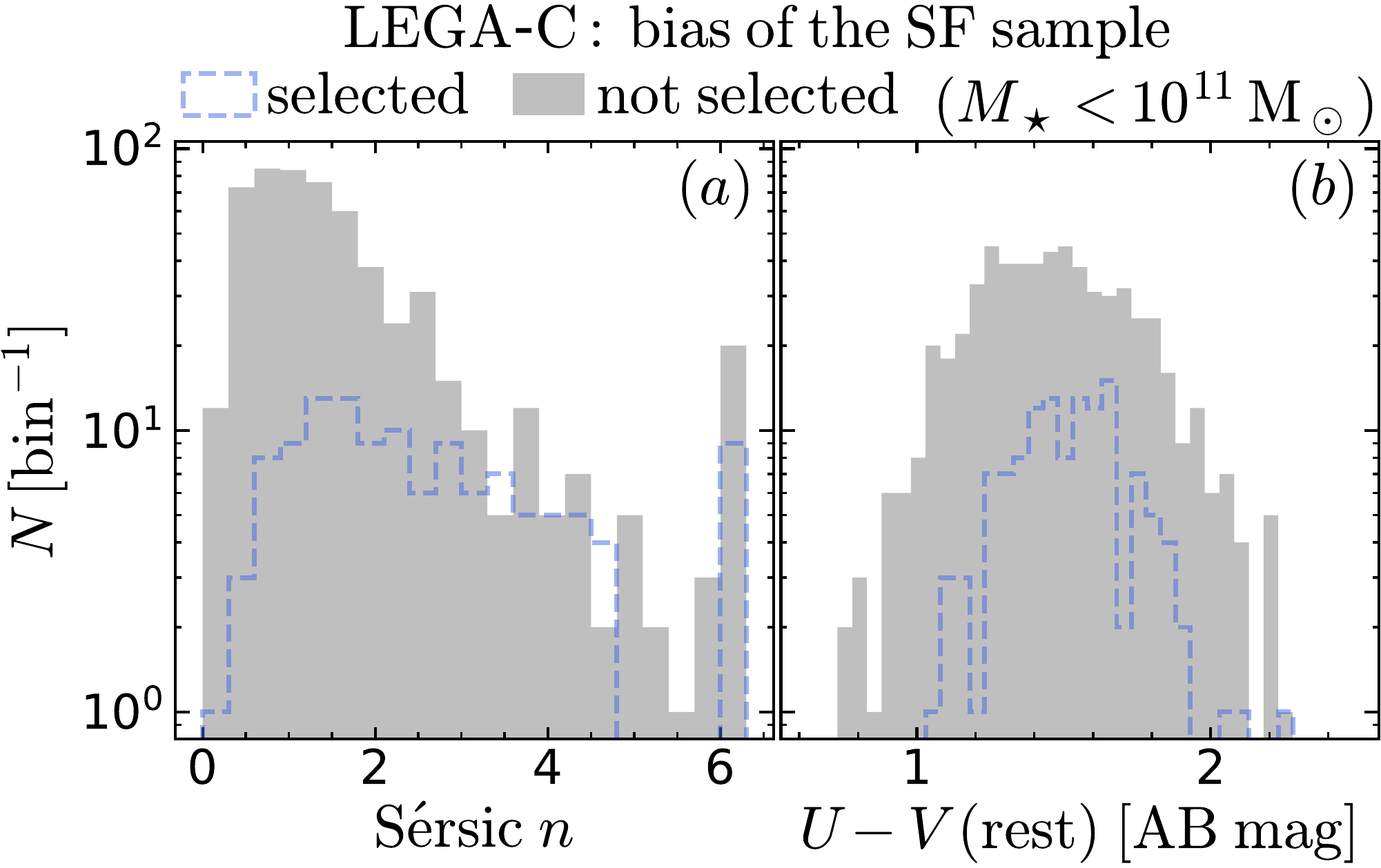}
  {\phantomsubcaption\label{f.app.sfbias.hist.a}
   \phantomsubcaption\label{f.app.sfbias.hist.b}}
  \caption{Below stellar mass $10^{11} \, \mathrm{M_\odot}$, the star-forming 
  galaxies we select from LEGA-C differ systematically from their parent sample,
  in both structure (as highlighted by the S\'ersic index $n$,
  panel~\subref{f.app.sfbias.hist.a}) and stellar-population properties (expressed
  by the rest-frame $U$-$V$ colour, panel~\subref{f.app.sfbias.hist.b}).
  Based on the trends between \h4 and \mstar, we expect not-selected star-forming
  galaxies to have lower \h4 than selected star-forming galaxies; a higher
  completeness sample would probably reinforce our results. Differences in
  concentration and colour appear to be second-order effects, at best}\label{f.app.sfbias.hist}
\end{figure}

In Fig.~\ref{f.app.sfbias.hist} we show that --- below $10^{11}
\, \mathrm{M_\odot}$ --- selected and not-selected star-forming galaxies differ
in both structure (panel~\subref{f.app.sfbias.hist.a}) and
stellar populations/dust (panel~\subref{f.app.sfbias.hist.b}). In the first
panel~\subref{f.app.sfbias.hist.a}, the distribution of S\'ersic index $n$ of
not-selected galaxies (grey histogram) clearly peaks at $n=1$, with a small tail
to higher $n$'s; in contrast, selected star-forming galaxies (dashed blue
histogram) peak at $n=1.5$ and have a broader distribution; our selection is
biased against low-concentration, discy systems. In 
panel~\subref{f.app.sfbias.hist.b}, the distribution of not-selected galaxies is
bluer, and extends also to significantly redder objects; so our
selection is biased against both the youngest and most dust-obscured star-forming
galaxies. However, using PCCs (\S~\ref{s.r}), we find that --- after controlling
for \mstar --- $n$ correlates with \h4 and $U$-$V$ shows no independent correlation
with \h4. This means that --- had we accounted for the bias against low-$n$ galaxies
--- the reported differences between star-forming and quiescent galaxies would be
even larger.


\bsp	
\label{lastpage}
\end{document}